\documentclass{article} 
\usepackage[final]{colm2026_conference}

\usepackage{microtype}
\usepackage{hyperref}
\usepackage{url}
\usepackage{booktabs}

\usepackage{lineno}
\usepackage{amsmath}
\usepackage{amssymb}
\usepackage{mathtools}
\usepackage{amsthm}

\usepackage{amsmath,amsfonts,bm}

\newcommand{\start}[1]{\vspace{.3mm}\noindent{{\bf #1}.}}

\def\eqref#1{equation~\ref{#1}}

\def\1{\bm{1}}

\DeclareMathAlphabet{\mathsfit}{\encodingdefault}{\sfdefault}{m}{sl}
\SetMathAlphabet{\mathsfit}{bold}{\encodingdefault}{\sfdefault}{bx}{n}

\usepackage{hyperref}
\usepackage{url}
\usepackage{amsmath}
\usepackage{graphicx}
\usepackage{booktabs}
\usepackage{multirow}
\usepackage{amssymb}
\usepackage{amsmath}
\usepackage{amssymb}
\usepackage{dsfont}
\usepackage{amsfonts}
\usepackage{caption} 
\usepackage{enumitem}
\usepackage{subcaption}
\usepackage{tipa}
\usepackage{tabularray}
\usepackage{pgfplots}
\usepackage{tcolorbox}
\usepgfplotslibrary{groupplots} 
\pgfplotsset{compat=1.18}
\usepackage{xcolor} 

\definecolor{mygreen}{RGB}{0, 128, 0}   
\definecolor{myred}{RGB}{180, 0, 0}

\definecolor{darkblue}{rgb}{0, 0, 0.5}
\hypersetup{colorlinks=true, citecolor=darkblue, linkcolor=darkblue, urlcolor=darkblue}

\theoremstyle{plain}

\theoremstyle{definition}

\theoremstyle{remark}

\title{MetaLint: Easy-to-Hard Generalization for Code Linting}

\author{Atharva Naik, Lawanya Baghel, Dhakshin Govindarajan \\ \textbf{Darsh Agrawal, Yiqing Xie, Daniel Fried, Carolyn Rose} \\ Carnegie Mellon University \\ \texttt{\{arnaik, lbaghel, dkundego, darsha, yiqingxi, dfried, cprose\}@cs.cmu.edu}}

\begin{document}

\ifcolmsubmission
\linenumbers
\fi

\maketitle

\begin{abstract}
Large language models excel at code generation but struggle with code linting, particularly in generalizing to unseen or evolving best practices beyond those observed during training. 
We introduce \textsc{MetaLint}, a meta-learning framework that formulates code linting as an instruction-following task, where a model evaluates whether code adheres to a natural language specification of best practices. 
In contrast to prior work that trains models to detect violations from a fixed set of best practices, \textsc{MetaLint} evaluates code against a provided natural language specification, enabling test-time control over which practices to enforce and generalization to unseen or evolving rules without retraining. 
We demonstrate that models trained solely on synthetic data generated from automatic linters still generalize to harder, context-dependent best practices for which such linters are not available.
To evaluate generalization beyond such easy signals, we introduce a human-curated benchmark of hard best practices inspired by Python Enhancement Proposals (PEPs). 
On this benchmark, \textsc{MetaLint} substantially improves performance without explicit fine-tuning on target best practices and exhibits strong easy-to-hard generalization. 
Qwen3-4B achieves a 2.7$\times$ detection F-score gain (25.9\% → 70.4\%), the highest recall, and a 26.7\% localization F-score, matching larger models such as o3-mini. 
These gains generalize across programming languages, model families, scales, reasoning settings, and linter sources. 
We release the code\footnote{https://github.com/atharva-naik/MetaLint/} and benchmark to support reproducibility and future work.
\end{abstract}

\section{Introduction}
\textit{Code linting}, or static analysis to ensure code complies with best practices, is a fundamental tool for improving software quality and reliability. 
Recent work has explored the use of large language models (LLMs) for best-practice-oriented linting \citep{vijayvergiya2024ai, fang2025lintllmopensourceveriloglinting, holden2024code, khare2023understanding, zhang2024refactoring}, demonstrating improvements over traditional rule-based linters by enabling more contextual and semantic reasoning. 

However, existing approaches typically train LLM linters to predict violations from a fixed set of best practices embedded in model parameters. 
In practice, best practices evolve over time as languages, libraries, and security standards change, leading models trained on static rule sets to over-flag outdated patterns \cite{vijayvergiya2024ai} and underperform on rare or emerging ones \cite{holden2024code}. 
This challenge is compounded by the varying difficulty of identifying violations: some are \textbf{easy to detect}, relying on surface-level cues (e.g., Ruff rules S104--S108 that flag likely hard-coded secrets), while others require \textbf{contextual and semantic reasoning} that cannot be reliably captured by rule-based matching. 
For example, PEP 506 \citep{pep506} recommends using \texttt{secrets.choice} instead of \texttt{random.choice} for security-sensitive applications, but detecting such violations requires reasoning about code intent, as not every use of \texttt{random.choice} is unsafe.

\begin{figure*}[t]
    \centering
    \vspace{-0.5cm}
    \includegraphics[width=\textwidth]{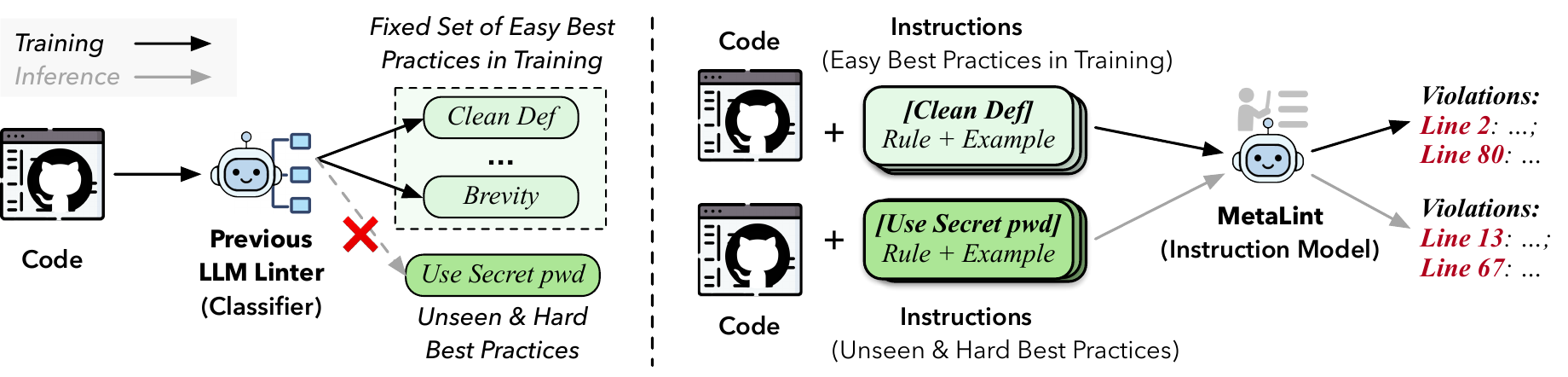}
\caption{MetaLint frames linting as instruction following. It is trained on synthetic linter data and enables generalization to novel and hard best practices without retraining.}
\label{fig:meta_lint_approach}
\end{figure*}

To address these limitations, we introduce \textsc{MetaLint}, a meta-learning framework that \textbf{formulates code linting as an instruction-following task, instead of static classification}, where best-practice specifications are provided in natural language as part of the input rather than fixed labels (Figure~\ref{fig:meta_lint_approach}). 
This enables \textbf{test-time control} over which best practices to enforce, allowing models to adapt to unseen or evolving rules without retraining. 
Unlike fixed-rule classification, this formulation encourages moving beyond memorizing rule-specific patterns to generalization across \textit{code idioms}, or recurring patterns that capture semantic intent beyond surface syntax. 

\textsc{MetaLint} is designed as a \textbf{lightweight specialist} suitable for use within agentic coding pipelines rather than as a replacement for frontier models.
A growing body of work on multi-agent systems \citep{belcak2025small, sharma2025small, liu2026caster, gandhi2024budgetmlagent} advocates decomposing complex tasks across specialized models of varying capability and cost; code linting is a natural fit for this decomposition, requiring little repository context or code execution.
As we show in Section~\ref{sec:results:pep_hard_idiom_benchmark}, MetaLint (4B) is 9.5--21$\times$ cheaper than agentic frontier linting on a real-world codebase while matching or exceeding frontier recall, making it a practical drop-in critic for larger pipelines.

\textsc{MetaLint} constructs scalable synthetic data by leveraging existing linters (e.g., Ruff \citep{ruff_linter}, PMD \citep{pmd_analyzer}) as sources of supervision and reward. Specifically, we use linters to generate labeled data for supervised fine-tuning (SFT) and to score outputs for constructing preference optimization (PO) data via rejection sampling. 
Models are trained using this abundant synthetic supervision via both SFT and PO, enabling evaluation of generalization beyond the best practices seen during training. 
We find that SFT learns the task format but tends to reinforce rule memorization, while PO with linter-derived rewards encourages generalization beyond seen best practices. 
While the training data is derived from rule-based best practices, we empirically observe that models trained in this manner generalize to harder, context-dependent violations.

To evaluate this generalization, we introduce and publicly release a human-curated benchmark of challenging best practices inspired by widely adopted Python Enhancement Proposals (PEPs). 
Unlike prior work using automatically generated or unreleased datasets \citep{holden2024code, zhang2024refactoring, jiang2025coupjava, vijayvergiya2024ai}, our benchmark is human-curated, publicly available, and targets hard, context-dependent violations beyond rule-based detection, explicitly measuring generalization to unseen rules.
Empirically, \textsc{MetaLint}-trained models substantially improve performance on hard best practices without explicit fine-tuning on target rules, exhibiting strong easy-to-hard generalization. 
For example, Qwen3-4B achieves a 2.7$\times$ detection F-score gain (25.9\% $\rightarrow$ 70.4\%),  the highest detection recall among all evaluated models, and a 26.7\% localization F-score, matching larger models like o3-mini. 
These gains generalize across programming languages (Python and Java), model families (Qwen and LLama), scales (3B-8B), reasoning settings, and linters (Ruff, PMD, Tree-Sitter).

\section{Related Work}
\start{Code Linting and Large Language Models}
A large body of prior work has explored using LLMs for code linting by interfacing them with static analysis tools or fixed rule sets.
\citet{agentic2024gptlint, lintrule_2023} treat LLMs as rule-guided linters via prompting or fine-tuning.
While \citet{blyth2025staticanalysisfeedbackloop} proposes a static analysis-driven prompting framework to improve LLM-generated code, \citet{du2025minimizingfalsepositivesstatic} conversely uses LLMs to enhance static analysis tools by reducing false-positives.
\citet{fang2025lintllmopensourceveriloglinting, Shin2025QuantumPL, khare2023understanding} leverage LLMs for linting and show they can outperform traditional static analysis tools. 
\citet{vijayvergiya2024ai} train LLMs for best practice violation detection and localization.
\citet{naik2024crscore, kapadnis2025crscore++, jaoua2025combining} provide linter results to LLMs for more informative code reviews.
\citet{ridiom_2024, zhang2024refactoring} explore AST rewrite rules and hybrid approaches combining LLMs and rules for code linting and refactoring. 
Finally, CoUpJava \citep{jiang2025coupjava} proposes a Java version upgrade benchmark, mined using AST refactoring rules.
Across these approaches, training, when employed, typically targets a fixed set of best practices, making it difficult to adapt to evolving or previously unseen rules without retraining or additional rule-based post-processing. 
Furthermore, existing datasets are either automatically constructed using linters \citep{holden2024code, zhang2024refactoring, jiang2025coupjava} or are not publicly released \citep{vijayvergiya2024ai}, limiting their ability to capture context-dependent violations.
In contrast, we propose a training framework designed to generalize beyond fixed rule sets, along with a publicly released benchmark of manually curated, hard-to-detect best practices that require contextual reasoning.

\start{Easy-to-Hard Generalization}
Recent work has explored easy-to-hard generalization in math and coding adjacent domains.
In mathematical reasoning, models trained on easier problems (levels 1–3) generalize better to harder benchmarks (levels 4–5) \citep{Bai2024EfficientMAB, Shafayat2025CanLRG, Parashar2025CurriculumRLC}, with high-quality supervision particularly important for difficult instances \citep{He2024GuidingTCE}. Reward models trained on simple code and math problems also improve performance on complex tasks \citep{Sun2024EasytoHardGSA}. More broadly, training on structurally simpler instances yields robustness to longer or more complex reasoning problems, including code \citep{Hu2025BeyondSRD, Gaunt2016DifferentiablePWH}, and reward models exhibit transfer from simpler to more complex algorithmic tasks \citep{Zhang2024GenerativeVRI}. 
Most closely, \citet{chang2025model} argue that easy-to-hard generalization is driven primarily by the learning paradigm and hypothesis class rather than data scarcity, model scale, or inference limits. 
While prior work studies this in domains such as counting, mathematics, or contest coding, code linting offers a complementary setting where correctness is largely objective but supervision becomes increasingly incomplete along the difficulty progression. Additional discussion on leveraging instruction following for generalization is in Appendix~\ref{sec:appendix:more_related_work}.

\section{The \textsc{MetaLint} Framework} 
\begin{figure*}[t]
    \centering
    \vspace{-0.5cm}
    \includegraphics[width=\textwidth]{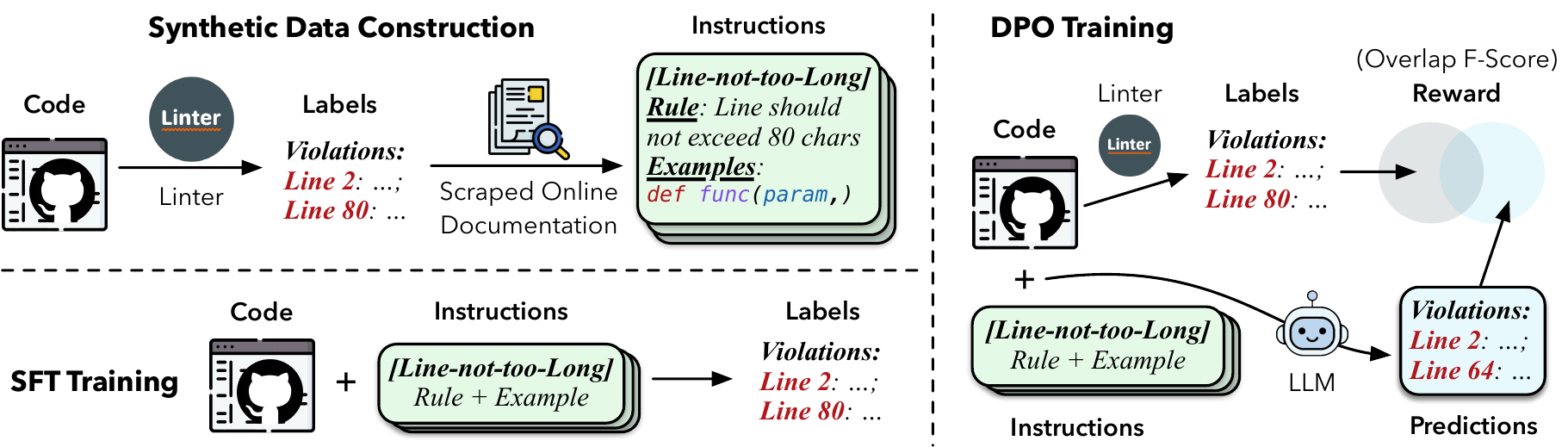}
    \caption{\textbf{\textsc{MetaLint}:} (1) Synthetic data generation with linters/tools, (2) SFT Training on this data, and (3) Verifiable Reward derived from the linter for DPO Training.}
    \vspace{-0.2cm}
\label{fig:meta_lint_training_framework_part1}
\end{figure*}

We design the \textsc{MetaLint} framework to reorganize supervision for linting by shifting from memorizing fixed rules to identifying violations from high-level natural language descriptions of problematic code idioms, in line with recent work showing that such reorganization is critical for generalization \citep{chang2025model}.
At a high level, \textsc{MetaLint} treats each best practice as a natural language task and trains models on a diverse set of such tasks derived from verifiable rule-based idioms.
This design decouples best practices from model parameters and represents them in the input, while leveraging abundant supervision on easy idioms to learn reusable patterns that can transfer to harder, context-dependent cases.
We describe the components of our training framework below (Figure~\ref{fig:meta_lint_training_framework_part1}, \ref{fig:meta_lint_training_framework_part2}).

\subsection{Problem Formulation: Code Linting as a Meta-Learning Instruction-Following Task}
\label{sec:definition}
We formulate best practice violation detection as an instruction-following \textit{meta-task} $M_I$, where for a given problematic idiom $I$, the prompt includes a natural language description $D_I$ and illustrative examples $E_I$, denoted as $M_I = \{D_I, E_I\}$.
The LLM must identify all and only code fragments matching idiom $I$.
This setup discourages rote memorization by construction, as correctness is defined relative to the given specification, and flagging violations of any other idiom $I' \neq I$ is penalized during $M_I$.
By framing best practices as meta-tasks, this approach supports flexibility under evolving best practices. 
This formulation differs from classification-based linting, where best practices are implicitly encoded in model parameters. 
Conditioning on explicit specifications requires the model to interpret and apply them at inference time, supporting adaptation to unseen best practices.

Under the easy-to-hard generalization setting, we train LLMs on a set of ``easy'' best practice idioms $I_{\mathcal{L}}$ that are detectable by existing linters $\mathcal{L}$, and evaluate them on a harder set $I_{\mathcal{L'}}$ consisting of idioms that linters cannot detect (where $\mathcal{L'}$ denotes the complement of $\mathcal{L}$, i.e., all idioms not detectable by a linter). 

\subsection{Synthetic Data Generation}
\label{sec:method:synth_data_gen}
We hypothesize that training on $I_{\mathcal{L}}$ exposes the model to a diverse family of best practice specifications expressed in natural language, enabling it to learn reusable patterns for applying such specifications. 
This is analogous to instruction tuning, where models trained on many tasks generalize to unseen instructions \citep{Sanh2021MultitaskPT, Wang2022SuperNaturalInstructionsGVA, chung2024scaling}. 
In our setting, linter-detectable idioms provide scalable supervision for such task-conditioned learning.
Since idioms in $I_{\mathcal{L}}$ are covered by linters, we leverage them to generate large-scale synthetic data.

For Python, we use the Ruff linter (800+ rules), and for Java, PMD (269 idioms) along with tree-sitter queries inspired by 8 JEPs (Table~\ref{tab:jeps_included}). 
We run these tools on Python and Java source code files $f \in \mathcal{F}$ from the STACK \citep{lozhkov2024starcoder} dataset, which contains code from a diverse range of GitHub repositories. 
This allows us to collect files with either no violations or one or more violations for each idiom in $I_{\mathcal{L}}$.
Ruff also incorporates rules from other linters such as PyFlakes, Bandit, and autoPEP8, making it well-suited for producing diverse and representative synthetic data.

To construct meta-task prompts $M_{I_{\mathcal{L}}}$, we scrape rule documentation from Ruff and PMD, including descriptions and examples.
An example prompt, along with a code file containing lines that violate the idiom, is shown in Appendix~\ref{sec:appendix:method:metalint_prompt_io}.
For the JEP tree-sitter queries, since they are few in number, we manually write the meta-task prompts.
We balance positive (violation) and negative (no-violation) examples to avoid biasing the model toward over or under-flagging, supporting a better precision–recall tradeoff. 
Due to the rarity of some Ruff idioms, the final distribution is approximately 3:1 no-violation to violation for Python, and roughly balanced for Java (PMD and JEP tree-sitter). 
This yields 53k Python instances (50 idioms), 96.8k Java PMD instances (269 idioms), and 127.3k tree-sitter instances (15 idioms).

\subsection{Instruction Supervised Fine-Tuning}
\label{sec:method:ift}
Instruction fine-tuning aligns model outputs with natural language specifications using linter outputs as supervision for easy cases.
We train the target LLM $\Phi$ on a set of linter-detectable, easy best practice idioms $I_{\mathcal{L}}$, using the corresponding meta-task specifications $M_{I_{\mathcal{L}}}$ and a set of source code files $\mathcal{F}$. 
The input prompt $p$ combines a meta-task $M_I$ with a code file $f \in \mathcal{F}$.
The model's output is a list of best practice violations in the file, denoted as $V_{f,I}$, formatted as a JSON list with one violation per line (see example output in Appendix~\ref{sec:appendix:method:metalint_prompt_io}). 
In cases where there are no violations ($|V_{f,I}| = 0$), the model is expected to output the phrase \texttt{NO VIOLATIONS FOUND}. 
This penalizes hallucinated or memorized violations, encouraging adherence to the provided specification.
By training on many such meta-tasks, the \textbf{model learns to align outputs with task specifications} rather than fixed rules, which supports applying unseen best practices at inference time.

\subsection{Verifiable Reward and Preference Optimization}
We use preference optimization to improve consistent application of specifications by contrasting accurately localized violations with inaccurate ones based on linter results.
While SFT aligns outputs with meta-task specifications, preference optimization enforces precise localization and reduces spurious and partial matches.
We adopt the RS-DPO approach \citep{khaki2024rs}, which combines rejection sampling (RS) \citep{touvron2023llama} with Direct Preference Optimization (DPO) \citep{rafailov2023direct} to generate on-policy data from a supervised fine-tuned (SFT) policy model. 
It samples $k$ outputs per input, scores them, and constructs contrastive pairs based on a threshold $\eta$ (Figure~\ref{fig:meta_lint_training_framework_part2}).
We detail the verifiable linter-based reward and contrastive pair sampling procedure below.

\start{Reward Function Design}  
The reward function evaluates model outputs by comparing predicted violations against those flagged by the linter, treating the linter's line numbers (blue circle in ``Verifiable Reward'', Figure~\ref{fig:meta_lint_training_framework_part1}) as ground truth and the model’s predicted lines (yellow circle) as predictions. 
Reward is the set-based F1-score (visualized via the Venn diagram in the same figure), based on line-level overlap. 
Since each meta-task $M_I$ corresponds to a single best practice idiom $I$, we compute one reward value per instance. 
This per-idiom setup allows tracking difficulty of individual idioms; in practice, multiple idioms could be processed in parallel.

\start{Sampling Contrastive Pairs}  
\label{sec:model:sampling_contrastive_pairs}
We begin with an SFT policy model $\Phi^{SFT}$ and sample $k=5$ outputs $y_i$, $i \in \{1, \dots, k\}$ for each input $x$, using a range of temperature values $\tau = \{0, 0.3, 0.5, 0.7, 1.0\}$ to promote output diversity. 
Each response $y_i$ receives a reward $r_{y_i}$, and for each pair $(y_i, y_j)$, we compute the reward gap $|r_{y_i} - r_{y_j}|$. 
Pairs with a gap greater than the threshold $\eta = 0.2$ (based on results in Table~\ref{tab:non_cot_eta_ablations}) are added to the preference dataset $\mathcal{D}_p$. 
For any such pair where $r_{y_i} \geq r_{y_j} + \eta$, we assign $y_{w} = y_i$, $y_{l} = y_j$, and store the instance $(x, y_{w}, y_{l}) \in \mathcal{D}_p$.
By constructing contrastive pairs within the same meta-task, RS-DPO encourages the model to apply a given specification consistently across variations in code structure, rather than across best practices.
Following \citet{khaki2024rs}, we train the preference-tuned model $\Phi^{RL}$ using the DPO objective:
\[
\begin{aligned}
\Phi^{RL}
= \arg\max \sum_{(x,y_{w},y_{l}) \in \mathcal{D}_p}
\log \sigma \Big(
&\beta \log \frac{\Phi^{RL}(y_{w}\mid x)}{\Phi^{SFT}(y_{w}\mid x)} - \beta \log \frac{\Phi^{RL}(y_{l}\mid x)}{\Phi^{SFT}(y_{l}\mid x)}
\Big)
\end{aligned}
\]
Here, $\sigma$ denotes the sigmoid function, and $\beta = 0.1$ is the KL penalty coefficient, corresponding to low-to-moderate regularization.

\subsection{Training with Reasoning Traces}
Reasoning traces provide intermediate representations that may support transfer across idioms.
We generate CoT traces via rejection sampling (Figure~\ref{fig:meta_lint_training_framework_part2}). 
For each input $x$, we sample $k=5$ responses $y_i$ from a CoT-capable LLM (e.g., Qwen3-4B) and compute rewards as in RS-DPO. 
We discard any $y_i$ with $r_{y_i} < \gamma$, where $\gamma = 1$, i.e., response after parsing the CoT is incorrect or improperly formatted. 
We also remove cases where the CoT fails to terminate or yield an answer.
If no valid $y_i$ is found for an input $x$, we skip it. 
To promote meta-task diversity, we retain at most two valid responses per input: multiple $y_i$ only for violation cases and a single $y_i$ otherwise. This maintains a similar no-violation-to-violation ratio as the Ruff Python SFT data (for more controlled comparison), with the latter more likely to fit within token limits. 
When excess valid responses exist, we keep the shortest completions, as they typically reflect more concise reasoning (final answers are of similar token length across samples). 

Following this policy, we collect 52.7k Python training instances from Ruff data, which we use to train the reasoning-enabled base Qwen3-4B with SFT. 
This yields a CoT-capable SFT model $\Phi_{CoT}^{SFT}$ for Python code linting. 
We then apply the RS-DPO procedure in Section~\ref{sec:model:sampling_contrastive_pairs} and Figure~\ref{fig:meta_lint_training_framework_part2}, with the only change being that each $y_i$ now includes both the CoT trace and final response. 
These reasoning traces provide intermediate representations that may support transfer across idioms by making the decision process more explicit.

\section{PEP Hard Best Practice Benchmark for Code Linting}
\label{sec:methods:pep_idiom_benchmark}
To evaluate reasoning over hard-to-detect best practices, we construct a benchmark from 15 Python Enhancement Proposals (PEPs) that require abstract, context-dependent analysis. 

Unlike prior work, which either relies on automatically generated datasets from linters \citep{holden2024code, zhang2024refactoring, jiang2025coupjava} or does not release evaluation data \citep{vijayvergiya2024ai}, our benchmark is manually curated and publicly released.
We first design high-recall heuristics for each PEP (Appendix~\ref{tab:appendix:experiments:pep_benchmark_heuristics_part1}, \ref{tab:appendix:experiments:pep_benchmark_heuristics_part2}, \ref{tab:appendix:experiments:pep_benchmark_heuristics_part3}) to retrieve candidate violations from STACK-V2. Training and benchmark examples were deduplicated using file-level content hash IDs embedded in both dataset identifiers, ensuring no file-level overlap between the STACK training data and STACK-V2 benchmark candidates. 
These candidates are then manually filtered and verified to ensure correctness. 
Because these violations cannot be reliably captured through rule-based matching, they provide a challenging testbed for contextual reasoning.
Critically, the chosen instances are outside the scope of the Ruff linter and evaluating \textsc{MetaLint} on it corresponds to a regime, where rule-based linters have been applied and found nothing.
For each PEP, we select 15--20 representative files and annotate precise violating line ranges for localization. 
Negative examples are sampled from other PEPs to construct a balanced dataset (52\% violation, 48\% no-violation) comprising 536 examples. 
We further validate annotation reliability through statistical significance testing and inter-annotator agreement (Cohen's $\kappa = 0.95$; Appendix~\ref{sec:appendix:results:stat_signficance}, \ref{sec:appendix:pep_benchmark_creation_additional_details}), providing a direct test of generalization from linter-detectable idioms to harder, unseen best practices.
\section{Experiments}
We evaluate the effectiveness of \textsc{MetaLint} along three dimensions: (1) generalization to unseen easy best practices (Section~\ref{sec:results:gen_synth_data_easy_idioms}); (2) easy-to-hard generalization across best practices (Section~\ref{sec:results:gen_easy_to_hard_pep}); and (3) whether \textsc{MetaLint} enables a 4B model to achieve performance competitive with state-of-the-art code and reasoning models (Section~\ref{sec:results:pep_hard_idiom_benchmark}).

\subsection{Experimental Setup}
\start{Evaluation Metrics}
We evaluate the ability of code models to detect best practice violations through two tasks: \textit{detection}, which assesses whether a best practice is violated at least once in a code file, and \textit{localization}, which evaluates whether the model accurately identifies the lines with the problematic idiom. 
For both tasks, we report precision, recall, and F-score metrics. 
Detection metrics are calculated at the corpus level for each best practice idiom, treating each as a separate class, while localization metrics are computed at the instance level using set-based precision, recall, and F-score between ground-truth and predicted sets of violating line numbers. 
To handle potential class imbalance, we use macro-averaging across idioms and exclude \texttt{NO VIOLATION} as a class to penalize models that only predict \texttt{NO VIOLATIONS FOUND} (such models will score zero on all detection metrics). 
For localization, metrics are averaged only across instances with at least one violation in the ground truth.
Details of the formal definitions and exact computations of precision, recall, and F-scores for detection and localization are provided in Appendix~\ref{sec:appendix:experiments:eval_metrics_math}. 

\start{Generalization to Novel Best Practices}
To evaluate whether \textsc{MetaLint} training encourages transfer to novel, easy-to-detect best practices in Python and Java, we construct the following test sets with linters:
\\
\textbf{(1) Ruff Python Idioms.} We construct a 5.3k-instance test set with the Ruff linter spanning 50 idioms, generated as in Section~\ref{sec:method:synth_data_gen}. The set is approximately balanced, though some rare idioms result in a 3:1 no-violation-to-violation split. Idioms vary in overlap with training (Figure~\ref{fig:transfer_eval_design}) and fall into three categories: \textbf{In domain} (idioms seen during training, for in-domain evaluation), \textbf{Near transfer} (idioms with specifications similar but not identical to training, probing memorization), and \textbf{Far transfer} (idioms distinct from training, testing adaptation to novel specifications) as detailed in Table~\ref{tab:train_test_idioms}. We evaluate \textsc{MetaLint} trained Qwen3-4B (with and without CoT using $<$think$>$ tokens) and Llama-3.2-3B-Instruct to study the effect of test-time compute and model family.
\\
\textbf{(2) PMD and JEP Tree-Sitter Idioms.} For Java, we construct two test sets using PMD and Tree-Sitter: 5.1k instances spanning 269 PMD idioms and 6.4k instances spanning 15 JEP idioms (Table~\ref{tab:jep_idiom_specs}), both with near-balanced no-violation-to-violation splits. We evaluate in-domain performance by training base LLMs on the corresponding linter-generated training sets (Section~\ref{sec:method:ift}) and study bidirectional transfer between PMD and JEP. We train Llama-3.2-3B-Instruct and Llama-3.1-8B-Instruct to assess model-scale effects.

\start{Generalization from Easy to Hard Best Practices}
\label{sec:experiments:pep_idiom_benchmark}
We evaluate whether \textsc{MetaLint}-trained models exhibit easy-to-hard generalization using the \textbf{manually curated} PEP-based benchmark described in Section~\ref{sec:experiments:pep_idiom_benchmark}. 
Although some PEPs or code snippets may appear in pre-training, we compare against the corresponding pre-\textsc{MetaLint} base models to isolate gains beyond pre-training.
We compare base, SFT, and DPO-trained \textsc{MetaLint} models to assess the impact of training on easy best practices for performance on hard, context-dependent violations.
We further benchmark against state-of-the-art open- and closed-source code and reasoning LLMs, including instruction-tuned Qwen2.5 \citep{qwen2}, Qwen2.5Coder \citep{hui2024qwen2}, DeepSeek-R1-Distill-Qwen \citep{deepseekai2025deepseekr1incentivizingreasoningcapability}, Qwen3 \citep{qwen3technicalreport}, GPT-oss 20B/120B \cite{agarwal2025gpt}, GPT-4o \citep{hurst2024gpt}, o3 mini and o4 mini \citep{o3_and_o4_mini_system_card}, GPT-4.1 \citep{openai_gpt4.1}, and GPT-5 \cite{openai_2025}, spanning diverse model scales (3B--120B), training paradigms, and test-time compute.

\subsection{Results on Generalization across Easy Best Practices}
\label{sec:results:gen_synth_data_easy_idioms}
\start{Python Ruff Idioms} The performance of Qwen3-4B with and without reasoning and Llama3.2-3B-Instruct when trained on synthetic Ruff idioms and evaluated on the Ruff synthetic test set with varying transfer settings 
(section~\ref{sec:methods:pep_idiom_benchmark}) is shown in Table~\ref{tab:gen_synth_python_ruff_idioms} (full results in Table~\ref{tab:gen_synth_python_ruff_idioms_full}). 
While Table~\ref{tab:gen_synth_python_ruff_idioms} shows the overall performance, we also analyze the performance broken down by each transfer setting in Table~\ref{tab:gen_synth_python_ruff_idioms_per_transfer_setting} (full results in Table~\ref{tab:gen_synth_python_ruff_idioms_per_transfer_setting_full}) .
The results show that the SFT stage leads to modest gains in detection and localization performance in most cases (except for a detection recall drop for Llama3.2-3B-Instruct), but the DPO stage leads to huge gains in detection recall, F-score, and all localization metrics with a slight drop in detection precision. 
We identify that the drop in precision in the DPO stage is tightly controlled by the fraction of cases with no violations used in the DPO training and explore it in detail in Appendix~\ref{sec:appendix:experiments_dpo_nv_fraction_ablations}. 
Additionally, Table~\ref{tab:gen_synth_python_ruff_idioms_per_transfer_setting} and \ref{tab:gen_synth_python_ruff_idioms_per_transfer_setting_full} show that while SFT can lead to slight gains for the transfer settings, most gains emerge in the DPO stage, especially for non-reasoning models and detection recall. 
Overall, this suggests that SFT is prone to memorizing best practices used for training, while DPO generalizes to novel best practices.

\start{PMD and JEP Tree-Sitter Idioms} To evaluate the generality of \textsc{MetaLint} training across programming languages and linters, we present results from training on PMD and JEP Tree-Sitter synthetic data in Table~\ref{tab:gen_synth_jep_pmd_experiments_subset} (full results in Table~\ref{tab:gen_synth_jep_pmd_experiments}). 
Training on PMD shows the same overall pattern as before but with larger recall gains for both SFT and DPO, and notably stronger localization under DPO. 
For Llama3.1‑8B‑Instruct, SFT initially reduces detection precision, which DPO then recovers; the same precision dip‑and‑recovery appears when transferring PMD→JEP for Llama3.2‑3B‑Instruct. 
Despite never seeing JEP idioms during training, DPO models achieve strong detection and localization on JEP. 
In the untrained setting, Llama3.2‑3B‑Instruct (on PMD) and Llama3.1‑3B‑Instruct (on JEP) nearly always output the correct format but predict \texttt{NO VIOLATIONS FOUND}, yielding zero or near‑zero scores, since our metrics exclude that class for detection and only score positive cases for localization. 
Training on JEP yields high in‑domain performance for all metrics with minimal additional benefit from DPO, likely due to JEP’s smaller idiom set (15 vs 269 for PMD) and more precise instructions (Table~\ref{tab:jep_idiom_specs}). 
In the harder JEP→PMD transfer, DPO outperforms SFT, though overall transfer remains weaker than PMD→JEP, reflecting PMD's broader diversity and more challenging specifications (Appendix~\ref{sec:appendix:example_pmd_idiom_spec}).\\
Overall, \textsc{MetaLint} outperforms the base model on novel best practice violations, though gains depend on the diversity of training idioms and the gap in instruction quality between training and test data.

\begin{table*}[!tbh]
\centering
\resizebox{0.8\linewidth}{!}{%
\begin{tabular}{@{}lcccccc@{}}
\toprule
\multirow{2}{*}{\textbf{Model}} & \multicolumn{3}{c}{\textbf{Detection (\%)}} & \multicolumn{3}{c}{\textbf{Localization (\%)}} \\ 
\cmidrule(l){2-4} \cmidrule(l){5-7} 
 & \textbf{$P_{Det}$} & \textbf{$R_{Det}$} & \textbf{$F_{Det}$} & \textbf{$P_{Loc}$} & \textbf{$R_{Loc}$} & \textbf{$F_{Loc}$} \\ \midrule
Qwen3-4B & 53.80 & 26.37 & 35.39 & 13.96 & 14.79 & 14.36 \\
\quad + \textsc{MetaLint} (SFT) & \textbf{76.86} & 31.78 & 44.97 & 29.76 & 29.60 & 29.68 \\
\quad + \textsc{MetaLint} (SFT + RS-DPO) & 74.69 & \textbf{83.15} & \textbf{78.69} & \textbf{65.27} & \textbf{66.96} & \textbf{66.11} \\
\midrule
Qwen3-4B w CoT & 88.12 & 68.54 & 77.10 & 50.49 & 48.78 & 49.62 \\
\quad + \textsc{MetaLint} (RS-SFT) & \textbf{93.50} & 81.83 & 87.27 & 66.39 & 65.00 & 65.69 \\
\quad + \textsc{MetaLint} (RS-SFT + RS-DPO) & 92.34 & \textbf{86.43} & \textbf{89.29} & \textbf{77.10} & \textbf{75.71} & \textbf{76.40} 
\\ \bottomrule
\end{tabular}
}
\caption{\textbf{Cross-Idiom Generalization on Python Ruff Idioms:} Effect of different \textsc{MetaLint} training setups (SFT, RS-SFT, \& RS-DPO) on Qwen3-4B (w \& w/o CoT). Best scores bolded.}
\label{tab:gen_synth_python_ruff_idioms}
\end{table*}

\begin{table}[!tbh]
\centering

\resizebox{\textwidth}{!}{%
\begin{tabular}{@{}lccccccccc@{}}
\toprule
\multirow{2}{*}{\textbf{Model}} & \multicolumn{3}{c}{\textbf{In-Domain (\%)}} & \multicolumn{3}{c}{\textbf{Near Transfer (\%)}} & \multicolumn{3}{c}{\textbf{Far Transfer (\%)}} \\ \cmidrule(l){2-4} \cmidrule(l){5-7} \cmidrule(l){8-10}  
 & \textbf{$P_{Det}$} & \textbf{$R_{Det}$} & \textbf{$F_{Det}$} & \textbf{$P_{Det}$} & \textbf{$R_{Det}$} & \textbf{$F_{Det}$} & \textbf{$P_{Det}$} & \textbf{$R_{Det}$} & \textbf{$F_{Det}$} \\ \midrule

Qwen3-4B & 45.0 & 14.3 & 21.7 & 58.2 & 23.5 & 33.5 & 54.3 & 28.8 & 37.6 \\

\quad + \textsc{MetaLint} (SFT) & \textbf{93.2 {\color{mygreen}(+48.2)}} & 74.5 {\color{mygreen}(+60.2)} & 82.8 {\color{mygreen}(+61.1)} & \textbf{89.3 {\color{mygreen}(+31.1)}} & 23.7 {\color{mygreen}(+0.2)} & 37.5 {\color{mygreen}(+4.0)} & 72.0 {\color{mygreen}(+17.7)} & 26.8 {\color{myred}(-2.0)} & 39.0 {\color{mygreen}(+1.4)} \\

\quad + \textsc{MetaLint} (RS-DPO) & 71.6 {\color{mygreen}(+26.6)} & \textbf{99.8 {\color{mygreen}(+85.5)}} & \textbf{83.4 {\color{mygreen}(+61.7)}} & 76.0 {\color{mygreen}(+17.8)} & \textbf{79.9 {\color{mygreen}(+56.4)}} & \textbf{77.9 {\color{mygreen}(+44.4)}} & \textbf{74.9 {\color{mygreen}(+20.6)}} & \textbf{81.2 {\color{mygreen}(+52.4)}} & \textbf{77.9 {\color{mygreen}(+40.3)}} \\ \midrule

Qwen3-4B w CoT & 86.8 & 49.8 & 63.2 & 94.8 & 88.0 & 91.3 & 87.0 & 67.7 & 76.2 \\

\quad + \textsc{MetaLint} (RS-SFT) & \textbf{86.9 {\color{mygreen}(+0.1)}} & 73.3 {\color{mygreen}(+23.5)} & 79.5 {\color{mygreen}(+16.3)} & \textbf{97.3 {\color{mygreen}(+2.5)}} & 86.2 {\color{myred}(-1.8)} & 91.4 {\color{mygreen}(+0.1)} & \textbf{93.8 {\color{mygreen}(+6.8)}} & 82.3 {\color{mygreen}(+14.6)} & 87.7 {\color{mygreen}(+11.5)} \\

\quad + \textsc{MetaLint} (RS-DPO) & 86.2 {\color{myred}(-0.6)} & \textbf{84.5 {\color{mygreen}(+34.7)}} & \textbf{85.3 {\color{mygreen}(+22.1)}} & \textbf{97.0 {\color{mygreen}(+2.2)}} & \textbf{92.1 {\color{mygreen}(+4.1)}} & \textbf{94.4 {\color{mygreen}(+3.1)}} & 92.4 {\color{mygreen}(+5.4)} & \textbf{85.7 {\color{mygreen}(+18.0)}} & \textbf{88.9 {\color{mygreen}(+12.7)}} \\

\bottomrule
\end{tabular}
}
\caption{\textbf{Cross-Idiom Generalization on Python Ruff Best Practice Idioms by Transfer Setting:} We evaluate the effect of different \textsc{MetaLint} training setups (SFT, RS-SFT, and RS-DPO) on Qwen3-4B (w \& w/o CoT) and Llama3.2-3B (Table~\ref{tab:gen_synth_python_ruff_idioms_per_transfer_setting_full}). Models are trained on easy synthetic Python Ruff best practices, and the performance is reported on other Ruff best practices with varying levels of transfer - In-Domain, Near Transfer, and Far Transfer.}
\label{tab:gen_synth_python_ruff_idioms_per_transfer_setting}
\end{table}

\begin{table*}[!tbh]
\centering
\resizebox{0.9\textwidth}{!}{%
\begin{tabular}{@{}lccccccc@{}}
\toprule
\multirow{2}{*}{\textbf{Model}} & \multirow{2}{*}{\textbf{Transfer}} & \multicolumn{3}{c}{\textbf{Detection (\%)}} & \multicolumn{3}{c}{\textbf{Localization (\%)}} \\ \cmidrule(l){3-8} 
 &  & \textbf{$P_{Det}$} & \textbf{$R_{Det}$} & \textbf{$F_{Det}$} & \textbf{$P_{Det}$} & \textbf{$R_{Det}$} & \textbf{$F_{Det}$} \\ \midrule
Llama3.2-3B-Instruct & \multirow{3}{*}{PMD $\rightarrow$ PMD} & 4.57 & 0.79 & 1.34 & 0.15 & 0.22 & 0.17 \\
\quad + \textsc{MetaLint} (SFT) & & 22.51 & 44.21 & 29.83 & 28.22 & 27.78 & 28.00 \\
\quad + \textsc{MetaLint} (SFT + RS-DPO) &  & \textbf{43.95} & \textbf{89.08} & \textbf{58.86} & \textbf{59.30} & \textbf{59.69} & \textbf{59.49} \\ \midrule
Llama3.2-3B-Instruct & \multirow{3}{*}{PMD $\rightarrow$ JEP} & 38.55 & 0.96 & 1.87 & 0.05 & 0.04 & 0.05 \\
\quad + \textsc{MetaLint} (SFT) & & 22.86 & 40.72 & 29.28 & 16.26 & 13.36 & 14.67 \\
\quad + \textsc{MetaLint} (SFT + RS-DPO) & & \textbf{49.03} & \textbf{83.38} & \textbf{61.75} & \textbf{42.16} & \textbf{33.33} & \textbf{37.21} \\ \bottomrule
\end{tabular}
}
\caption{\textbf{Cross-Idiom Generalization on JEP \& PMD Idioms:} Effect of different \textsc{MetaLint} training setups (SFT \& RS-DPO) on Llama3.2-3B-Instruct (Table~\ref{tab:gen_synth_jep_pmd_experiments}). The transfer column indicates training and test data on the left and right side of the arrow. Best scores bolded.}
\label{tab:gen_synth_jep_pmd_experiments_subset}
\end{table*}

\subsection{Results on Easy-to-Hard Generalization}
\label{sec:results:gen_easy_to_hard_pep}
To evaluate whether \textsc{MetaLint} training on easy to detect Ruff best practices improves performance on hard, PEP best practices, we report results on our PEP hard best practice benchmark (Table~\ref{tab:gen_easy_to_hard_pep}, full results in Table~\ref{tab:gen_easy_to_hard_pep_full}). At the SFT stage, performance declines for Qwen3-4B (with and without CoT) but improves slightly for Llama3.2-3B-Instruct, suggesting that SFT is prone to memorization of the training distribution. In contrast, DPO yields clear improvements in detection and localization (except detection precision for Llama3.2-3B-Instruct), with statistically significant gains (Appendix~\ref{sec:appendix:results:stat_signficance}). 
An additional experiment training Qwen3-4B (CoT) directly with RS-DPO, bypassing SFT, resulted in near-zero performance because many generated DPO pairs violated the required output format, which the model inherited. 
Thus, SFT, despite its drawbacks, is essential for teaching format compliance and setting the stage for DPO to encourage easy-to-hard generalization.
Interestingly, the non-CoT model achieves substantially higher detection recall and slightly higher F-score than the CoT variant, despite lower precision. 
Our analysis attributes the CoT model's reduced recall to its more conservative interpretation of idiom specifications and to errors such as misinterpretation, overthinking, and skipped lines, as detailed in Appendix~\ref{sec:appendix:results:cot_model_failure_analysis}.
To directly test whether the instruction-following framing, rather than training data volume, is the decisive factor for generalization, we trained Qwen3-4B on the identical 50k Ruff examples but replaced every rule-specific instruction with a generic ``find code quality issues'' prompt (no PEP specification).
Under strict kind-matched scoring, this instruction-free model achieves 0.00\% $F_{\text{Det}}$ and 0.00\% $F_{\text{Loc}}$ on the PEP benchmark (Table~\ref{tab:instr_free_ablation} in Appendix~\ref{sec:appendix:more_results}), confirming that the specification is the mechanism of transfer, not data exposure or DPO training.
Finally we analyze the impact of the illustrative examples in the prompt in Table~\ref{tab:pep_example_ablation}.

\begin{table*}[!tbh]
\centering
\resizebox{0.8\textwidth}{!}{%
\begin{tabular}{@{}lcccccc@{}}
\toprule
\multirow{2}{*}{\textbf{Model}} & \multicolumn{3}{r}{\textbf{Detection (\%)}} & \multicolumn{3}{r}{\textbf{Localization (\%)}} \\ \cmidrule(l){2-7} 
 & \textbf{$P_{Det}$} & \textbf{$R_{Det}$} & \textbf{$F_{Det}$} & \textbf{$P_{Loc}$} & \textbf{$R_{Loc}$} & \textbf{$F_{Loc}$} \\ \midrule
Qwen3-4B & 52.67 & 17.15 & 25.87 & 9.54 & 8.24 & 8.84 \\
\quad + \textsc{MetaLint} (SFT) & 43.33 & 8.21 & 13.81 & 4.32 & 2.21 & 2.92 \\
\quad + \textsc{MetaLint} (SFT + RS-DPO) & \textbf{70.31} & \textbf{70.43} & \textbf{70.37} & \textbf{35.36} & \textbf{19.30} & \textbf{24.97} \\
\midrule
Qwen3-4B w CoT & 81.54 & 39.86 & 53.54 & 26.25 & 14.67 & 18.82 \\
\quad + \textsc{MetaLint} (RS-SFT) & 76.15 & 36.89 & 49.70 & 27.85 & 14.37 & 18.96 \\
\quad + \textsc{MetaLint} (RS-SFT + RS-DPO) & \textbf{93.03} & \textbf{49.58} & \textbf{64.68} & \textbf{34.82} & \textbf{21.69} & \textbf{26.73} \\ \bottomrule
\end{tabular}
}
\caption{\textbf{Easy-to-Hard Generalization:} We evaluate the effect of different \textsc{MetaLint} training setups (SFT, RS-SFT, \& RS-DPO) on Qwen3-4B (w \& w/o CoT) and Llama3.2-3B (Table~\ref{tab:gen_easy_to_hard_pep_full}), trained on easy synthetic Ruff idioms and tested on hard manually curated PEP idioms (section~\ref{sec:methods:pep_idiom_benchmark}). Best scores bolded.}
\label{tab:gen_easy_to_hard_pep}
\end{table*}

\begin{table*}[!tbh]
\centering
\resizebox{\textwidth}{!}{%
\begin{tabular}{@{}lcccccc@{}}
\toprule
 & \multicolumn{3}{c}{\textbf{Detection (\%)}} & \multicolumn{3}{c}{\textbf{Localization (\%)}} \\ \cmidrule(l){2-4} \cmidrule(l){5-7} 
\multirow{-2}{*}{\textbf{Model}} &\textbf{$P_{Det}$} & \textbf{$R_{Det}$} & \textbf{$F_{Det}$} & \textbf{$P_{Loc}$} & \textbf{$R_{Loc}$} & \textbf{$F_{Loc}$} \\ \midrule
\textbf{Qwen3-4B + \textsc{MetaLint} (SFT+RS-DPO)} & 70.31 & \textbf{70.43} & 70.37 & 35.36 & 19.30 & 24.97 \\
\textbf{Qwen3-4B with CoT + \textsc{MetaLint} (RS-SFT + RS-DPO)} & 93.03 & 49.58 & 64.68 & 34.82 & 21.69 & 26.73 \\
\midrule
Qwen3-8B & 82.67 & 35.72 & 49.88 & 18.06 & 12.85 & 15.01 \\
Qwen3-8B with CoT & 88.86 & 46.72 & 61.24 & 31.22 & 20.29 & 24.59 \\
Qwen3-14B & 90.21 & 46.12 & 61.03 & 28.90 & 25.21 & 26.93 \\
Qwen3-14B with CoT & 91.16 & 48.57 & 63.37 & 39.93 & 29.15 & 33.69 \\
Qwen3-32B & 90.21 & 52.05 & 66.01 & 28.07 & 27.11 & 27.58 \\
Qwen3-32B with CoT & {\underline{93.77}} & 56.45 & {70.48} & 41.52 & 30.86 & 35.40 \\
Qwen2.5-32B-Instruct & 86.67 & 26.56 & 40.66 & 16.30 & 14.77 & 15.50 \\
Qwen2.5Coder-32B-Instruct & 89.61 & 53.28 & 66.83 & 34.32 & 30.77 & 32.45 \\
DeepSeek-R1-Distill-Qwen-32B with CoT & 90.08 & 58.99 & 71.30 & 40.15 & 34.03 & 36.84 \\
GPT-oss-20b & 83.77 & 35.31 & 49.68 & 25.10 & 16.95 & 20.24 \\
GPT-oss-120b & 91.57 & 64.56 & \underline{75.73} & 39.91 & 33.31 & 36.31 \\
\midrule
o3-mini & 89.39 & 58.45 & 70.68 & 31.69 & 23.61 & 27.06 \\
o4-mini & \textbf{96.67} & 59.43 & 73.61 & 41.31 & 31.64 & 35.84 \\ 
GPT-4o & 89.38 & 67.88 & \textbf{77.16} & {\underline{44.61}} & 33.20 & 38.07 \\
GPT-4.1 & 90.70 & 64.60 & 75.46 & \textbf{46.32} & \textbf{46.73} & \textbf{46.53} \\

GPT-5 (high) & 91.30 & 56.73 & 69.98 & 43.97 & \underline{42.57} & \underline{43.26} \\
\bottomrule
\end{tabular} 
}
\caption{\textbf{Benchmarking on Hard Best Practices:} Results comparing state of the art code and reasoning models on the hard PEP benchmark to contextualize the gains achieved with \textsc{MetaLint} training. The best scores are bolded and second best and underlined.
}
\label{tab:hard_pep_benchmark_evaluation}
\end{table*}

\subsection{Benchmarking on Hard Best Practices}
\label{sec:results:pep_hard_idiom_benchmark}
Table~\ref{tab:hard_pep_benchmark_evaluation} compares the best-performing Qwen3-4B \textsc{MetaLint} DPO models against state-of-the-art code and reasoning models (full results in Table~\ref{tab:hard_pep_benchmark_evaluation_full}).
For detection F-score, the non-CoT \textsc{MetaLint} model is competitive with o3-mini and GPT-5 but is outperformed by larger open and closed-source models (e.g., Qwen3-32B w CoT, DeepSeek-R1-Distill-Qwen-32B w CoT, GPT-4o, GPT-4.1, and o4-mini) but achieves the highest detection recall, while the CoT variant ranks third in precision, behind Qwen3-32B and o4-mini.
For localization, the \textsc{MetaLint} models trail larger 32B/120B and GPT models, but perform comparably to o3-mini (statistical significance analysis in Appendix~\ref{sec:appendix:results:stat_signficance}) and outperform GPT-oss-20B. 
This is notable given that the \textsc{MetaLint} models are much smaller (4B parameters), trained only on synthetic data derived from easy best practices, and that the non-CoT model does not use test-time compute. 
We also identify localization errors and their frequency in Table~\ref{tab:egregious_localization_errors}. \\
Overall, our models perform comparably to state-of-the-art systems, with the non-CoT Qwen3-4B variant achieving the highest recall, and both \textsc{MetaLint}-trained Qwen and LLaMA models exhibiting evidence of easy-to-hard generalization.

\paragraph{Cost comparison.}
Tables~\ref{tab:pep_benchmark_cost} and~\ref{tab:real_repo_cost} (Appendix~\ref{sec:appendix:more_results}) report API costs on the PEP benchmark (536 examples) and on a real repository (flask, 83 files $\times$ 15 PEPs) respectively.
\textsc{MetaLint} (non-CoT) is 175$\times$ cheaper than GPT-5, with GPT-5 costs dominated by hidden reasoning tokens billed at output rates.
On the real-repo setting (the Flask github repo is used as an illustrative example), \textsc{MetaLint} CoT is 9.5-21$\times$ cheaper than agentic frontier linting (Claude Sonnet 4.6 / Opus 4.7 via Claude Code), and this gap widens at scale since frontier agent costs grow with conversation context while \textsc{MetaLint}'s cost is strictly linear in file-rule pairs.
Teams can also self-host a 4B model on a single GPU, reducing per-token cost to near zero.

\paragraph{False positive rates.}
Table~\ref{tab:pep_fpr} (Appendix~\ref{sec:appendix:more_results}) reports False Positive Rates (FPR) across all models on the 257 no-violation instances.
High recall universally incurs elevated FPR: GPT-4o reaches 67.9\% recall at 9.7\% FPR; \textsc{MetaLint} non-CoT achieves the highest recall (70.4\%) at 24.5\% FPR.
The CoT variant reaches 49.6\% recall at only 5.1\% FPR, outperforming all open-source models up to Qwen3-14B w/ CoT at 4B parameters.
FPR is a controllable operating point via the NV fraction hyperparameter in RS-DPO (Table~\ref{tab:non_cot_dpo_ablations}).
\section{Conclusion and Future Work}
Our results show that \textsc{MetaLint} encourages inductive transfer over best practice specifications rather than memorization. 
Models generalize to unseen best practices across Python and Java, linters, model families, reasoning settings, and scales, and exhibit easy-to-hard transfer from linter-detectable cases to harder PEP violations. 
With only 4B parameters, \textsc{MetaLint}-trained Qwen models achieve detection performance comparable to strong code and reasoning systems, attaining the highest recall with competitive precision. 
Localization remains more challenging but is competitive with o3-mini and surpasses GPT-oss-20B without test-time compute. 
\textsc{MetaLint} is not a replacement for static analysis, but a proof of concept that an instruction following framing of code linting improves generalization to unseen and hard best practices.
Future work will explore patch generation and automated refactoring (a natural downstream application of \textsc{MetaLint}'s violation localizations), as well as improved RL objectives such as GRPO \citep{shao2024deepseekmath}, and broader connections to principle-conditioned generalization in reward modeling \citep{yu2025rewardanything}.

\bibliography{colm2026_conference}
\bibliographystyle{colm2026_conference}

\appendix
\section*{Acknowledgments}
This work was partially supported by a generous collaborative research grant from Oracle Labs.  
We are grateful to Jack Sullivan, Qinlan Shen, Adam Pocock, and Nidhi Chandra at Oracle Labs, and Emmy Liu, Zora Wang, Saujas Vaduguru, Pranjal Aggarwal, Andre He from Carnegie Mellon University for their guidance, feedback, and insightful discussions that played a key role in shaping the core ideas behind \textsc{MetaLint} and advancing the project.  

\section*{Appendix}
We organize the Appendix in five parts: 1) Limitations of our work (Appendix~\ref{sec:appendix:limitations}), 2) Clarifications and definitions of terms used in the paper (Appendix~\ref{sec:appendix:term_definitions}), 3) Generative AI usage (Appendix~\ref{sec:appendix:gen_ai_usage}), 4) Additional related work (Appendix~\ref{sec:appendix:more_related_work}), 5) Method details like prompts and hyperparameters (Appendix~\ref{sec:appendix:method_details}), 6) Additiomal experimental details (Appendix~\ref{sec:appendix:experimental_details}), 7) Additional results and ablations (Appendix~\ref{sec:appendix:more_results}), and 8) Code Repository Details (Appendix~\ref{sec:appendix:code_repository}).

\section{Limitations}
\label{sec:appendix:limitations}
Despite promising results, including evidence of easy-to-hard generalization, \textsc{MetaLint} has several limitations that suggest directions for future work:
\begin{itemize}[itemsep=0pt, leftmargin=*, parsep=0pt, topsep=-5pt, partopsep=0pt]
    \item \textbf{Pre-training Data Contamination Risk:} While the Python and Java best practices are publicly available and may appear in model pre-training, our evaluation focuses on inductive transfer, testing whether models can systematically apply learned specifications to unseen or harder violations. Our human-curated hard PEP benchmark further reduces the risk of contamination and provides stronger evidence of generalization.
    \item \textbf{Chain-of-Thought (CoT) reasoning:} We did not explore whether non-CoT models can be trained to generate CoT-style reasoning using supervision from a teacher model.  
    \item \textbf{Self-improvement and data generation:} We attempted self-improvement strategies for RS-SFT data generation (e.g., STaR \citep{zelikman2022star}) when the base model failed. However, generating CoTs that do not directly reference provided hints proved challenging and risks contaminating the training data. We therefore adopted a simpler rejection sampling (RS-SFT) strategy.  
    \item \textbf{Preference optimization methods:} We only experimented with DPO for preference optimization. Alternative approaches, such as Proximal Policy Optimization (PPO) \cite{schulman2017proximal} or Group Relative Policy Optimization (GRPO) \citep{shao2024deepseekmath}, combined with our verifiable linter-based reward model, could be explored.  
    \item \textbf{Difficulty progression and curriculum:} Our current setup does not implement a fine-grained difficulty progression or a curriculum over meta-task abstractions. Incorporating such structure could encourage more systematic induction over code idioms and code linting concepts.  
    \item \textbf{Multilingual training:} Experiments focus on one programming language at a time (Python or Java). Future work could explore joint training across multiple languages, cross-language transfer, and structured learning that encourages abstraction across similar language constructs (e.g., Python’s PEP 557: Data Classes vs. Java’s JEP 395: Records).  
    \item \textbf{Task scope:} We focus exclusively on best practice violation detection and do not yet explore inductive transfer over code refactoring or other code linting tasks.  
    \item \textbf{Depolyment Optimizations for Industry:} Our per-idiom evaluation is designed for clean, interpretable benchmarking. In industrial settings, it could be parallelized or batched to detect multiple categories simultaneously, but we do not explore those optimizations here.
\end{itemize}

\section{Clarification and Term Definitions}
\label{sec:appendix:term_definitions}
Because this paper studies a software engineering–heavy problem setting, we clarify the key technical terms used throughout.

\paragraph{Code Idiom / Programming Idiom.}
A \emph{code idiom} is a recurring, language-specific pattern of code that conveys a particular semantic intent or programming practice beyond what is expressed by its literal syntax.
\paragraph{Code Quality Analysis.}
\emph{Code Quality Analysis} refers to the systematic evaluation of source code with respect to software engineering attributes such as readability, complexity, maintainability, and adherence to best practices, using automated tools or model-based methods. 

\paragraph{Code Linting.}
\emph{Code Linting} is the automated process of analyzing source code using a lint tool (or linter) to detect programmatic errors, bugs, stylistic inconsistencies, and other potential issues without executing the code. It can fall under the broader arena of code quality analysis as an execution free method of finding opportunities to improve code quality.

\paragraph{Best Practice Violation Detection.}
\emph{Best practice violation detection} is a type of code quality analysis that involves identifying and localizing code fragments that violate established or recommended programming practices.
In this work, we operationalize best practice violation detection as the task of identifying \emph{problematic code idioms}, i.e., recurring patterns that correspond to non-recommended practices. While this formulation is more restrictive than the full space of best practices, it enables a systematic task structure that supports controlled supervision and generalization.
Additionally we often use the shorthand best practice idiom or idiom to refer to the problematic idiom that needs to be detected to flag a best practice violation.

\section{Generative AI Usage}
\label{sec:appendix:gen_ai_usage}
We used generative AI tools for limited writing assistance, such as improving phrasing, grammar, and clarity. The tools were not used for generating technical content or experimental results. All outputs were reviewed and edited by the authors to ensure accuracy.
\section{More Related Work}
\label{sec:appendix:more_related_work}
\textbf{Instruction Following and Generalization.}
Instruction tuning has emerged as a powerful form of meta-learning, enabling cross-task generalization by training models to interpret and follow natural language instructions rather than memorizing fixed tasks. Prior work shows that exposure to diverse task instructions allows models to extract underlying task abstractions and apply them to unseen settings \citep{Mishra2021CrossTaskGVA,Wang2022SuperNaturalInstructionsGVA}. Large-scale instruction tuning further improves zero- and few-shot generalization across tasks and modalities \citep{Wei2021FinetunedLMA,Gao2021MultitaskPTA,Iyer2022OPTIMLSLA,Brown2020LanguageMAA,chung2024scaling}.  
Instructions serve as high-density task representations, substituting for explicit supervision \citep{Puri2022HowMDA} and enabling generalization even with minimal labeled data or pseudo-labeled examples \citep{Gu2022LearningIWA}. Studies also show that instruction diversity drives generalization: varied instructions outperform repeated exposure to identical formats \citep{Charton2024InstructionDDA}. This effect holds across domains, including program synthesis, where task-level prompting facilitates generalization in code generation models \citep{Niu2023CrossCodeBenchBCA}. SELF-GUIDE \citep{zhao2024self} performs task-specific instruction following using synthetic data, demonstrating effectiveness but relying entirely on LLM-generated data without verifiers.  
These results suggest that instruction tuning functions as task-level meta-learning, enabling models to adapt to new tasks through natural language. 
Building on this insight, we model individual code idioms as separate meta-tasks and generate large-scale synthetic data for each task to support cross-idiom generalization. This approach allows the trained model to adapt to new idioms and evolving best practices.

\section{Method Additional Details}
\label{sec:appendix:method_details}

\begin{figure*}[t]
    \centering
    \includegraphics[width=0.9\textwidth]{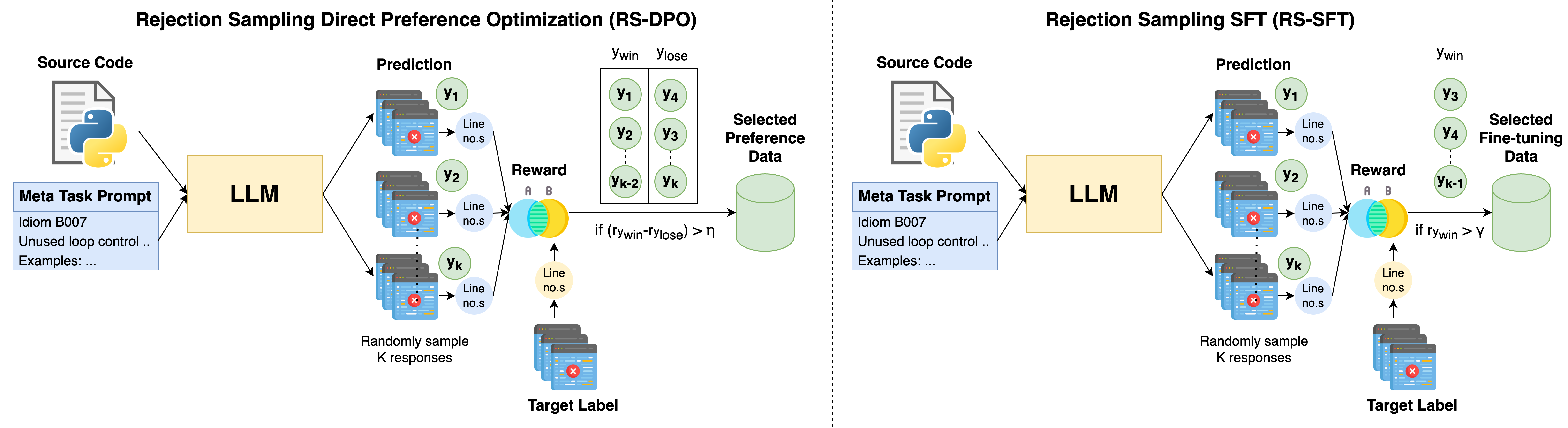}
    \caption{\textbf{\textsc{MetaLint}:} Preference Optimization using reward function: (4) Rejection Sampling Direct Preference Optimization (RS-DPO), and (5) Rejection Sampling Supervised Fine-Tuning (RS-SFT).}
    \label{fig:meta_lint_training_framework_part2}
\end{figure*}

\subsection{\textsc{MetaLint} Instruction Following Prompt}
\label{sec:appendix:method:metalint_prompt_io}
We used the following instruction following style prompt to train the model with synthetic Ruff best practice idiom data for the meta-linting task:

\begin{tcolorbox}[
    colback=orange!5,
    colframe=orange!40!black,
    title={\textbf{\textsc{MetaLint} Instruction Following Prompt}},
    coltitle=white,
    colbacktitle=orange!75!black,
]
Look at the following list of code idiom specifications with definitions and examples:
\{LIST\_OF\_IDIOM\_SPECS\}
\\ \\ 
Given these idioms, your task is to look at a code file and detect violations of the above idioms, and flag them like a linter. You should also suggest a fix if possible. Report the results per idiom specification mentioned above and just say \texttt{NO VIOLATIONS FOUND} if no violations are found for a given idiom. Do not detect any idioms not specified above.
\\ \\ 
Code file:
\{CODE\_FILE\}
\\ \\ 
Violations per idiom:
\end{tcolorbox}

An example input with the code file and best practice idiom spec populated as well as the expected JSON style output is shown below:

\begin{tcolorbox}[
    colback=orange!5,
    colframe=orange!40!black,
    title={\textbf{Example Ruff Meta-Task Input}},
    coltitle=white,
    colbacktitle=orange!75!black,
]
Look at the following list of code idiom specifications with definitions and examples:
\# Idiom ANN202 (missing-return-type-private-function)

Definition: Checks that private functions and methods have return type annotations.

Rationale: Type annotations are a good way to document the return types of functions. They also
help catch bugs, when used alongside a type checker, by ensuring that the types of
any returned values, and the types expected by callers, match expectation.

Example:
\begin{verbatim}
def _add(a, b):
    return a + b
\end{verbatim}

Use instead:

\begin{verbatim}
def _add(a: int, b: int) -> int:
    return a + b
\end{verbatim}

Given these idioms, your task is to look at a code file and detect violations of the above idioms, and flag them like a linter. You should also suggest a fix if possible. Report the results per idiom specification mentioned above and just say 'NO VIOLATIONS FOUND' if no violations are found for a given idiom. Do not detect any idioms not specified above.

Code file:
\begin{verbatim}
  1 # -*- coding: utf-8 -*-
  2 # pragma pylint: disable=unused-argument, no-self-use
  ...
 86     def _reload(self, event, opts):
 87         """Configuration options have changed, 
    save new values"""
 88         self.options = opts.get("fn_cisco_amp4ep", {})
 89         validate_opts(self)
 90 
 91     @function("fn_amp_move_computer")
 92     def _fn_amp_move_computer_function(self, event, *args, 
    **kwargs):
 93         """Function: Move computer to a group with given 
    connector guid and group guid."""
 94         try:
 ...
\end{verbatim}

Violations per idiom:
\end{tcolorbox}

\begin{tcolorbox}[
    colback=orange!5,
    colframe=orange!40!black,
    title={\textbf{Example Ruff Meta-Task Output}},
    coltitle=white,
    colbacktitle=orange!75!black,
]
**Idiom ANN202 Violations:**

\begin{verbatim}
{"line": " 86     def _reload(self, event, opts):", "fix": null}
{"line": " 92     def _fn_amp_move_computer_function(self, 
event, *args, **kwargs):", "fix": null}
\end{verbatim}

\end{tcolorbox}

\subsection{DPO Contrastive Pair and RS-SFT Sampling Details}
To generate RS-DPO contrastive samples (or RS-SFT outputs) from the baseline SFT (or untrained) models, we used the following hyperparameters: nucleus sampling with a maximum of 2048 new tokens, $k=5$ sampled outputs per input, temperatures picked cyclically from \{0, 0.3, 0.5, 0.7, 1\}, a top-$p$ (cumulative probability threshold) of 0.95, and a seed of $42 + i$, where $i \in \{1, \dots, k\}$, to encourage both reproducibility and output diversity.

For RS-DPO sampling (in both CoT and non-CoT settings), we used the standard \textsc{MetaLint} instruction-following prompt with the SFT models. In contrast, for RS-SFT output sampling from the untrained model, we employed the expanded ``Baseline Inference Prompt'' described in Section~\ref{sec:appendix:method:baseline_inference_details}.

\subsection{Training Hyperparameters and Computational Environment}
\textbf{Python SFT/RS-SFT hyperparameters:} \\
We fine-tune the \texttt{Qwen3-4B} model using \texttt{flash\_attention\_2} and \texttt{bfloat16} precision. The model is trained for 2 epochs with a learning rate of 2e-5, cosine learning rate schedule, and a warmup ratio of 0.1. We use a maximum sequence length of 3000 tokens, a per-device batch size of 2, and gradient accumulation steps of 4. Gradient checkpointing is enabled to reduce memory usage, with non-reentrant mode. Evaluation is performed every 2000 steps, and checkpoints are saved at the same interval. Special tokens are manually handled in the chat template without automatic insertion. The training uses 12 preprocessing workers and is seeded with 42 for reproducibility.

\textbf{Python RS-DPO parameters:} \\
We fine-tune the model using RS-DPO with \texttt{bfloat16} precision and a reward shaping parameter \(\beta = 0.1\). Training is performed for 1 epoch with a learning rate of 5e-7, cosine learning rate scheduling, and a warmup ratio of 0.1. We use a maximum input length of 3500 tokens, a per-device batch size of 2, and gradient accumulation steps of 4. Gradient checkpointing is enabled with non-reentrant mode to optimize memory usage. The optimizer is \texttt{AdamW}, and evaluation is conducted every 200 steps with checkpoints saved at the same interval. The training is seeded with 42 for reproducibility.

\textbf{Java SFT hyperparameters:} \\
For Java experiments, we fine-tune \texttt{Llama-3.1-8B-Instruct} and \texttt{Llama-3.2-3B-Instruct} with \texttt{bfloat16} precision. Both models are trained for 2 epochs with a learning rate of 2e-5, cosine learning rate schedule, and warmup ratio of 0.1. We use a maximum sequence length of 3000 tokens, per-device batch size of 2, and gradient accumulation steps of 4. Gradient checkpointing (non-reentrant) is enabled. Evaluation and checkpoint saving occur every 5000 steps. Special tokens are manually handled in the chat template. Training is seeded with 42.

\textbf{Java RS-DPO parameters:} \\
RS-DPO training is performed on \texttt{Llama-3.1-8B-Instruct} and \texttt{Llama-3.2-3B-Instruct} using \texttt{bfloat16} precision. Training runs for 1 epoch with a learning rate of 5e-7, cosine learning rate scheduling, and warmup ratio of 0.1. We use a maximum input length of 3500 tokens, a per-device batch size of 2, and gradient accumulation steps of 4. Gradient checkpointing (non-reentrant) is enabled. Evaluation and checkpoints are recorded every 200 steps. Reward shaping parameters vary across settings, with \(\beta \in \{0.1, 0.5, 1\}\). Seeds are fixed at 42 for reproducibility.

\textbf{Computational Environment:} \\
All SFT, RS-SFT, and RS-DPO experiments (Python and Java) were conducted on a Linux server equipped with NVIDIA A100 80GB GPUs (Ampere architecture), CUDA 12.9, and driver version 575.51.03. Each job had access to 100 GB of CPU memory and 2 CPU cores. Training used mixed-precision (\texttt{bfloat16}) with gradient checkpointing to optimize memory usage. Inference used a similar setup with GPU allocation varying by model size.

\subsection{Inference Details}
\label{sec:appendix:method:baseline_inference_details}
We use the following hyperparameters for performing inference with the baseline LLMs and \textsc{MetaLint} trained models: \\
\textbf{Open Source LLMs:}
We perform nucleus sampling with 8192 max-new tokens, temperature of 0.7, top-p (cumulative probability threshold) of 0.95, and seed of 42 (to promote reproducibility). \\
\textbf{Closed Source LLMs:} We use the chat completion OpenAI API with max tokens of 1024 for GPT-4.1 and GPT-4o and max completion tokens of 3000 for o3-mini and o4-mini. 
We use default parameters for everything else (temperature of 1 and top-p of 1, no presence penalty).
For GPT-5, we use 8192 max completion tokens and high reasoning effort.

Additionally, we use an expanded prompt (Baseline Inference Prompt) compared to the one used for \textsc{MetaLint}, specifically adding more details about output formatting to ensure all baselines have a fair chance and do not suffer performance drops due to formatting mismatches. 
For the same reason, we also allow certain relaxations in output formatting during evaluation on the PEP Hard Best Practice Benchmark.

\begin{tcolorbox}[
    colback=orange!5,
    colframe=orange!40!black,
    title={\textbf{Baseline Inference Prompt}},
    coltitle=white,
    colbacktitle=orange!75!black,
]
Look at the following list of code idiom specifications with definitions and examples:
\{LIST\_OF\_IDIOM\_SPECS\}
\\ \\ 
Given these idioms, your task is to look at a code file and detect violations of the above idioms, and flag them like a linter. You should also suggest a fix if possible. Report the results per idiom specification mentioned above and just say \texttt{NO VIOLATIONS FOUND} if no violations are found for a given idiom. Do not detect any idioms not specified above.
\\ \\ 
Code file:
\{CODE\_FILE\}
\\ \\ 
\# OUTPUT FORMAT
\\ \\ 
I want you to generate your output under a section called ``\#\#\# Final Idiom Violations Found''.
\\ \\ 
Structure you response for a given idiom XYZ as follows for cases with violations:
\\ \\ 
\#\#\# Final Idiom Violations Found
\\ \\ 
**Idiom XYZ Violations:**

\begin{verbatim}
{"line": " 12 \\t\\t#event = forms.ModelChoiceField(queryset=
Inquiry.objects.filter(owner=kwargs.pop('user')))", "fix": null}
{"line": "  1 from django import forms\\n  
2 from django.forms.models import inlineformset_factory\\n  
3 from .models import Request\\n  
4 from inquiry.models import *", 
"fix": [{"before": "from django import forms\\n
from django.forms.models import inlineformset_factory\\n
from .models import Request\\n
from inquiry.models import *\\n\\n\\n\\n", 
"after": "from django import forms\\n
from django.forms.models import inlineformset_factory\\n
from inquiry.models import *\\n\\n
from .models import Request\\n\\n\\n"}]}
\end{verbatim}

and as follows for cases with violations:
\\ \\ 
\#\#\# Final Idiom Violations Found
\\ \\
**Idiom XYZ Violations:**
\\ \\
\texttt{NO VIOLATIONS FOUND}
\\ \\
Violations per idiom:
\end{tcolorbox}


\begin{table}[!tbh]
\centering
\resizebox{\textwidth}{!}{%
\begin{tabular}{@{}lllll@{}}
\toprule
\textbf{JEP\#} & \textbf{JEP Title} & \textbf{Definition} & \textbf{Example(s)} & \textbf{Tree Sitter Queries} \\ \midrule
394 & \begin{tabular}[c]{@{}l@{}}PatternMatching\\ InstanceOf\\ (Before)\end{tabular} & \begin{tabular}[c]{@{}l@{}}Usage of the old pattern \\ of testing with instanceof \\ followed by a manual cast \\ to extract and operate on \\ the object. This pattern is \\ verbose and repetitive. Flag \\ the instanceof expression \\ check within a conditional \\ statement and the \\ accompanying cast \\ expression in the body of \\ the conditional statement.\end{tabular} & \begin{tabular}[c]{@{}l@{}}public class ShapeExample \{\\     static double getPerimeter(Object obj) \{\\         if (obj instanceof Rectangle) \{\\             Rectangle r = (Rectangle) obj;\\             return 2 * r.length() + 2 * r.width();\\         \} else if (obj instanceof Circle) \{\\             Circle c = (Circle) obj;\\             return 2 * c.radius() * Math.PI;\\         \} else \{\\             throw new IllegalArgumentException(\\ "Unrecognized shape");\\         \}\\     \}\\ \}\end{tabular} & \begin{tabular}[c]{@{}l@{}}(if\_statement \\   condition: (parenthesized\_expression \\     (instanceof\_expression \\       left: (identifier) @H1\\       right: (type\_identifier) @H2\\     )\\   ) @jep\_394\_before\_instanceof\_expression.part1\\   consequence: (block \\     ((local\_variable\_declaration\\             type: (type\_identifier) @H3\\         declarator: (variable\_declarator\\                         value: (cast\_expression\\                     type: (type\_identifier) @H4\\                 value: (identifier) @H5\\             )       \\         )\\     )(\#eq? @H1 @H5) (\#eq? @H2 @H3) (\\ \#eq? @H3 @H4)\\ ) @jep\_394\_before\_instanceof\_expression.part2\\   )  \\ )\end{tabular} \\
394 & \begin{tabular}[c]{@{}l@{}}PatternMatching\\ InstanceOf\\ (After)\end{tabular} & \begin{tabular}[c]{@{}l@{}}Replaces verbose instanceof \\ tests plus manual casting into \\ a concise form that tests and \\ declares a typed variable in \\ one step, for example, \\ "if (obj instanceof String s)" \\ which improves readability, \\ reduces boilerplate, and \\ introduces flow-scoped pattern \\ variables. Flag only the line \\ containing the combined \\ instancesof test and casting \\ within the conditional \\ statement.\end{tabular} & \begin{tabular}[c]{@{}l@{}}public class ShapeExample \{\\     static double getPerimeter(Object obj) \{\\         if (obj instanceof Rectangle r) \{\\             return 2 * r.length() + 2 * r.width();\\         \} else if (obj instanceof Circle c) \{\\             return 2 * c.radius() * Math.PI;\\         \} else \{\\             throw new IllegalArgumentException(\\ "Unrecognized shape");\\         \}\\     \}\\ \}\end{tabular} & \begin{tabular}[c]{@{}l@{}}{[}\\         (instanceof\_expression\\       left: (\_)\\       right: (type\_identifier) \\       name: (identifier)\\         ) @jep\_394\_after\_instanceof\_expression\\ {]}\end{tabular} \\
378 & \begin{tabular}[c]{@{}l@{}}TextBlocks\\ (Before)\end{tabular} & \begin{tabular}[c]{@{}l@{}}Multiline strings represented \\ using concatenated string \\ literals, requiring explicit \\ newline escape sequences \\ (\textbackslash{}n) and manual concatenation \\ with the + operator. This \\ approach is verbose and \\ error-prone. Flag cases \\ where a variable \\ declaration or method \\ invocation uses \\ concatenated string literals \\ instead of multiline strings.\end{tabular} & \begin{tabular}[c]{@{}l@{}}String html = "\textless{}html\textgreater{}\textbackslash{}n" +\\               "  \textless{}body\textgreater{}\textbackslash{}n" +\\               "    \textless{}p\textgreater{}Hello, world!\textless{}/p\textgreater{}\textbackslash{}n" +\\               "  \textless{}/body\textgreater{}\textbackslash{}n" +\\               "\textless{}/html\textgreater{}\textbackslash{}n";\end{tabular} & \begin{tabular}[c]{@{}l@{}}{[}\\         (local\_variable\_declaration\\                 declarator: (variable\_declarator\\                 name: (identifier)\\             value: {[}\\               (binary\_expression\\ ...\\     )\\ {]} @jep\_378\_before\_concatenated\_string\_literals\end{tabular} \\
378 & \begin{tabular}[c]{@{}l@{}}TextBlocks\\ (After)\end{tabular} & \begin{tabular}[c]{@{}l@{}}Use of multiline string literal \\ enclosed by triple \\ double-quote marks ("""), \\ allowing for cleaner and more \\ readable representation of \\ multiline strings without \\ explicit escape sequences. \\ Flag cases that use triple \\ double-quote marks for \\ multiline strings in variable \\ declarations or \\ method invocations.\end{tabular} & \begin{tabular}[c]{@{}l@{}}String html = """\\                \textless{}html\textgreater\\                  \textless{}body\textgreater\\                    \textless{}p\textgreater{}Hello, world!\textless{}/p\textgreater\\                  \textless{}/body\textgreater\\                \textless{}/html\textgreater\\                """;\end{tabular} & \begin{tabular}[c]{@{}l@{}}{[}\\ (string\_literal) @jep\_378\_after\_text\_block\\     (\#match? \\ @jep\_378\_after\_text\_block "\textasciicircum{}\textbackslash{}"\textbackslash{}"\textbackslash{}"")\\ {]}\end{tabular} \\
361 & \begin{tabular}[c]{@{}l@{}}Switch\\ Expressions\\ (Before)\end{tabular} & \begin{tabular}[c]{@{}l@{}}Misuse of switch statement \\ with fall-through behavior \\ 'for pattern matching. This \\ pattern is verbose and error \\ prone. You should flag case \\ statements with empty \\ bodies that are misusing \\ fall-through behavior.\end{tabular} & \begin{tabular}[c]{@{}l@{}}int numLetters;\\ switch (day) \{\\     case MONDAY:\\     case FRIDAY:\\     case SUNDAY:\\         numLetters = 6;\\         break;\\     case TUESDAY:\\         numLetters = 7;\\         break;\\ ...\\         throw new IllegalStateException(\\ "Unexpected value: " + day);\\ \}\end{tabular} & jep\_361\_before\_custom\_detectors \\
 &  &  &  &  \\
 &  &  &  &  \\
 &  &  &  &  \\ \bottomrule
\end{tabular}
}
\caption{\textbf{JEP Best Practice Idiom Specifications (1/3):} This table presents 15 best practice idioms across 8 JEPs, including both ``before'' (old best practice) and ``after'' (updated best practice) patterns. The JEP\# column lists the JEP number, the JEP title specifies the idiom topic, and the parenthesized value indicates whether it is a before or after pattern. The Definition, Example, and Tree-Sitter Queries columns provide the idiom definition, minimal Java examples shown to the LLM as instructions, and the queries used to flag problematic idioms for synthetic data creation.}
\label{tab:jep_idiom_specs}
\end{table}

\begin{table}[!tbh]
\centering
\resizebox{\textwidth}{!}{%
\begin{tabular}{@{}lllll@{}}
\toprule
\textbf{JEP\#} & \textbf{JEP Title} & \textbf{Definition} & \textbf{Example(s)} & \textbf{Tree Sitter Queries} \\ \midrule
361 & \begin{tabular}[c]{@{}l@{}}Switch\\ Expressions\\ (After)\end{tabular} & \begin{tabular}[c]{@{}l@{}}Use of switch expressions, allowing \\ a return value. Employs the -\textgreater \\ syntax for case labels, eliminating \\ fall-through behavior. Flag \\ statements that use the arrow \\ operator "-\textgreater{}" or "yield" syntax.\end{tabular} & \begin{tabular}[c]{@{}l@{}}Example 1:\\ \\ int numLetters = switch (day) \{\\     case MONDAY, FRIDAY, SUNDAY -\textgreater 6;\\     case TUESDAY                -\textgreater 7;\\     case THURSDAY, SATURDAY     -\textgreater 8;\\     case WEDNESDAY              -\textgreater 9;\\ ...\\ Example 7:\\ \\ String category = switch (age) \{\\     case 0, 1, 2, 3, 4, 5 -\textgreater "Toddler";\\     case 6, 7, 8, 9, 10, 11, 12 -\textgreater "Child";\\     case 13, 14, 15, 16, 17, 18, 19 -\textgreater "Teenager";\\     default -\textgreater "Adult";\\ \};\\ \\ Example 8:\\ \\ String response = switch (input) \{\\     case "yes" -\textgreater "Affirmative";\\     case "no"  -\textgreater "Negative";\\     default -\textgreater "Unrecognized input";\\ \};\end{tabular} & \begin{tabular}[c]{@{}l@{}}{[}\\ (yield\_statement\\ ) @jep\_361\_after\_yield\\ (switch\_rule\\     (switch\_label)\\     "-\textgreater{}" @jep\_361\_after\_arrow\\ ) \\ (switch\_rule\\     (switch\_label)\\     "-\textgreater{}"            ;; ensures it's not arrow\\     (block (yield\_statement\\ ) @jep\_361\_after\_yield)\\   )\\ {]}\end{tabular} \\
314 & \begin{tabular}[c]{@{}l@{}}UnicodeLang\\ TagExtensions\\ (After)\end{tabular} & \begin{tabular}[c]{@{}l@{}}Use java.util.Locale with additional \\ BCP 47 Unicode extensions \\ (cu, fw, rg, tz) in Java 10 to customize \\ locale behavior like currency \\ (java.util.Currency), first-day-of-week \\ (java.time.temporal.WeekFields), \\ region override \\ (java.text.NumberFormat.getInstance), \\ and time zone \\ (java.time.format.DateTimeFormatter). \\ Flag imports and function calls \\ related to these.\end{tabular} & \begin{tabular}[c]{@{}l@{}}Example 1: Currency Type (cu)\\ \\ import java.util.Locale;\\ import java.util.Currency;\\ \\ public class Foo \{\\     void bar() \{\\         Locale locale = Locale.forLanguageTag(\\ "en-US-u-cu-EUR");\\         Currency c = Currency.getInstance(locale);\\         System.out.println(c);\\     \}\\ \}\\ ...\\ Example 4: Time Zone (tz)\\ \\ import java.util.Locale;\\ import java.time.format.DateTimeFormatter;\\ import java.time.ZonedDateTime;\\ \\ public class Foo \{\\     void bar() \{\\         Locale locale = Locale.forLanguageTag(\\ "en-US-u-tz-Asia-Tokyo");\\         DateTimeFormatter fmt = \\ DateTimeFormatter.ofPattern(\\ "yyyy-MM-dd HH:mm z").withLocale(locale);\\         System.out.println(\\ fmt.format(ZonedDateTime.now()));\\     \}\\ \}\end{tabular} & \begin{tabular}[c]{@{}l@{}}{[}\\         (import\_declaration\\             ((scoped\_identifier\\                 scope: (scoped\_identifier\\ ) @H2\\             name: (identifier) @H1\\         ) (\#eq? @H2 "java.util") (\\ \#eq? @H1 "Currency"))\\     ) @jep\_314\_after\_currency\_import\\     \\         ((method\_invocation\\             object: (identifier) @H7\\         name: (identifier) @H8\\     ) (\#eq? @H7 "Currency") (\\ \#eq? @H8 "getInstance")\\ ) @jep\_314\_after\_currency...\\     \\ ...            \\         ((method\_invocation\\             object: (identifier) @H13\\         name: (identifier) @H14\\     ) (\\ \#eq? @H13 "NumberFormat") (\\ \#eq? @H14 "getInstance")\\ ) @jep\_314\_after\_number\_format...\\ {]}\end{tabular} \\
395 & \begin{tabular}[c]{@{}l@{}}RecordClass\\ (Before)\end{tabular} & \begin{tabular}[c]{@{}l@{}}Use of simple data aggregates with \\ traditional classes which could be \\ replaced with a record class. This \\ approach requires explicit \\ declarations of fields, constructors, \\ and accessor methods, leading to \\ verbose and repetitive code. Flag \\ non record classes containing equals(), \\ hashCode(), and toString() methods.\end{tabular} & \begin{tabular}[c]{@{}l@{}}public class Point \{\\     private final int x;\\     private final int y;\\ ...\\ \\     public int x() \{\\         return x;\\     \}\\ ...\\     @Override\\     public String toString() \{\\         return "Point\{x=" + x + ", y=" + y + "\}";\\     \}\\ \\     @Override\\     public boolean equals(Object obj) \{\\ ...\\     \}\\ \\     @Override\\     public int hashCode() \{\\         return Objects.hash(x, y);\\     \}\\ \}\end{tabular} & \begin{tabular}[c]{@{}l@{}}{[}\\         (class\_declaration\\             body: (class\_body\\                 (constructor\_declaration\\ ) @H1\\             (method\_declaration\\                     name: (identifier) @H2\\             )\\         ) (\#match? @H2 "\textasciicircum{}(\\ hashCode|equals|toString)\$")\\ ) @jep\_395\_before\_record\_like\_class\\ {]}\end{tabular} \\
395 & \begin{tabular}[c]{@{}l@{}}RecordClass\\ (After)\end{tabular} & \begin{tabular}[c]{@{}l@{}}Use of record class. Record classes \\ introduce a concise syntax for \\ defining immutable data aggregates, \\ automatically generating canonical \\ constructors, accessors, equals(), \\ hashCode(), and toString() methods, \\ thereby reducing boilerplate code \\ and enhancing readability.\end{tabular} & \begin{tabular}[c]{@{}l@{}}Example 1 (Record Declaration):\\ \\ record Point(int x, int y) \{\}\\ \\ Example 2 (Record Declaration):\\ \\ record Rectangle(double length, double width) \{\}\\ ...\end{tabular} & \begin{tabular}[c]{@{}l@{}}{[}\\   (record\_declaration\\ ) @jep\_395\_after\_record\_...\\ {]}\end{tabular} \\
 &  &  &  &  \\
 &  &  &  &  \\
 &  &  &  &  \\
 &  &  &  &  \\ \bottomrule
\end{tabular}
}
\caption{\textbf{JEP Best Practice Idiom Specifications (2/3)}}
\label{tab:jep_idiom_specs_page2}
\end{table}

\begin{table}[!tbh]
\centering
\resizebox{\textwidth}{!}{%
\begin{tabular}{@{}lllll@{}}
\toprule
\textbf{JEP\#} & \textbf{JEP Title} & \textbf{Definition} & \textbf{Example(s)} & \textbf{Tree Sitter Queries} \\ \midrule
409 & \begin{tabular}[c]{@{}l@{}}Sealed\\ Class\\ (Before)\end{tabular} & \begin{tabular}[c]{@{}l@{}}Use of abstract classes with \\ private constructors to simulate \\ sealed classes using \\ package-private visibility to \\ restrict subclassing. This \\ approach lacks explicit language \\ support and is error-prone. \\ Switch to sealed classes. Flag \\ abstract classes with private \\ constructors.\end{tabular} & \begin{tabular}[c]{@{}l@{}}public abstract class Shape \{\\     private Shape() \{\}\\ \}\\ public class Circle extends Shape \{ \\ /* Implementation */ \\ \}\\ public class Square extends Shape \{\\ /* Implementation */ \\ \}\end{tabular} & \begin{tabular}[c]{@{}l@{}}{[}\\     ((class\_declaration\\             (modifiers) @H1\\         name: (identifier) @H4\\         body: (class\_body\\                 (constructor\_declaration\\                     (modifiers) @H2\\                 name: (identifier) @H3\\ ...    )(\#match? @H1 "abstract")\\ (\#eq? @H3 @H4)\\ (\#match? @H2 "private")\\ ) @jep\_409\_before\_abstract\_class...\\ {]}\end{tabular} \\
409 & \begin{tabular}[c]{@{}l@{}}Sealed\\ Class\\ (After)\end{tabular} & \begin{tabular}[c]{@{}l@{}}Use of sealed classes to \\ explicitly define which classes \\ or interfaces can extend or \\ implement them using the \\ sealed modifier and the permits \\ clause. This feature enhances \\ type safety and exhaustiveness \\ checking. Flag class declarations \\ with the sealed or non-sealed \\ modifiers and lines with the \\ permit clause.\end{tabular} & \begin{tabular}[c]{@{}l@{}}public sealed class Shape\\     permits Circle, Square \{ \\ /* Implementation */ \\ \}\\ public final class Circle extends Shape \{\\  /* Implementation */ \\ \}\\ public final class Square extends Shape \{ \\ /* Implementation */ \\ \}\end{tabular} & \begin{tabular}[c]{@{}l@{}}{[}\\     (permits\\ ) @jep\_409\_after\_permits\_clause\\         ((class\_declaration\\             (modifiers) @H1\\     )(\#match? @H1 "sealed")\\ ) @jep\_409\_after\_sealed\_modifier\\ {]}\end{tabular} \\
406 & \begin{tabular}[c]{@{}l@{}}Pattern\\ Matching\\ Switch\\ (Before)\end{tabular} & \begin{tabular}[c]{@{}l@{}}Use of a sequence of if-else \\ if statements to test an object's \\ type via instanceof, with a \\ manual cast, to handle each case \\ separately. This approach is \\ verbose, error-prone, and lacks \\ exhaustiveness checking or \\ compiler assistance for missing \\ cases. Flag if or else-if statements \\ that contain instanceof statements \\ with a manual cast in the \\ statement body.\end{tabular} & \begin{tabular}[c]{@{}l@{}}static String formatter(Object o) \{\\     if (o instanceof Integer) \{\\         Integer i = (Integer) o;\\         return String.format("int \%d", i);\\     \} else if (o instanceof Long) \{\\         Long l = (Long) o;\\         return String.format("long \%d", l);\\     \} else if (o instanceof String) \{\\         String s = (String) o;\\         return String.format("String \%s", s);\\     \} else \{\\         return o.toString();\\     \}\\ \}\end{tabular} & \begin{tabular}[c]{@{}l@{}}(if\_statement \\   condition: (parenthesized\_expression \\     (instanceof\_expression \\       left: (identifier) @H1\\       right: (type\_identifier) @H2\\ ...\\   ) @jep\_406\_before\_if\_else\_if\_...\\   consequence: (block \\     ((local\_variable\_declaration\\             type: (type\_identifier) @H3\\         declarator: (variable\_declarator\\                         value: (cast\_expression\\                     type: (type\_identifier) @H4\\                 value: (identifier) @H5\\ ...\\ (\#eq? @H1 @H5) \\ (\#eq? @H2 @H3) \\ (\#eq? @H3 @H4)\\ ) @jep\_406\_before\_if\_else\_if...\\   )  \\ )\end{tabular} \\
406 & \begin{tabular}[c]{@{}l@{}}Pattern\\ Matching\\ Switch\\ (After)\end{tabular} & \begin{tabular}[c]{@{}l@{}}Use of a switch expression or \\ statement with case labels \\ containing type patterns (and \\ optionally a guard), binding \\ the matched variable within \\ the branch. This style is more \\ concise, expressive, and opens \\ opportunities for compiler-checked \\ exhaustiveness and performance \\ optimizations. Flag switch labels \\ (case statements) with patterns, null \\ literals or paranthesized expressions \\ but skip default switch labels/cases.\end{tabular} & \begin{tabular}[c]{@{}l@{}}Example 1:\\ \\ static String formatter(Object o) \{\\     return switch (o) \{\\         case Integer i -\textgreater String.format("int \%d", i);\\         case Long l    -\textgreater String.format("long \%d", l);\\         case String s  -\textgreater String.format("String \%s", s);\\         default        -\textgreater o.toString();\\     \};\\ \}\\ \\ Example 2:\\ \\ static String checkShape(Object o) \{\\     return switch (o) \{\\ ...\end{tabular} & \begin{tabular}[c]{@{}l@{}}{[}\\     (switch\_label \\         (null\_literal) \\     ) @jep\_406\_after\_null\_case\\     (switch\_label \\         (pattern) \\     ) @jep\_406\_after\_switch\_pattern\\     (switch\_label \\         (parenthesized\_expression) \\     ) @jep\_406\_after\_paranthesized\_pattern    \\     (switch\_label \\         (binary\_expression) \\     ) @jep\_406\_after\_binary\_expression\\ {]}\end{tabular} \\
323 & \begin{tabular}[c]{@{}l@{}}LocalVar\\ Syntax\\ Lambda\\ Params\\ (Before)\end{tabular} & \begin{tabular}[c]{@{}l@{}}Use of implicitly typed lambda \\ expressions with omitted type \\ declarations. These lambda \\ expressions rely solely on parameter \\ names. This approach prioritizes \\ brevity but lacks explicit type \\ information. Flag full lambda \\ expressions without type \\ declarations.\end{tabular} & \begin{tabular}[c]{@{}l@{}}Example 1:\\ xs.stream().filter((a, b) -\textgreater a \textless b).forEach(\\ System.out::println);\\ ...\\ Example 4:\\ xs.stream().filter((a) -\textgreater a \textgreater 10).forEach(\\ System.out::println);\end{tabular} & jep\_323\_before\_custom\_detector \\
323 & \begin{tabular}[c]{@{}l@{}}LocalVar\\ Syntax\\ Lambda\\ Params\\ (After)\end{tabular} & \begin{tabular}[c]{@{}l@{}}Use of explicit type declarations for \\ lambda parameters, enhancing code \\ clarity and enabling better static \\ analysis tools. Flag full lambda \\ expressions with explicit type \\ declarations using formal \\ parameters (var).\end{tabular} & \begin{tabular}[c]{@{}l@{}}Example 1:\\ xs.stream().filter(\\ (var a, var b) -\textgreater a.compareTo(b) \textless 0).forEach(\\ System.out::println);\\ ...\\ Example 4:\\ xs.stream().filter((var a) -\textgreater a \textgreater 10).forEach(\\ System.out::println);\end{tabular} & \begin{tabular}[c]{@{}l@{}}(lambda\_expression\\   parameters: (formal\_parameters\\     (formal\_parameter\\       type: (type\_identifier) @H1 \\     (\#eq? @H1 "var"))\\   )) @jep\_323\_after\_local\_var\_lambda\end{tabular} \\ \bottomrule
\end{tabular}
}
\caption{\textbf{JEP Best Practice Idiom Specifications (3/3)}}
\label{tab:jep_idiom_specs_page3}
\end{table}

\subsection{PMD Best Practice Idiom Specifications}
\label{sec:appendix:example_pmd_idiom_spec}
We scrape PMD idioms specification from the Java section of the PMD rules documentation\url{https://docs.pmd-code.org/latest/pmd_rules_java.html}. The PMD instructions are more complex and more ambiguous than our handcrafted JEP specifications because the examples are more verbose and don't pinpoint the specific lines that should be flagged as best practice idiom violations, as can be seen in the example below.

\begin{tcolorbox}[
    colback=green!5,
    colframe=green!40!black,
    title={\textbf{PMD Rule Specification:} UnitTestShouldIncludeAssert},
    coltitle=white,
    colbacktitle=green!75!black,
]
\begin{verbatim}
Since: PMD 2.0
Priority: Medium (3)
Unit tests should include at least one assertion. This makes 
the tests more robust, and using assert with messages provide 
the developer a clearer idea of what the test does. This rule 
checks for JUnit (3, 4 and 5) and TestNG Tests. Note: This rule 
was named JUnitTestsShouldIncludeAssert before PMD 7.7.0. This 
rule is defined by the following Java class: 
net.sourceforge.pmd.lang.java.rule.bestpractices.
UnitTestShouldIncludeAssertRule

Example(s):
public class Foo {
   @Test 
   public void testSomething() {
      Bar b = findBar();
      // This is better than having a NullPointerException
      // assertNotNull("bar not found", b);
      b.work();
   }
}

This rule has the following properties:

Name
Default Value
Description

extraAssertMethodNames
 
Extra valid assertion methods names

Use this rule with the default properties by just referencing 
it:
<rule ref="category/java/bestpractices.xml/
UnitTestShouldIncludeAssert" />

Use this rule and customize it:
<rule ref="category/java/bestpractices.xml/
UnitTestShouldIncludeAssert">
    <properties>
        <property name="extraAssertMethodNames" value="" />
    </properties>
</rule>
\end{verbatim}
\end{tcolorbox}

\begin{table}[!tbh]
\centering
\begin{tabular}{@{}lllllr@{}}
\toprule
{\color[HTML]{333333} \textbf{JEP \#}} & {\color[HTML]{000000} \textbf{Before}} & {\color[HTML]{000000} \textbf{After}} & {\color[HTML]{000000} \textbf{Title}} & {\color[HTML]{000000} \textbf{JDK\#}} & {\color[HTML]{000000} \textbf{Release Date}} \\ \midrule
409 & Yes & Yes & Sealed Classes & 17 &  \\
406 & Yes & Yes & Pattern Matching for switch & 17 & \multirow{-2}{*}{14 Sept 2021} \\
395 & Yes & Yes & Records & 16 &  \\
394 & Yes & Yes & Pattern Matching for instanceof & 16 & \multirow{-2}{*}{16 Mar 2021} \\
378 & Yes & Yes & Text Blocks & 15 & 15 Sept 2020 \\
361 & Yes & Yes & Switch Expressions & 14 & 17 Mar 2020 \\
323 & Yes & Yes & \begin{tabular}[c]{@{}l@{}}Local-Variable Syntax for \\ Lambda Parameters\end{tabular} & 11 & 25 Sept 2018 \\
314 & No & Yes & \begin{tabular}[c]{@{}l@{}}Additional Unicode \\ Language-Tag Extensions\end{tabular} & 10 & 20 Mar 2018 \\ \bottomrule
\end{tabular}
\caption{List of JEPs addressed by our tree-sitter synthetic data. The JEP\# and Title column indicate the number and title of the JEP while JDK\# and Release Date indicate the JDK needed for compilation to be able to use the JEP features. The Before and After columns indicate whether we include rules/patterns to flag the old problematic idiom or new recommended idiom introduced by the JEP.}
\label{tab:jeps_included}
\end{table}

\begin{tcolorbox}[
    colback=red!5,
    colframe=red!40!black,
    title={Java \textsc{MetaLint} Instruction Following Prompt},
    coltitle=white,
    colbacktitle=red!75!black,
]
\textbf{Task Instructions (1/2):} \\
Look at the following code idiom specification with definitions and examples: \\
\{IDIOM\_SPEC\}
\\ \\
\textbf{Task Instructions (2/2):} \\
Given this idiom, your task is to look at a code file and detect violations of the above idiom, and flag them like a linter. 
You should also suggest a fix if possible. Report the results for only the idiom specification mentioned above and just say \texttt{NO VIOLATIONS FOUND} if no violations are found for the given idiom. 
Do not detect violations of any idiom not specified above.
\\ \\
\textbf{Code file:} \\
\{CODE\_FILE\}
\\ \\
\textbf{Violations per idiom:}
\end{tcolorbox}

\section{Additional Experimental Details}
\label{sec:appendix:experimental_details}

\subsection{Evaluation Metrics}
\label{sec:appendix:experiments:eval_metrics_math}

Let \( I \) denote an best practice idiom, \( M_I \) its corresponding meta task specification, \( f \in \mathcal{F} \) a code file, \( V_{f,I} \) the ground truth set of violating line numbers, and \( \hat{y} = V^\Phi_{f,I} \) the model predicted violations. For each dataset instance with input prompt $x$ and ground truth set of line numbers $y$, \( (x, y) = (\{f, M_I\}, V_{f,I}) \in \mathcal{D} \). 

We define the indicator variable:

\[
\mathds{1}[x] =
\begin{cases}
1 & \text{if } x \text{ is true} \\
0 & \text{otherwise}
\end{cases}
\]

\textbf{Detection Metrics:} \\
\[
P_I = \frac{\sum_{(x,y) \in \mathcal{D}} \mathds{1}[|y|>0] \cdot \mathds{1}[|\hat{y}|>0]}{\sum_{(x,y) \in \mathcal{D}} \left( \mathds{1}[|y|>0] \cdot \mathds{1}[|\hat{y}|>0] + \mathds{1}[|y|=0] \cdot \mathds{1}[|\hat{y}|>0] \right)}
\]

\[
R_I = \frac{\sum_{(x,y) \in \mathcal{D}} \mathds{1}[|y|>0] \cdot \mathds{1}[|\hat{y}|>0]}{\sum_{(x,y) \in \mathcal{D}} \left( \mathds{1}[|y|>0] \cdot \mathds{1}[|\hat{y}|>0] + \mathds{1}[|y|>0] \cdot \mathds{1}[|\hat{y}|=0] \right)}
\]

Macro-averaged detection metrics:

\[
P_{\text{Det}} = \frac{1}{|I|} \sum_{I} P_I, \quad R_{\text{Det}} = \frac{1}{|I|} \sum_{I} R_I, \quad F_{\text{Det}} = \frac{2 P_{\text{Det}} R_{\text{Det}}}{P_{\text{Det}} + R_{\text{Det}}}
\]

\textbf{Localization Metrics:} \\
\[
P_{\text{Loc}} = \frac{1}{|\mathcal{D}|} \sum_{(x,y) \in \mathcal{D}} \frac{|y \cap \hat{y}|}{|\hat{y}|}, \quad
R_{\text{Loc}} = \frac{1}{|\mathcal{D}|} \sum_{(x,y) \in \mathcal{D}} \frac{|y \cap \hat{y}|}{|y|}, \quad
F_{\text{Loc}} = \frac{2 P_{\text{Loc}} R_{\text{Loc}}}{P_{\text{Loc}} + R_{\text{Loc}}}
\]

\subsection{Idioms Chosen for Ruff Best Practice Transfer Dataset}
Table~\ref{tab:train_test_idioms} lists the Ruff best practice idioms used in the SFT training and synthetic transfer evaluation test sets. Idioms are grouped by their source linter and cover a range of syntax, semantics, naming, and upgrade-related rules.

\begin{table}[h]
\centering
\small
\begin{tabular}{@{}ll@{}}
\toprule
\textbf{Training Set Idioms} & \textbf{Test Set Idioms} \\
\midrule
\textbf{PyFlakes:} & \textbf{PyFlakes:} \\
F405, F501, F502, F601, F621 & F403, F406, F503, F602, F622 \\[4pt]

\textbf{pycodestyle:} & \textbf{pycodestyle:} \\
E402, E701, E721, E741, E743 & E401, E702, E722, E731, E742 \\[4pt]

\textbf{Naming:} & \textbf{Miscellaneous:} \\
N801, N802, N803, N804, N805, & ERA001, C901, I001, I002, BLE001 \\
N806, N807, N811, N812, N813 & (shared with training) \\[4pt]

\textbf{pyupgrade:} & \textbf{flake8 annotations:} \\
UP001, UP002, UP003, UP004, UP005, UP006, & ANN001, ANN002, ANN003, ANN201, ANN202, \\
UP007, UP008, UP009, UP010, UP011, & ANN204, ANN205, ANN206 \\
UP040, UP044, UP045, UP046, UP047 & \\[4pt]

\textbf{Miscellaneous:} & \textbf{flake8 async:} \\
ERA001, C901, I001, I002, BLE001 & ASYNC100, ASYNC105, ASYNC109, ASYNC110, \\
& ASYNC115, ASYNC116, ASYNC210, ASYNC220, \\
& ASYNC221, ASYNC222, ASYNC230, ASYNC251 \\[4pt]

\textbf{Bugbear:} & \textbf{flake8 bandit:} \\
B002, B003, B004, B005, B006, & S102, S103, S104, S105, S106, \\
B007, B008, B009, B010, B012 & S107, S108, S110, S112, S113, \\
& S201, S202, S301, S302, S303 \\
\bottomrule
\end{tabular}
\caption{Ruff best practice idioms included in the supervised training and transfer evaluation test sets. Test set idioms span both overlapping linters and novel ones not seen during training.}
\label{tab:train_test_idioms}
\end{table}

\begin{figure}[!tbh]
    \centering
    \includegraphics[width=0.8\textwidth]{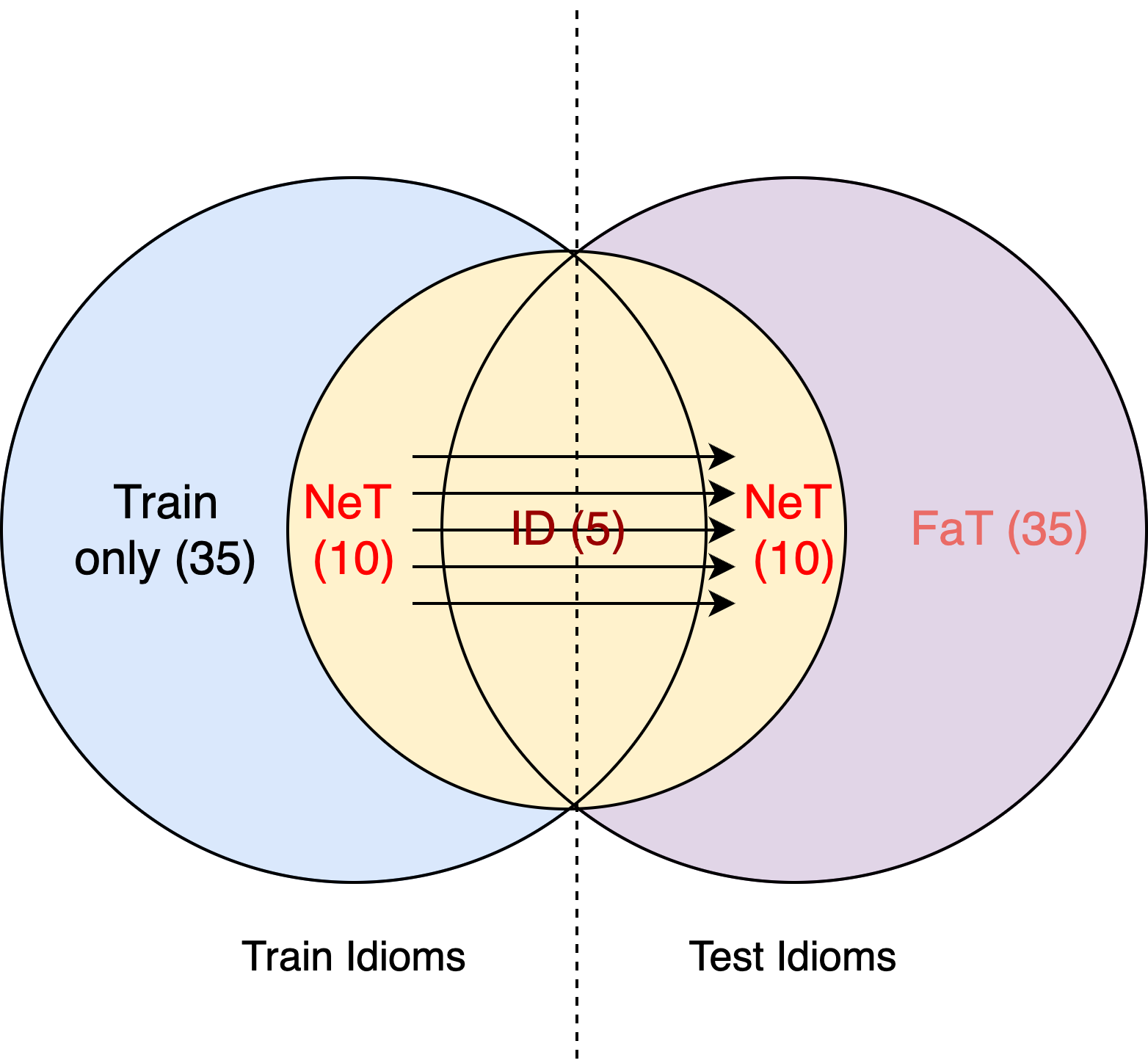}
    \caption{ID: In-Domain, NeT: Near Transfer, FaT: Far Transfer.}
    \label{fig:transfer_eval_design}
\end{figure}

\subsection{DPO No Violation Fraction Ablations}
\label{sec:appendix:experiments_dpo_nv_fraction_ablations}
We analyze the impact of varying the amount of samples with zero violations used for RS-DPO training. These experiments were motivated by initial findings comparing models trained only on data with at least one violation to those trained on the full dataset. By design, RS-DPO generates significantly more training data for cases with at least one violation, due to greater variance in reward signals. This is further amplified by the fact that the initial SFT policy/checkpoint is already quite accurate in handling cases with \texttt{NO VIOLATIONS FOUND} leading to low variance in reward across responses.

Our early experiments showed that excluding all \texttt{NO VIOLATIONS FOUND} cases led to notable gains in recall and line-level localization. However, this came at the cost of a significant drop in precision compared to the SFT policy/base model. Further analysis revealed a sharp decline in the accuracy of predicting \texttt{NO VIOLATIONS FOUND}, from nearly 99\% down to 70-80\%, with performance worsening monotonically over training steps. Conversely, training on the full dataset (i.e., including 100\% of the \texttt{NO VIOLATIONS FOUND} cases) improved precision but offered only modest gains in recall and localization, which also degraded with continued training. These findings suggest that while some \texttt{NO VIOLATIONS FOUND} data is necessary to maintain high precision, too much of it may hinder recall and localization.

To investigate this trade-off, we experimented with keeping only a fraction of the \texttt{NO VIOLATIONS FOUND} data during training. Specifically, we randomly sampled $k$\% of such data, varying $k$ across \{0\%, 2\%, 5\%, 10\%, 20\%, 40\%, 100\%\}. These percentages were selected based on observed trends: 20\%, 40\%, and 100\% yielded similar results, which discouraged further tests at 60\% or 80\%, while 2\% and 5\% were chosen due to a noticeable performance jump between 0\% and 10\%. We found that 5\% offered a favorable middle ground, largely retaining or slightly reducing precision, while preserving most of the recall (resulting in the highest detection F-score), and only modestly impacting line-level localization. Based on these insights, we conducted a limited ablation on the CoT model, evaluating 2\% and 5\% inclusion to determine the optimal setting for both detection and localization (as shown in Table~\ref{tab:cot_dpo_ablations}).

\begin{table}[!tbh]
\centering
\begin{tabular}{@{}lrrrrrr@{}}
\toprule
\multirow{2}{*}{\textbf{\begin{tabular}[c]{@{}l@{}}Fraction of \\ NV data\end{tabular}}} & \multicolumn{3}{r}{\textbf{Detection}} & \multicolumn{3}{r}{\textbf{Localization}} \\ \cmidrule(l){2-7} 
 & {$P_{Det}$} & {$R_{Det}$} & {$F_{Det}$} & {$P_{Loc}$} & {$R_{Loc}$} & {$F_{Loc}$} \\ \midrule
0\% & 0.6268 & 0.9577 & 0.7577 & 0.6777 & 0.6932 & 0.6854 \\
2\% & 0.671 & 0.9128 & 0.7734 & 0.6681 & 0.6812 & 0.6746 \\
5\% & 0.7469 & 0.8315 & 0.7869 & 0.6527 & 0.6696 & 0.6611 \\
10\% & 0.7584 & 0.8114 & 0.784 & 0.6263 & 0.6474 & 0.6367 \\
20\% & 0.8382 & 0.7227 & 0.7762 & 0.5721 & 0.5815 & 0.5768 \\
40\% & 0.8683 & 0.5618 & 0.6822 & 0.4683 & 0.4735 & 0.4709 \\
100\% & 0.8565 & 0.4152 & 0.5593 & 0.4041 & 0.4056 & 0.4048 \\ \bottomrule
\end{tabular}
\caption{Effect of varying the fraction of \texttt{NO VIOLATIONS FOUND} instances in the training data for \textsc{MetaLint} Qwen3-4B model without CoT. Including 0\% yields the highest recall and best line-level localization but reduces precision due to more false positives and lower accuracy in predicting \texttt{NO VIOLATIONS FOUND}. Conversely, including 100\% improves precision but leads to reduced recall and localization performance. All rows report the performance at the best training step, selected based on a balance of detection and localization F-score on the Ruff Best Practice Transfer test set.}
\label{tab:non_cot_dpo_ablations}
\end{table}

\begin{table}[]
\centering
\begin{tabular}{@{}lrrrrrr@{}}
\toprule
\multirow{2}{*}{\begin{tabular}[c]{@{}l@{}}Impact \\ of $\eta$\end{tabular}} & \multicolumn{3}{r}{Detection} & \multicolumn{3}{r}{Localization} \\ \cmidrule(l){2-7} 
 & {$P_{Det}$} & {$R_{Det}$} & {$F_{Det}$} & {$P_{Loc}$} & {$R_{Loc}$} & {$F_{Loc}$} \\ \midrule
0.1 & 0.6982 & 0.8875 & 0.7816 & 0.6654 & 0.6704 & 0.6679 \\
0.2 & 0.671 & 0.9128 & 0.7734 & 0.6681 & 0.6812 & 0.6746 \\
0.5 & 0.5702 & 0.9611 & 0.7157 & 0.6917 & 0.7199 & 0.7055 \\ \bottomrule
\end{tabular}
\caption{Effect of varying the RS-DPO $\eta$ (reward gap) parameter at an NV fraction of 2\% for the Qwen3-4B model without CoT. Lowering $\eta$ to 0.1 improves precision and slightly enhances detection performance, whereas increasing $\eta$ to 0.5 improves localization but incurs a substantial drop in precision and overall detection. We select $\eta=0.2$ as a balanced trade-off between detection and localization performance.}
\label{tab:non_cot_eta_ablations}
\end{table}

\begin{table}[!tbh]
\centering
\begin{tabular}{@{}lrrrrrr@{}}
\toprule
\multirow{2}{*}{\textbf{\begin{tabular}[c]{@{}l@{}}Fraction of \\ NV data\end{tabular}}} & \multicolumn{3}{r}{\textbf{Detection}} & \multicolumn{3}{r}{\textbf{Localization}} \\ \cmidrule(l){2-7} 
 & {$P_{Det}$} & {$R_{Det}$} & {$F_{Det}$} & {$P_{Loc}$} & {$R_{Loc}$} & {$F_{Loc}$} \\ \midrule
2\% & 0.9226 & 0.8901 & 0.906 & 0.7688 & 0.7638 & 0.7663 \\
5\% & 0.9234 & 0.8643 & 0.8929 & 0.771 & 0.7571 & 0.764 \\ \bottomrule
\end{tabular}
\caption{Effect of varying the fraction of \texttt{NO VIOLATIONS FOUND} instances in the training data for \textsc{MetaLint} Qwen3-4B model with CoT. We perform limited ablations because of the insights from the non CoT model training.}
\label{tab:cot_dpo_ablations}
\end{table}

\begin{table}[!tbh]
\centering
\begin{tabular}{@{}lrrrrrr@{}}
\toprule
\multirow{2}{*}{\textbf{\begin{tabular}[c]{@{}l@{}}Fraction of \\ NV data\end{tabular}}} & \multicolumn{3}{r}{\textbf{Detection}} & \multicolumn{3}{r}{\textbf{Localization}} \\ \cmidrule(l){2-7} 
 & {$P_{Det}$} & {$R_{Det}$} & {$F_{Det}$} & {$P_{Loc}$} & {$R_{Loc}$} & {$F_{Loc}$} \\ \midrule
1\% & 0.654 & 0.6468 & 0.6504 & 0.491 & 0.4788 & 0.4848 \\
2\% & 0.6636 & 0.6057 & 0.6333 & 0.4869 & 0.4745 & 0.4806 \\ \bottomrule
\end{tabular}
\caption{Effect of varying the fraction of \texttt{NO VIOLATIONS FOUND} instances in the training data for \textsc{MetaLint} Llama3.2-3B-Instruct model. We perform limited ablations because of the insights from the non CoT model training.}
\label{tab:llama3.2_3b_instruct_dpo_ablations}
\end{table}

\begin{table}[!tbh]
\centering
\resizebox{\textwidth}{!}{%
\begin{tabular}{@{}lrrrrrr@{}}
\toprule
\multirow{2}{*}{\textbf{Ensembling Strategy}} & \multicolumn{3}{r}{\textbf{Detection}} & \multicolumn{3}{r}{\textbf{Localization}} \\ \cmidrule(l){2-7}
 & \textbf{$P_{Det}$} & \textbf{$R_{Det}$} & \textbf{$F_{Det}$} & \textbf{$P_{Loc}$} & \textbf{$R_{Loc}$} & \textbf{$F_{Loc}$} \\ \midrule
Qwen3-4B DPO Intersection (non-CoT \& CoT) & 0.0667 & 0.0033 & 0.0063 & 0.0000 & 0.0000 & 0.0000 \\
Qwen3-4B DPO Union (non-CoT \& CoT) & 0.6188 & 0.4187 & 0.4995 & 0.1952 & 0.1251 & 0.1524 \\
\bottomrule
\end{tabular}
}
\caption{\textbf{Effect of Simple Ensembling Strategies on Detection and Localization Performance:} Comparison of intersection and union-based ensembling of SFT and RS-SFT models on Qwen3-4B. The intersection strategy retains only line numbers predicted by both models, while the union strategy aggregates all predicted line numbers from either model. Both ensembling approaches underperform the individual models, with the intersection strategy exhibiting particularly poor precision and recall, suggesting that the observed precision–recall differences are driven by training setup rather than a fundamental trade-off.}
\label{tab:ensemble_strategies}
\end{table}

\begin{table}[!tbh]
\centering
\resizebox{\textwidth}{!}{%
\begin{tabular}{@{}lrrrrrr@{}}
\toprule
\multirow{2}{*}{\textbf{Model}} & \multicolumn{3}{r}{\textbf{Detection}} & \multicolumn{3}{r}{\textbf{Localization}} \\ \cmidrule(l){2-7}
 & \textbf{$P_{Det}$} & \textbf{$R_{Det}$} & \textbf{$F_{Det}$} & \textbf{$P_{Loc}$} & \textbf{$R_{Loc}$} & \textbf{$F_{Loc}$} \\ \midrule
Qwen3-4B (SFT + RS-DPO) & 0.7031 & 0.7043 & 0.7037 & 0.3536 & 0.1930 & 0.2497 \\
Qwen3-4B (SFT + RS-DPO, 0\% NV) & 0.6658 & \textbf{0.8037} & \textbf{0.7283} & \textbf{0.3797} & \textbf{0.2063} & \textbf{0.2674} \\
Qwen3-4B w CoT (RS-SFT + RS-DPO) & \textbf{0.9303} & 0.4958 & 0.6468 & 0.3482 & 0.2169 & 0.2673 \\
\bottomrule
\end{tabular}
}
\caption{\textbf{Effect of Removing NO VIOLATIONS (NV) Data on Hard PEP Benchmark Performance:} Full detection and localization results for Qwen3-4B trained with SFT+RS-DPO under different NV data settings. The 0\% NV variant achieves substantially higher recall than the model reported in the main paper, with only a moderate reduction in precision. Since the precision drop is smaller than the recall gain, this leads to an improved overall F score, confirming that the originally reported recall was an underestimate and further strengthening our state-of-the-art recall claim.}
\label{tab:hard_pep_no_nv}
\end{table}

\subsection{PEP Benchmark Creation Additional Details}
\label{sec:appendix:pep_benchmark_creation_additional_details}
As discussed in section~\ref{sec:experiments:pep_idiom_benchmark} we use some high recall heuristics to find promising candidates for detecting the selected hard PEP best practice idioms.
These are summarized in Table~\ref{tab:appendix:experiments:pep_benchmark_heuristics_part1}, \ref{tab:appendix:experiments:pep_benchmark_heuristics_part2} and \ref{tab:appendix:experiments:pep_benchmark_heuristics_part3}.

\paragraph{Annotation protocol.}
Three annotators labeled the benchmark, all domain experts and co-authors of the paper who were not involved in any of the training experiments.
Annotators A and B split the full 551 file-PEP pairs \emph{disjointly}, each independently responsible for a non-overlapping portion.
Annotator C independently re-annotated \emph{all} 551 pairs as a blind verification pass, without access to A or B's labels.
Annotators were instructed to consult the official PEP webpage for the relevant PEP when making each judgment, grounding labels in the authoritative specification.
Expert annotators were deliberately chosen over crowdworkers because accurate PEP violation annotation requires deep Python knowledge; we acknowledge that shared author background is a limitation, but the high agreement reflects genuine clarity of the annotation scheme.

We compute inter-annotator agreement using Cohen's $\kappa$. $\kappa$ was computed pairwise between annotator C and the primary annotator (A or B) for each example. We converted each annotated range (for example, lines 13 to 15) into per-line binary labels, where a line containing a violation is marked as 1 and all others as 0.
Cohen's $\kappa$ was then computed over all lines for each file and averaged across the dataset, yielding $\kappa = 0.95$.
The high agreement reflects the clarity and near objectivity of the annotation scheme.

\begin{table}[!tbh]
\centering
\resizebox{\textwidth}{!}{%
\begin{tabular}{@{}rlll@{}}
\toprule
\multicolumn{1}{l}{PEP} &
  Description &
  Heuristics &
  Example \\ \midrule
506 &
  \begin{tabular}[c]{@{}l@{}}Adds secrets module \\ to the standard library \\ for cryptographically secure \\ random value generation\end{tabular} &
  \begin{tabular}[c]{@{}l@{}}Conjunction of 2 conditions: \\ 1. Presence of "random" module imports\\ 2. Presence of "random" function usage\end{tabular} &
  \begin{tabular}[c]{@{}l@{}}characters = string.ascii\_letters + \\ string.punctuation  + string.digits\\ password =  "".join(random.choice\\ (characters) for x in range(16))\\ Use instead:\\ characters = string.ascii\_letters + \\ string.punctuation  + string.digits\\ password =  "".join(secrets.choice\\ (characters) for x in range(16))\end{tabular} \\ \midrule
557 &
  \begin{tabular}[c]{@{}l@{}}Introduces the dataclasses \\ module, enabling automatic \\ generation of common boilerplate \\ methods for classes\end{tabular} &
  \begin{tabular}[c]{@{}l@{}}Conjunction of 2 conditions: \\ 1. There is a class with manual \\ implementation of "\_\_init\_\_" method \\ 2. On the same class there is manual \\ implementation of common \\ special methods or comparison methods \\ that follow standard data storage patterns.\end{tabular} &
  \begin{tabular}[c]{@{}l@{}}"class Point:\\     def \_\_init\_\_(self, x, y):\\         self.x = x\\         self.y = y\\     def \_\_repr\_\_(self):\\         return f""Point(x=\{self.x\}, y=\{self.y\})""\\ Use instead:\\ from dataclasses import dataclass\\ @dataclass\\ class Point:\\     x: int\\     y: int"\end{tabular} \\ \midrule
655 &
  \begin{tabular}[c]{@{}l@{}}Introduces Required{[}{]} and \\ NotRequired{[}{]} type qualifiers \\ to replaces cumbersome \\ TypedDict inheritance patterns.\end{tabular} &
  \begin{tabular}[c]{@{}l@{}}Conjuction of:\\ 1. "TypedDict" defined with inheritance pattern.\\ 2. total=False parameter usage in class definition\end{tabular} &
  \begin{tabular}[c]{@{}l@{}}class \_MovieBase(TypedDict):  \# implicitly total=True\\     title: str\\ class Movie(\_MovieBase, total=False):\\     year: int\\ Use instead:\\ class Movie(TypedDict):\\     title: str\\     year: NotRequired{[}int{]}\end{tabular} \\ \midrule
634 &
  \begin{tabular}[c]{@{}l@{}}Introduced structural pattern \\ matching, enabling more \\ expressive and concise ways\\ to match data structures \\ and control flow.\end{tabular} &
  \begin{tabular}[c]{@{}l@{}}Multiple consecutive if-elif-else \\ statements that compare a \\ single variable against different \\ values with dysjunction of 2 conditions:\\ 1. Length of ladder (number of \\ conditons at the "top level" + one level in) \textgreater{}= 6\\ 2. Depth of ladder (degree of nesting) \textgreater{}=3\end{tabular} &
  \begin{tabular}[c]{@{}l@{}}"def handle\_response(response):\\     if isinstance(response, dict):\\         if ""error"" in response:\\             print(f""Error: \{response{[}'error'{]}\}"")\\         elif ""data"" in response:\\             print(f""Data: \{response{[}'data'{]}\}"")\\         else:\\             print(""Unknown response format"")\\     elif isinstance(response, list):\\         print(""List of items:"", response)\\     else:\\         print(""Invalid response type"")\\ Use instead:\\ def handle\_response(response):\\     match response:\\         case \{""error"": error\_message\}:\\             print(f""Error: \{error\_message\}"")\\         case \{""data"": data\_content\}:\\             print(f""Data: \{data\_content\}"")\\         case list(items):\\             print(""List of items:"", items)\\         case \_:             print(""Invalid response type"")"\end{tabular} \\ \midrule
614 & 
\begin{tabular}[c]{@{}l@{}}Removes previous restrictions \\ on decorator syntax. Before, \\ only simple names or dotted \\ names were valid decorators. \\ After 614, any valid  expression \\ can be used as a decorator\end{tabular}
   &
     \begin{tabular}[c]{@{}l@{}}
     Conjunction of 2 conditions: \\
1. A decorator is applied using a name \\ (e.g., @decorator) where that name is \\ assigned earlier in the code. \\
2. The assignment value is an expression \\ of type Call, Attribute, or Subscript (e.g., \\deco = factory(), deco = module.decorator, \\ deco = decorators[i]).
     \end{tabular}
   &
   \begin{tabular}[c]{@{}l@{}}\#
def uppercase(func):\\
    def wrapper(*args, **kwargs):\\
        return func(*args, **kwargs).upper()\\
    return wrapper\\
@uppercase\\
def greet():\\
    return "hello"\\
Use Instead:\\
deco = [uppercase]\\
@deco[0]\\
def greet2():\\
    return "hi"\end{tabular}
   \\ \midrule
616 &
  \begin{tabular}[c]{@{}l@{}}Replaces manual slicing \\ with dedicated methods\end{tabular} &
  \begin{tabular}[c]{@{}l@{}}dysjunction of 2 conditions:\\ 1. There is a "check" with \\ startswith or endswith on a \\ given variable x.\\ 2. On the same variable x \\ check if there is an "edit" using a \\ program slicing syntax or using "replace()".\end{tabular} &
  \begin{tabular}[c]{@{}l@{}}if s.startswith(prefix): s = s{[}len(prefix):{]}\\ Use instead:\\ s = s.removeprefix(prefix)\\ OR\\ s{[}:-len(suffix){]}\\ Use instead:\\ s.removesuffix(suffix)\end{tabular} \\ \midrule
584 &
  \begin{tabular}[c]{@{}l@{}}Introduces the binary operators \\ | (merge) and |= (update) on \\ dict (and other built-in mapping\\  types), providing an expressive, \\ in-place-or-new‐object way to \\ combine dictionaries.\end{tabular} &
  \begin{tabular}[c]{@{}l@{}}disjunction of two conditions:\\ 1. A copy-and-update sequence \\ on the same variable \\ or in close proximity: d = d1.copy() \\ followed by d.update(d2)\\ 2. A dictionary literal using multiple \\ unpackings \{**d1, **d2\}, \\ indicating ad-hoc merging rather \\ than the new operators\end{tabular} &
  \begin{tabular}[c]{@{}l@{}}d1 = \{'a': 1, 'b': 2\} d2 = \{'c': 3, 'd': 4\}\\ merged = d1.copy()  \\ merged.update(d2)\\ d1 = \{'a': 1, 'b': 2\} d2 = \{'c': 3, 'd': 4\}\\ merged = \{**d1, **d2\}\\ d1 = \{'a': 1, 'b': 2\} d2 = \{'c': 3, 'd': 4\}\\ merged = dict(list(d1.items()) + list(d2.items()))\\ Use instead:\\ d1 = \{'a': 1, 'b': 2\} d2 = \{'c': 3, 'd': 4\}\\ merged = d1 | d2\\ d1 = \{'a': 1, 'b': 2\} d2 = \{'c': 3, 'd': 4\}\\ d1 |= d2  \# d1 is now \{'a': 1, 'b': 2, 'c': 3, 'd': 4\}\end{tabular}
\\ \bottomrule
\end{tabular}%
}
\caption{High recall heuristics used to find instances of PEP violations that human annotators vet}
\label{tab:appendix:experiments:pep_benchmark_heuristics_part1}
\end{table}

\begin{table}[!tbh]
\centering
\resizebox{\textwidth}{!}{%
\begin{tabular}{@{}rlll@{}}
\toprule
\multicolumn{1}{l}{PEP} &
  Description &
  Heuristics &
  Example \\ \midrule
570 &
  \begin{tabular}[c]{@{}l@{}}Introduces new syntax \\ (the / marker) in Python \\ function signatures to \\ specify positional-only \\ parameters, ensuring that \\ certain arguments can \\ only be supplied by their \\ position and not as keywords\end{tabular} &
  \begin{tabular}[c]{@{}l@{}}Conjunction of the \\ following conditions:\\ 1. Have only positional-or-\\ keyword parameters \\ (without *args, **kwargs, keyword\\ -only parameters, or the '/' marker),\\ 2. Include 2 to 4 parameters, all of \\ which have no default values\end{tabular} &
  \begin{tabular}[c]{@{}l@{}}def compute\_area(width, height):\\     return width * height\\ area = compute\_area(width=5, height=10)\\ print("Area:", area)\\ Use instead:\\ def compute\_area(width, height, /):\\     return width * height\\ area = compute\_area(5, 10)\\ print("Area:", area)\end{tabular} \\ \midrule
567 &
  \begin{tabular}[c]{@{}l@{}}Adds the contextvars module, \\ enabling context-local variables \\ for managing dynamic state.\end{tabular} &
  \begin{tabular}[c]{@{}l@{}}Dysjunction of the following conditions:\\ 1. Look for import threading together \\ with threading.local() object creation and use.\\ 2. Find global statements or assignment \\ to variables at the module \\ level that are  accessed or mutated in functions, \\ especially as shared state.\\ 3. Identify async functions or classes \\ where context or state \\ variables are passed as parameters \\ (e.g., def func(context, ...) or async \\ def func(context, ...)), not as context-local variables.\end{tabular} &
  \begin{tabular}[c]{@{}l@{}}import threading\\ \_thread\_local = threading.local()\\ def set\_context(value):\\     \_thread\_local.value = value\\ def get\_context():\\     return getattr(\_thread\_local, 'value', None)\\ Use instead:\\ from contextvars import ContextVar\\ context\_var = ContextVar('value')\\ def set\_context(value):\\     context\_var.set(value)\\ def get\_context():\\     return context\_var.get()\end{tabular} \\ \midrule
530 &
  \begin{tabular}[c]{@{}l@{}}Enables the use of "async for" \\ and "await" in list, set, and dict \\ comprehensions as well as in \\ generator expressions, providing \\ concise asynchronous data \\ processing within comprehensions\end{tabular} &
  \begin{tabular}[c]{@{}l@{}}Dysjunction of the following conditions:\\ 1. "async" def functions that uses "async for" \\ loops to build lists, sets, or dicts.\\ 2. "async for" loops, followed by methods like \\ result.append(...), result.extend(...), or result{[}key{]} = ....\\ 3. Comprehensions written without the "async for" \\ clause despite being inside an "async def"\end{tabular} &
  \begin{tabular}[c]{@{}l@{}}result = {[}{]}\\ async for i in aiter():\\ if i \% 2:\\ result.append(i)\\ Use instead:\\ result = {[}i async for i in aiter() if i \% 2{]}\end{tabular} \\ \midrule
525 &
  \begin{tabular}[c]{@{}l@{}}Introduces the ability to define \\ asynchronous generator functions\\ using the async def and yield \\ syntax, enabling concise, native \\ support for asynchronous iteration.\end{tabular} &
  \begin{tabular}[c]{@{}l@{}}Dysjunction of the following conditions:\\ 1.  classes defining both "\_\_aiter\_\_" and "\_\_anext\_\_" \\ methods, especially where the class is used solely to \\ produce a sequence of values asynchronously.\\ 2. async def functions that create and return custom \\ iterator classes instead of using async def with yield.\end{tabular} &
  \begin{tabular}[c]{@{}l@{}}class Ticker:\\     """Yield numbers from 0 to `to`\\  every `delay` seconds."""\\     def \_\_init\_\_(self, delay, to):\\         self.delay = delay\\         self.i = 0\\         self.to = to\\     def \_\_aiter\_\_(self):\\         return self\\     async def \_\_anext\_\_(self):\\         i = self.i\\         if i \textgreater{}= self.to:\\             raise StopAsyncIteration\\         self.i += 1\\         if i:\\             await asyncio.sleep(self.delay)\\         return i\\ Use instead:\\ async def ticker(delay, to):\\     """Yield numbers from 0 to `to` \\ every `delay` seconds."""\\     for i in range(to):\\         yield i\\         await asyncio.sleep(delay)\end{tabular} \\ \midrule
520 &
  \begin{tabular}[c]{@{}l@{}}Ensures that the order in \\ which attributes are defined within \\ a class body is preserved in the \\ resulting class object, making \\ the attribute order predictable \\ and consistent.\end{tabular} &
  \begin{tabular}[c]{@{}l@{}}Dysjunction of the following conditions:\\ 1. Uses sorted() or otherwise processes \\ class.\_\_dict\_\_.keys() to impose attribute order.\\ 2. Attribute names are tracked in a list or similar\\ structure solely to maintain definition order.\\ 3. Custom metaclass logic or "\_\_prepare\_\_" impleme-\\ ntations created to preserve the order of class attributes.\end{tabular} &
  \begin{tabular}[c]{@{}l@{}}class Person:\\ name = "Alice" \\ age = 30\\ city = "Wonderland"\\ def display\_attributes(self):\\ \# Manually sorting keys \\ for key in sorted(self.\_\_class\_\_.\_\_dict\_\_.keys()):\\     if not key.startswith("\_\_"):\\           print(key, getattr(self, key))\\ Use instead:\\ class Person:\\ name = "Alice"\\ age = 30\\ city = "Wonderland"\\ def display\_attributes(self):\\ \# Directly iterate over the preserved definition order\\ for key in self.\_\_class\_\_.\_\_definition\_order\_\_:\\     print(key, getattr(self, key))\end{tabular} \\ \midrule
498 &
  \begin{tabular}[c]{@{}l@{}}Introduces f-strings (formatted \\ string literals) as a new, \\ concise, and efficient way \\ to embed Python \\ expressions inside string \\ literals using the f'' prefix.\end{tabular} &
  \begin{tabular}[c]{@{}l@{}}Dysjunction of the following conditions:\\ 1. Occurrences of string literals with \\ .format(...) applied, \\ especially where keys or variables match \\ braces in the string \\ 2. String literals concatenated using \\ "+" with variables.\\ 3. Uses of the "\%" operator for string formatting,\end{tabular} &
  \begin{tabular}[c]{@{}l@{}}name = "Alice"\\ age = 30\\ greeting = "Hello, " + name + \\ "! You are " + str(age) + " years old."\\ Use instead:\\ name = "Alice"\\ age = 30\\ greeting = f"Hello, \{name\}! \\ You are \{age\} years old."\\ OR\\ value = 12.3456\\ formatted = "The value is \{:.2f\}".format(value)\\ Use instead:\\ value = 12.3456\\ formatted = f"The value is \{value:.2f\}"\end{tabular} \\ \\ \bottomrule
\end{tabular}%
}
\caption{High recall heuristics used to find instances of PEP violations that human annotators vet}
\label{tab:appendix:experiments:pep_benchmark_heuristics_part2}
\end{table}

\begin{table}[!tbh]
\centering
\resizebox{\textwidth}{!}{%
\begin{tabular}{@{}rlll@{}}
\toprule
\multicolumn{1}{l}{PEP} &
  Description &
  Heuristics &
  Example \\ \midrule
487 &
  \begin{tabular}[c]{@{}l@{}}Makes customizing class \\ creation and subclass initialization \\ easier by introducing \_\_init\_\\ subclass\_\_ and \_\_set\_name\_\_, \\ eliminating the need for \\ most custom metaclasses\end{tabular} &
  \begin{tabular}[c]{@{}l@{}}Disjunction of the following conditions:\\ 1. Custom metaclasses defined to \\ execute code during class creation \\ or subclassing (e.g., overriding \_\_new\_\_, \\ \_\_init\_\_, or \_\_call\_\_ in metaclasses) \\ instead of using \_\_init\_subclass\_\_.\\ 2. Descriptor classes lacking \_\_set\_\\ name\_\_ method and employing \\ manual workarounds to determine \\ their assigned attribute names.\\ 3. Classes or frameworks manually \\ tracking or registering sub-\\ classes via metaclass hooks instead \\ of leveraging \_\_init\_subclass\_\_.\end{tabular} &
  \begin{tabular}[c]{@{}l@{}}class Meta(type):\\     def \_\_new\_\_(meta, name, bases, namespace):\\         for key, value in namespace.items():\\             if isinstance(value, Descriptor):\\                 value.name = key\\         return super().\_\_new\_\_\\ (meta, name, bases, namespace)\\ class MyClass(metaclass=Meta):\\     attr = Descriptor()\\ Use instead:\\ class Descriptor:\\     def \_\_set\_name\_\_(self, owner, name):\\         self.name = name\\ class MyClass:\\     attr = Descriptor()\\ OR\\ class PluginBase(type):\\     plugins = \{\}\\     def \_\_new\_\_(meta, name, bases, namespace):\\         if name != 'Plugin':\\             meta.plugins{[}name{]} = namespace{[}'priority'{]}\\         return super().\_\_new\_\_\\ (meta, name, bases, namespace)\\ class Plugin(metaclass=PluginBase):\\     priority = 0\\ class HighPriority(Plugin):\\     priority = 10\\ Use instead:\\ class Plugin:\\     plugins = \{\}\\     priority = 0\\     def \_\_init\_subclass\_\_(cls, **kwargs):\\         super().\_\_init\_subclass\_\_(**kwargs)\\         cls.plugins{[}cls.\_\_name\_\_{]} = cls.priority\\ class HighPriority(Plugin):\\     priority = 10\end{tabular} \\ \midrule
  593 &
 \begin{tabular}[c]{@{}l@{}}
Introduces flexible function \\ and variable annotations via \\ typing.Annotated, which lets you \\attach context-specific metadata \\to type hints (e.g., validation \\ constraints, units)
 \end{tabular}
   &
   \begin{tabular}[c]{@{}l@{}}
   Conjunction of 2 conditions:\\
1. Type hints are already present \\ in function arguments, return \\ types, or variable annotations.\\
2. Nearby comments/docstrings (within \\ ±2 lines) contain metadata-like \\ patterns such as "min", "max", \\ "nullable", "regex", "enum", \\ "unit", "deprecated", etc.
   \end{tabular}
   & 
  \begin{tabular}[c]{@{}l@{}}\# max 100, min 1\\
def set\_age(age: int) -\textgreater None:\\
    pass\\
\\
Use instead:\\
\\
from typing import Annotated\\
Age = Annotated[int, "min=1", "max=100"]\\
def set\_age(age: Age) -\textgreater None:\\
    pass\\
  \end{tabular} 
   \\ \midrule
526 &
  \begin{tabular}[c]{@{}l@{}}introduces explicit variable \\ annotations, allowing type \\ hints directly on variable \\ declarations for local, global, \\ and class variables in Python\end{tabular} &
  \begin{tabular}[c]{@{}l@{}}Disjunction of the following conditions:\\ 1. Variables assigned values with a \\ type comment (e.g., x = 0 \# type: int) \\ instead of using annotation syntax.\\ 2. Identify variable assignments, \\ especially class and \\ instance attributes, that lack any \\ type annotation (e.g., name = "" in class bodies).\\ 3.Module-level variables assigned \\ values without accompanying type hints—\\ especially in type-annotated codebases.\end{tabular} &
  \begin{tabular}[c]{@{}l@{}}\# type: List{[}int{]}\\ numbers = {[}{]}\\ Use instead:\\ numbers: List{[}int{]} = {[}{]}\\ OR\\ class Player:\\ \# type: str\textless{}br\textgreater    name = "Guest"\\ Use instead:\\ class Player:\\     name: str = "Guest"\end{tabular} \\ \midrule
589 &
  \begin{tabular}[c]{@{}l@{}}Introduces TypedDict, enabling \\ precise type hints for \\ dictionaries with a fixed \\ set of string keys, improving \\ static type checking and \\ readability in Python code.\end{tabular} &
  \begin{tabular}[c]{@{}l@{}}Disjunction of the following conditions:\\ 1. Dictionary literals or variables \\ consistently using the same fixed set of \\ string keys without accompanying \\ TypedDict annotations.\\ 2. Functions annotated with broad \\ dictionary types like Dict{[}str, Any{]}, dict, \\ or untyped parameters/returns that actually \\ expect dictionaries with a known fixed set of keys.\\ 3. Explicit key presence checks or \\ accessing dictionary keys repeatedly that suggest \\ a structured dictionary shape.\end{tabular} &
  \begin{tabular}[c]{@{}l@{}}movie = \{'name': 'Blade Runner',\\          'year': 1982\}\\ Use instead:\\ from typing import TypedDict\\ class Movie(TypedDict):\\     name: str\\     year: int\\ movie: Movie = \{'name': 'Blade Runner',\\                 'year': 1982\}\end{tabular} \\ \midrule
572 &
  \begin{tabular}[c]{@{}l@{}}Introduces the assignment \\ expression operator := \\ (the "walrus operator"), \\ allowing assignment to \\ variables within expressions,\end{tabular} &
  \begin{tabular}[c]{@{}l@{}}1. Patterns where a value is first assigned \\ to a variable, and then immediately checked \\ or used in the next line or inside a \\ loop, list comprehension, or condition.\\ 2.  separate assignment and conditional test statements\end{tabular} &
  \begin{tabular}[c]{@{}l@{}}match = pattern.search(data)  \\ if match is not None:  \\     process(match)\\ Use instead:\\ if (match := pattern.search(data)) is not None:  \\     process(match)\end{tabular} \\ \bottomrule
\end{tabular}%
}
\caption{High recall heuristics used to find instances of PEP violations that human annotators vet}
\label{tab:appendix:experiments:pep_benchmark_heuristics_part3}
\end{table}
\section{More Results}
\label{sec:appendix:more_results}

\begin{table}[!tbh]
\centering

\resizebox{\textwidth}{!}{%
\begin{tabular}{@{}lrrrrrrrrr@{}}
\toprule
\multirow{2}{*}{\textbf{Model}} & \multicolumn{3}{r}{\textbf{In-Domain}} & \multicolumn{3}{r}{\textbf{Near Transfer}} & \multicolumn{3}{r}{\textbf{Far Transfer}} \\ \cmidrule(l){2-10} 
 & \textbf{$P_{Det}$} & \textbf{$R_{Det}$} & \textbf{$F_{Det}$} & \textbf{$P_{Det}$} & \textbf{$R_{Det}$} & \textbf{$F_{Det}$} & \textbf{$P_{Det}$} & \textbf{$R_{Det}$} & \textbf{$F_{Det}$} \\ \midrule
Qwen3-4B & 0.45 & 0.14 & 0.22 & 0.58 & 0.24 & 0.34 & 0.54 & 0.29 & 0.38 \\

\multicolumn{1}{r}{+SFT} & \textbf{0.93 {\color{mygreen}(+0.48)}} & 0.74 {\color{mygreen}(+0.6)} & 0.83 {\color{mygreen}(+0.61)} & \textbf{0.89 {\color{mygreen}(+0.31)}} & 0.24 {\color{mygreen}(+0)} & 0.38 {\color{mygreen}(+0.04)} & 0.72 {\color{mygreen}(+0.18)} & 0.27 {\color{myred}(-0.02)} & 0.39 {\color{mygreen}(+0.01)} \\

\multicolumn{1}{r}{+RS-DPO} & 0.72 {\color{mygreen}(+0.27)} & \textbf{1 {\color{mygreen}(+0.86)}} & \textbf{0.83 {\color{mygreen}(+0.61)}} & 0.76 {\color{mygreen}(+0.18)} & \textbf{0.8 {\color{mygreen}(+0.56)}} & \textbf{0.78 {\color{mygreen}(+0.44)}} &\textbf{ 0.75 {\color{mygreen}(+0.21)}} & \textbf{0.81 {\color{mygreen}(+0.52)}} & \textbf{0.78 {\color{mygreen}(+0.4)}} \\ \midrule
Qwen3-4B w CoT & 0.87 & 0.5 & 0.63 & 0.95 & 0.88 & 0.91 & 0.87 & 0.68 & 0.76 \\

\multicolumn{1}{r}{+RS-SFT} & \textbf{0.87 {\color{mygreen}(+0)}} & 0.73 {\color{mygreen}(+0.23)} & 0.8 {\color{mygreen}(+0.17)} & \textbf{0.97 {\color{mygreen}(+0.02)}} & 0.86 {\color{myred}(-0.02)} & 0.91 {\color{mygreen}(+0)} & \textbf{0.94 {\color{mygreen}(+0.07)}} & 0.82 {\color{mygreen}(+0.14)} & 0.88 {\color{mygreen}(+0.12)} \\

\multicolumn{1}{r}{+RS-DPO} & 0.86 {\color{myred}(-0.1)} & \textbf{0.85 {\color{mygreen}(+0.35)}} & \textbf{0.85 {\color{mygreen}(+0.22)}} & \textbf{0.97 {\color{mygreen}(+0.02)}} & \textbf{0.92 {\color{mygreen}(+0.04)}} & \textbf{0.94 {\color{mygreen}(+0.03)}} & 0.92 {\color{mygreen}(+0.05)} & \textbf{0.86 {\color{mygreen}(+0.18)}} & \textbf{0.89 {\color{mygreen}(+0.13)}} \\
\midrule

Llama3.2-3B-Instruct & 0.54 & 0.43 & 0.48 & 0.69 & 0.68 & 0.69 & 0.47 & 0.51 & 0.49 \\

\multicolumn{1}{r}{+SFT} & \textbf{0.88 {\color{mygreen}(+0.34)}} &\textbf{ 0.87 {\color{mygreen}(+0.44)}} & \textbf{0.88 {\color{mygreen}(+0.4)}} & \textbf{0.89 {\color{mygreen}(+0.2)}} & 0.44 {\color{myred}(-0.24)} & 0.59 {\color{myred}(-0.1)} & 0.61 {\color{mygreen}(+0.14)} & 0.27 {\color{myred}(-0.24)} & 0.37 {\color{myred}(-0.12)} \\

\multicolumn{1}{r}{+RS-DPO} & 0.75 {\color{mygreen}(+0.21)} & 0.92 {\color{mygreen}(+0.49)} & 0.83 {\color{mygreen}(+0.35)} & 0.81 {\color{mygreen}(+0.12)} & \textbf{0.71 {\color{mygreen}(+0.03)}} & \textbf{0.76 {\color{mygreen}(+0.07)}} & \textbf{0.61 {\color{mygreen}(+0.14)}} & \textbf{0.59 {\color{mygreen}(+0.08)}} & \textbf{0.60 {\color{mygreen}(+0.11)}} \\
\bottomrule
\end{tabular}
}
\caption{\textbf{Cross-Idiom Generalization on Python Ruff Best Practice Idioms by Transfer Setting:} We evaluate the effect of different \textsc{MetaLint} training setups (SFT, RS-SFT, and RS-DPO) on Qwen3-4B (with and without reasoning) and Llama3.2-3B. Models are trained on easy synthetic Python Ruff best practice idioms, and the performance is reported on other Ruff best practice idioms with varying levels of transfer - In-Domain, Near Transfer, and Far Transfer.}
\label{tab:gen_synth_python_ruff_idioms_per_transfer_setting_full}
\end{table}

\begin{table*}[!tbh]
\centering
\resizebox{\textwidth}{!}{%
\begin{tabular}{@{}llrrrrrr@{}}
\toprule
\multirow{2}{*}{\textbf{Model}} & \multirow{2}{*}{\textbf{Examples?}} & \multicolumn{3}{r}{\textbf{Detection}} & \multicolumn{3}{r}{\textbf{Localization}} \\ \cmidrule(l){3-8}
 &  & \textbf{$P_{Det}$} & \textbf{$R_{Det}$} & \textbf{$F_{Det}$} & \textbf{$P_{Loc}$} & \textbf{$R_{Loc}$} & \textbf{$F_{Loc}$} \\ \midrule
Qwen3-4B & Yes & 0.5267 & 0.1715 & 0.2587 & 0.0954 & 0.0824 & 0.0884 \\
Qwen3-4B w CoT & Yes & 0.8154 & 0.3986 & 0.5354 & 0.2625 & 0.1467 & 0.1882 \\
Qwen3-4B (SFT + RS-DPO) & Yes & 0.7031 & 0.7043 & 0.7037 & 0.3536 & 0.1930 & 0.2497 \\
Qwen3-4B w CoT (RS-SFT + RS-DPO) & Yes & \textbf{0.9303} & 0.4958 & 0.6468 & \textbf{0.3482} & 0.2169 & \textbf{0.2673} \\
\midrule
Qwen3-4B & No & 0.5556 & 0.1736 & 0.2645 & 0.0802 & 0.0733 & 0.0766 \\
Qwen3-4B w CoT & No & 0.8304 & 0.3669 & 0.5089 & 0.2572 & 0.1542 & 0.1928 \\
Qwen3-4B (SFT + RS-DPO) & No & 0.7167 & \textbf{0.7268} & \textbf{0.7217} & 0.3396 & 0.1909 & 0.2444 \\
Qwen3-4B w CoT (RS-SFT + RS-DPO) & No & 0.8215 & 0.4249 & 0.5601 & 0.2860 & \textbf{0.1983} & 0.2342 \\
\bottomrule
\end{tabular}
}
\caption{\textbf{Ablation on In-Context Examples for the Hard PEP Benchmark:} Effect of including before/after code examples in the prompt for base Qwen3-4B models and final \textsc{MetaLint} models trained with SFT and DPO. Across most settings, removing examples results in minimal changes in detection and localization performance, indicating that the natural-language descriptions alone are often sufficient. The only notable degradation occurs for the CoT-trained \textsc{MetaLint} model, which exhibits a modest drop in detection performance without examples due to an increased tendency to predict \emph{NO VIOLATIONS}, suggesting that this more conservative variant benefits more from additional in-context guidance.}
\label{tab:pep_example_ablation}
\end{table*}

\subsection{Instruction-Following Framing as the Mechanism of Transfer}
Table~\ref{tab:instr_free_ablation} compares the zero-shot base model, an instruction-free SFT variant trained on the same 50,028 Ruff examples but with every rule-specific prompt replaced by a generic ``find code quality issues'' instruction, and the full \textsc{MetaLint} model.
Under strict kind-matched evaluation (a detection counts as a true positive only if the predicted violation kind matches the target PEP number), the instruction-free model collapses to 0.00\% on both metrics despite being fine-tuned on more task-relevant data than the zero-shot baseline.
This directly demonstrates that the instruction-following framing---not data volume or DPO---is what enables cross-rule generalization.

\begin{table}[!tbh]
\centering
\begin{tabular}{@{}lcc@{}}
\toprule
\textbf{Model} & \textbf{$F_{\text{Det}}$ (\%)} & \textbf{$F_{\text{Loc}}$ (\%)} \\ \midrule
Qwen3-4B (zero-shot, with PEP spec) & 25.9 & 8.8 \\
Instruction-free SFT (same 50K Ruff data, no spec) & 0.0 & 0.0 \\
\textsc{MetaLint} (SFT+RS-DPO, with PEP spec) & 70.4 & 25.0 \\
\bottomrule
\end{tabular}
\caption{\textbf{Instruction-Following Framing Ablation:} Training on the same 50K Ruff examples without rule specifications collapses PEP benchmark performance to exactly zero under strict kind-matched scoring, confirming that the specification is the mechanism of cross-rule generalization.}
\label{tab:instr_free_ablation}
\end{table}

\subsection{False Positive Rates on the PEP Benchmark}
Table~\ref{tab:pep_fpr} reports the False Positive Rate (FPR) on the 257 no-violation instances alongside detection recall and F-score.
High recall universally incurs elevated FPR across all models evaluated.
FPR is a controllable operating point: the NV fraction hyperparameter in RS-DPO (Table~\ref{tab:non_cot_dpo_ablations}) directly governs this tradeoff and can be tuned for deployment requirements.

\begin{table}[!tbh]
\centering
\resizebox{\linewidth}{!}{%
\begin{tabular}{@{}lrrr@{}}
\toprule
\textbf{Model} & \textbf{FPR (\%)} & \textbf{$R_{\text{Det}}$ (\%)} & \textbf{$F_{\text{Det}}$ (\%)} \\ \midrule
Qwen3-4B + \textsc{MetaLint} (SFT+RS-DPO) & 24.5 & 70.4 & 70.4 \\
Qwen3-4B w/ CoT + \textsc{MetaLint} (RS-SFT+RS-DPO) & 5.1 & 49.6 & 64.7 \\
Qwen3-8B & 2.7 & 35.7 & 49.9 \\
Qwen3-14B w/ CoT & 1.9 & 48.6 & 63.4 \\
Qwen3-32B w/ CoT & 3.9 & 56.5 & 70.5 \\
DeepSeek-R1-Distill-Qwen-32B w/ CoT & 7.8 & 59.0 & 71.3 \\
GPT-4o & 9.7 & 67.9 & 77.2 \\
GPT-4.1 & 7.8 & 64.6 & 75.5 \\
GPT-5 (high) & 8.2 & 56.7 & 70.0 \\
\bottomrule
\end{tabular}
}
\caption{\textbf{False Positive Rates on the PEP Benchmark:} FPR is the fraction of the 257 no-violation instances where the model predicts at least one violation. High recall universally incurs elevated FPR across all models. FPR is a controllable operating point via the NV fraction hyperparameter in RS-DPO (Table~\ref{tab:non_cot_dpo_ablations}).}
\label{tab:pep_fpr}
\end{table}

\subsection{Cost Comparison}

\begin{table}[!tbh]
\centering
\resizebox{\linewidth}{!}{%
\begin{tabular}{@{}llrrr@{}}
\toprule
\textbf{Model} & \textbf{Approach} & \textbf{Total cost} & \textbf{\$/example} & \textbf{Notes} \\ \midrule
GPT-5 (high reasoning) & API & \$18.64 & \$0.035 & avg 3,052 reasoning tokens/call \\
Qwen3-4B + \textsc{MetaLint} (SFT+RS-DPO) & API & \$0.11 & \$0.00020 & Qwen3-8B pricing (upper bound) \\
Qwen3-4B w/ CoT + \textsc{MetaLint} (RS-SFT+RS-DPO) & API & \$0.24 & \$0.00045 & avg 762 output tokens/call \\
\bottomrule
\end{tabular}
}
\caption{\textbf{API cost on PEP benchmark (536 examples).} GPT-5 costs 175$\times$ more than \textsc{MetaLint} (non-CoT) and 77$\times$ more than \textsc{MetaLint} CoT, driven by hidden reasoning tokens billed at output rates.}
\label{tab:pep_benchmark_cost}
\end{table}

\begin{table}[!tbh]
\centering
\resizebox{\linewidth}{!}{%
\begin{tabular}{@{}llrr@{}}
\toprule
\textbf{Model} & \textbf{Approach} & \textbf{Total cost} & \textbf{Ratio vs \textsc{MetaLint}} \\ \midrule
Claude Opus 4.7 (Claude Code) & Agentic --- 60 turns, 20 min & \$8.64 & 21$\times$ \\
Claude Sonnet 4.6 (Claude Code) & Agentic --- 11 turns, 9 min & \$3.88 & 9.5$\times$ \\
Qwen3-4B w/ CoT + \textsc{MetaLint} & API (OpenRouter, upper bound) & \$0.41 & 1$\times$ \\
Qwen3-4B w/ CoT + \textsc{MetaLint} & Self-hosted (local GPU) & $\sim$\$0 & --- \\
\bottomrule
\end{tabular}
}
\caption{\textbf{Real-repo linting cost (flask, 83 files $\times$ 15 PEPs).} Claude Code costs are exact (API-reported, including prompt cache discounts). \textsc{MetaLint} API cost uses Qwen3-8B OpenRouter pricing as an upper bound for Qwen3-4B. Self-hosted cost reflects a 4B model fitting on a single GPU.}
\label{tab:real_repo_cost}
\end{table}

\subsection{Expanded Results on the PEP Hard Best Practice Benchmark}
We show the expanded results across various model sizes for the evaluated model families in Table~\ref{tab:hard_pep_benchmark_evaluation_full}. We note that most results follow the expected trends with more parameters or CoT usage leading to better performance but there are soem exceptions to the trend.
We mainly see this for cases like Qwen2.5 and Qwen2.5Coder families.
We note that Qwen2.5Coder-7B-Instruct has almost zero metrics because it always predicts \texttt{NO VIOLATIONS FOUND} for all instances and Qwen2.5Coder-14B-Instruct has really low scores because of similar reasons.
for Qwen2.5 family we notice that 32B variant performs a bit worse than 32B.

We also analyze \textsc{MetaLint} SFT models on the hard PEP benchmark and observe that they perform similarly or slightly worse than the base untrained models. 
This suggests that SFT alone may lead to overfitting on the Ruff best practice idiom distribution and struggles to generalize from easy to hard cases without DPO training. 
These findings highlight the importance of the DPO (preference-tuning) stage in the \textsc{MetaLint} pipeline. 
However, we also emphasize that while the SFT stage can limit generalization, it remains essential for effective DPO training, as it teaches the LLM to follow the correct output format and establishes a strong base policy. 
This is supported by our experiments with the CoT model, where applying RS-DPO directly to the Qwen/Qwen3-4B model (without SFT) led to near-zero performance across all metrics, as the model consistently failed to produce outputs in the required format.

\begin{table}[!tbh]
\centering
\resizebox{\textwidth}{!}{%
\begin{tabular}{@{}lrrrrrr@{}}
\toprule
 & \multicolumn{3}{r}{\textbf{Detection}} & \multicolumn{3}{r}{\textbf{Localization}} \\ \cmidrule(l){2-7} 
\multirow{-2}{*}{\textbf{Model}} & {$P_{Det}$} & {$R_{Det}$} & {$F_{Det}$} & {$P_{Loc}$} & {$R_{Loc}$} & {$F_{Loc}$} \\ \midrule
Llama3.2-3B-Instruct & 0.7042 & 0.214 & 0.3283 & 0.0691 & 0.0798 & 0.0741 \\ \midrule
Qwen3-4B & 0.5267 & 0.1715 & 0.2587 & 0.0954 & 0.0824 & 0.0884 \\
Qwen3-4B with CoT & 0.8154 & 0.3986 & 0.5354 & 0.2625 & 0.1467 & 0.1882 \\
Qwen3-8B & 0.8267 & 0.3572 & 0.4988 & 0.1806 & 0.1285 & 0.1501 \\
Qwen3-8B with CoT & 0.8886 & 0.4672 & 0.6124 & 0.3122 & 0.2029 & 0.2459 \\
Qwen3-14B & 0.9021 & 0.4612 & 0.6103 & 0.289 & 0.2521 & 0.2693 \\
Qwen3-14B with CoT & 0.9116 & 0.4857 & 0.6337 & 0.3993 & 0.2915 & 0.3369 \\
Qwen3-32B & 0.9021 & 0.5205 & 0.6601 & 0.2807 & 0.2711 & 0.2758 \\
Qwen3-32B with CoT & {\underline{0.9377}} & 0.5645 & {0.7048} & 0.4152 & 0.3086 & 0.354 \\
\midrule
Qwen2.5-3B-Instruct & 0.0667 & 0.0033 & 0.0063 & 0.0036 & 0.0036 & 0.0036 \\
Qwen2.5-7B-Instruct & 0.4333 & 0.1379 & 0.2092 & 0.0585 & 0.0518 & 0.0549 \\
Qwen2.5-14B-Instruct & 0.8017 & 0.4324 & 0.5618 & 0.2389 & 0.2158 & 0.2267 \\
Qwen2.5-32B-Instruct & 0.8667 & 0.2656 & 0.4066 & 0.163 & 0.1477 & 0.155 \\
\midrule
Qwen2.5Coder-3B-Instruct & 0.7802 & 0.411 & 0.5384 & 0.1257 & 0.0745 & 0.0936 \\
Qwen2.5Coder-7B-Instruct & 0.0667 & 0.0033 & 0.0063 & 0 & 0 & 0 \\
Qwen2.5Coder-14B-Instruct & 0.2 & 0.0443 & 0.0726 & 0.0294 & 0.0264 & 0.0278 \\
Qwen2.5Coder-32B-Instruct & 0.8961 & 0.5328 & 0.6683 & 0.3432 & 0.3077 & 0.3245 \\
\midrule
DeepSeek-R1-Distill-Qwen-7B with CoT & 0.7143 & 0.2841 & 0.4065 & 0.1064 & 0.1122 & 0.1092 \\
DeepSeek-R1-Distill-Qwen-14B with CoT & 0.69 & 0.2345 & 0.35 & 0.1856 & 0.1245 & 0.149 \\
DeepSeek-R1-Distill-Qwen-32B with CoT & 0.9008 & 0.5899 & 0.713 & 0.4015 & 0.3403 & 0.3684 \\
\midrule
GPT-oss-20b & 0.8377 & 0.3531 & 0.4968 & 0.251 & 0.1695 & 0.2024 \\
GPT-oss-120b & 0.9157 & 0.6456 & \underline{0.7573} & 0.3991 & 0.3331 & 0.3631 \\
\midrule
Qwen3-4B \textsc{MetaLint} (SFT) \textbf{(Ours)} & 0.4333 & 0.0821 & 0.1381 & 0.0432 & 0.0221 & 0.0292 \\
Qwen3-4B \textsc{MetaLint} (SFT+RS-DPO) \textbf{(Ours)} & 0.7031 & \textbf{0.7043} & 0.7037 & 0.3536 & 0.193 & 0.2497 \\
Qwen3-4B \textsc{MetaLint} w CoT (RS-SFT) \textbf{(Ours)} & 0.7615 & 0.3689 & 0.497 & 0.2785 & 0.1437 & 0.1896 \\
Qwen3-4B \textsc{MetaLint} w CoT (RS-SFT+RS-DPO) \textbf{(Ours)} & 0.9303 & 0.4958 & 0.6468 & 0.3482 & 0.2169 & 0.2673 \\
Llama3.2-3B-Instruct \textsc{MetaLint} (SFT) (\textbf{Ours}) & 0.5627 & 0.259 & 0.3547 & 0.1066 & 0.0509 & 0.0689 \\
Llama3.2-3B-Instruct \textsc{MetaLint} (SFT+RS-DPO) (\textbf{Ours}) & 0.6368 & 0.5614 & 0.5965 & 0.2364 & 0.1263 & 0.1647 \\
\midrule
o3-mini & 0.8939 & 0.5845 & 0.7068 & 0.3169 & 0.2361 & 0.2706 \\
o4-mini & \textbf{0.9667} & 0.5943 & 0.7361 & 0.4131 & 0.3164 & 0.3584 \\ 
GPT-4o & 0.8938 & \underline{0.6788} & \textbf{0.7716} & {\underline{0.4461}} & 0.332 & 0.3807 \\
GPT-4.1 & 0.907 & 0.646 & 0.7546 & \textbf{0.4632} & \textbf{0.4673} & \textbf{0.4653} \\
GPT-5 (high) & 0.913 & 0.5673 & 0.6998 & 0.4397 & \underline{0.4257} & \underline{0.4326} \\
\bottomrule
\end{tabular}
}
\caption{Results on the hard PEP benchmark to measure easy to hard generalization.}
\label{tab:hard_pep_benchmark_evaluation_full}
\end{table}

\begin{table}[!tbh]
\centering
\begin{tabular}{@{}lrrrrrr@{}}
\toprule
\multirow{2}{*}{\textbf{Model}} & \multicolumn{3}{r}{\textbf{Detection}} & \multicolumn{3}{r}{\textbf{Localization}} \\ \cmidrule(l){2-7} 
 & \textbf{$P_{Det}$} & \textbf{$R_{Det}$} & \textbf{$F_{Det}$} & \textbf{$P_{Loc}$} & \textbf{$R_{Loc}$} & \textbf{$F_{Loc}$} \\ \midrule
Qwen3-4B & 0.538 & 0.2637 & 0.3539 & 0.1396 & 0.1479 & 0.1436 \\
Qwen3-4B + SFT & \textbf{0.7686} & 0.3178 & 0.4497 & 0.2976 & 0.296 & 0.2968 \\
Qwen3-4B + SFT + RS-DPO & 0.7469 & \textbf{0.8315} & \textbf{0.7869} & \textbf{0.6527} & \textbf{0.6696} & \textbf{0.6611} \\
\midrule
Qwen3-4B w CoT & 0.8812 & 0.6854 & 0.771 & 0.5049 & 0.4878 & 0.4962 \\
Qwen3-4B w CoT + RS-SFT & \textbf{0.935} & 0.8183 & 0.8727 & 0.6639 & 0.65 & 0.6569 \\
Qwen3-4B w CoT + RS-SFT + RS-DPO & 0.9234 & \textbf{0.8643} & \textbf{0.8929} & \textbf{0.771} & \textbf{0.7571} & \textbf{0.764} \\
\midrule
Llama3.2-3B-Instruct & 0.5092 & 0.5286 & 0.5187 & 0.1371 & 0.3 & 0.1882 \\
Llama3.2-3B-Instruct + SFT & \textbf{0.6793} & 0.3598 & 0.4704 & 0.3424 & 0.3485 & 0.3454 \\
Llama3.2-3B-Instruct + SFT + RS-DPO & 0.654 & \textbf{0.6468} & \textbf{0.6504} & \textbf{0.491} & \textbf{0.4788} & \textbf{0.4848}
\\ \bottomrule
\end{tabular}
\caption{\textbf{Cross-Idiom Generalization on Python Ruff Best Practice Idioms:} We evaluate the effect of different \textsc{MetaLint} training setups (SFT, RS-SFT, and RS-DPO) on Qwen3-4B (with and without reasoning) and Llama3.2-3B. Models are trained on easy synthetic Python Ruff best practice idioms and tested on other Ruff best practice idioms with varying levels of transfer. Best score across the compared training setups per model is bolded.}
\label{tab:gen_synth_python_ruff_idioms_full}
\end{table}

\begin{table}[!tbh]
\centering
\begin{tabular}{@{}lrrrrrr@{}}
\toprule
\multirow{2}{*}{\textbf{Model}} & \multicolumn{3}{r}{\textbf{Detection}} & \multicolumn{3}{r}{\textbf{Localization}} \\ \cmidrule(l){2-7} 
 & \textbf{$P_{Det}$} & \textbf{$R_{Det}$} & \textbf{$F_{Det}$} & \textbf{$P_{Loc}$} & \textbf{$R_{Loc}$} & \textbf{$F_{Loc}$} \\ \midrule
Qwen3-4B & 0.5267 & 0.1715 & 0.2587 & 0.0954 & 0.0824 & 0.0884 \\
Qwen3-4B + SFT & 0.4333 & 0.0821 & 0.1381 & 0.0432 & 0.0221 & 0.0292 \\
Qwen3-4B + SFT + RS-DPO & \textbf{0.7031} & \textbf{0.7043} & \textbf{0.7037} & \textbf{0.3536} & \textbf{0.193} & \textbf{0.2497} \\
\midrule
Qwen3-4B w CoT & 0.8154 & 0.3986 & 0.5354 & 0.2625 & 0.1467 & 0.1882 \\
Qwen3-4B w CoT + RS-SFT & 0.7615 & 0.3689 & 0.497 & 0.2785 & 0.1437 & 0.1896 \\
Qwen3-4B w CoT + RS-SFT + RS-DPO & \textbf{0.9303} & \textbf{0.4958} & \textbf{0.6468} & \textbf{0.3482} & \textbf{0.2169} & \textbf{0.2673} \\
\midrule
Llama3.2-3B-Instruct & \textbf{0.7042} & 0.214 & 0.3283 & 0.0691 & 0.0798 & 0.0741 \\
Llama3.2-3B-Instruct + SFT & 0.5627 & 0.259 & 0.3547 & 0.1066 & 0.0509 & 0.0689 \\
Llama3.2-3B-Instruct + SFT + RS-DPO & 0.6368 & \textbf{0.5614} & \textbf{0.5965} & \textbf{0.2364} & \textbf{0.1263} & \textbf{0.1647} \\ \bottomrule
\end{tabular}
\caption{\textbf{Easy-to-Hard Generalization on PEP Best Practice Idioms:} We evaluate the effect of different \textsc{MetaLint} training setups (SFT, RS-SFT, and RS-DPO) on Qwen3-4B (with and without reasoning) and Llama3.2-3B. Models are trained on easy synthetic Python Ruff best practice idioms and tested on hard manually curated PEP best practice violation detection data which can't be handled by linters or static analyzers (section~\ref{sec:experiments:pep_idiom_benchmark}). Best score across the compared training setups per model are bolded.}
\label{tab:gen_easy_to_hard_pep_full}
\end{table}

\subsection{Statistical Significance of Results on the PEP Hard Best Practice Benchmark}
\label{sec:appendix:results:stat_signficance}
To analyze the statistical significance of performance differences over the PEP benchmark, we conduct Wilcoxon signed-rank tests comparing various \textsc{MetaLint} variants against each other and against baseline models. 
We evaluate instance-level detection accuracy (binary labels indicating whether the LLM correctly predicted the presence of a violation) as well as instance-level precision and recall for line-level localization. 
To control for multiple comparisons, we apply a Bonferroni correction to adjust the significance threshold $\alpha$ as $\alpha=\frac{0.05}{m}$ where $m$ is the number of comparisons (or rows in any given statistical significance table in this case).

Table~\ref{tab:metalint_stat_tests} reports the Wilcoxon signed-rank test statistic and corresponding $p$-value (in parentheses) for detection accuracy, localization precision, and localization recall when comparing various \textsc{MetaLint} variants to assess the effects of RS-DPO and CoT. We find that applying RS-DPO to the base SFT policy leads to statistically significant improvements in both detection and localization performance, with RS-DPO consistently outperforming the original SFT checkpoint across all three metrics with it being always better for localization. For the CoT variant, RS-DPO also yields consistent but less significant gains, likely because the RS-SFT CoT checkpoint is already relatively strong. Finally, we observe no statistically significant difference between the CoT (RS-SFT+RS-DPO) and the standard (SFT+RS-DPO) variant, suggesting that CoT does not provide a meaningful additional benefit in this setting.

Table~\ref{tab:metalint_vs_untrained} shows the statistical significance of comparing the base untrained model Qwen3-4B with its \textsc{MetaLint} variants (SFT and SFT+RS-DPO), and the Qwen3-4B CoT model with \textsc{MetaLint} w/ CoT (RS-SFT and RS-SFT+RS-DPO). The SFT variant yields significant gains in detection and localization recall, but not in localization precision. The SFT+RS-DPO model improves significantly across all three metrics. In contrast, training RS-SFT from the Qwen3-4B w/ CoT base does not yield significant improvements. However, the RS-SFT+RS-DPO variant produces significant gains in localization precision and recall, but not detection. These results suggest that while SFT alone offers limited generalization, combining it with DPO reliably improves localization and can significantly boost detection when starting from a weaker base model.

Table~\ref{tab:metalint_vs_baselines} shows the statistical significance results when comparing the \textsc{MetaLint} (SFT+RS-DPO) and \textsc{MetaLint} w CoT (RS-SFT+RS-DPO) variants against various baselines.
Here we want to higlight that \textsc{MetaLint} offers comparable performance across two out of three or all three metrics against several 32B models that outperform it like Qwen3-32B, Qwen3-32B w CoT, Qwen2.5Coder-32B and R1-Distill-Qwen-32B. Also the \textsc{MetaLint} non CoT (SFT+RS-DPO) variant has no singificant difference in performance compared to o3-mini, soldifying that \textbf{\textsc{MetaLint} without CoT has generalized to the point of being as capable as o3-mini} (even though the Qwen3-4B models without CoT and Qwen3-4B model with CoT perform worse than it with the difference being statistically singificant in Table~\ref{tab:qwen_vs_o3mini}).

\begin{table}[!tbh]
\centering
\begin{tabular}{@{}lccc@{}}
\toprule
\textbf{Model Comparison} & \textbf{Detection} & \textbf{Localization P} & \textbf{Localization R} \\
\midrule
Qwen3-4B vs o3-mini & 1266.5 (7.20e-21) & 743.5 (2.96e-11) & 739.0 (4.23e-09) \\
Qwen3-4B w CoT vs o3-mini & 921.5 (3.23e-09) & 2385.5 (9.61e-02) & 1891.0 (4.99e-04) \\
\bottomrule
\end{tabular}
\caption{Wilcoxon signed-rank test results comparing untrained Qwen3-4B variants with \texttt{o3-mini}, using Bonferroni-adjusted significance threshold $\alpha = 0.025$. Each cell reports the test statistic (p-value).}
\label{tab:qwen_vs_o3mini}
\end{table}

Table~\ref{tab:cot_effect} shows the effect of using a CoT for the Qwen3 model families and we notice that using a CoT leads to singificant gains for all metrics for the 4B and 8B models indicating that for smaller models CoTs might be essential for good performance on this task.
However the 14B and 32B model only show statistically significant improvement in localization precision with the CoT indicating that the CoT might offer limited benefit for larger models.

Table~\ref{tab:model_scale_effect} shows the effect of varying model scale for the Qwen3, Qwen2.5, Qwen2.5Coder, and DeepSeek-R1-Distill-Qwen families.
For Qwen3 we see benefits moving from 4B to 8B abd 8B to 14B but no statistically significant difference moving from 14B to 32B when not using a CoT.
Wehn using a CoT for Qwen3 we notice that the performance differences are rarely different in terms of statistical significant except for localizaiton performance between 4B and 8B and 8B and 14B.
For R1-Distill-Qwen family we notice a significant difference moving from 14B to 32B but not for 7B to 14B.
For the Qwen2.5Coder family we notice difference across all model scales, but the trend is weird with a big drop in performance from 3B to 7B and then a slow climb back to great performance around 32B.
We notice that for the Qwen2.5 family which shows relatively reasonable trends with model scale, the performance differences are statistically singificant execpt for the performance gain from 14B to 32B being significant only for recall.
To conclude the trends across model scales vary a lot across model families but in general the model size does help but differences may be smaller if the models are capable of reasoning and use a CoT.

Table~\ref{tab:gpt_model_comparison} shows comparison between the GPT models. We only compared GPT-4o and its successor GPT-4.1 and o3-mini against o4-mini and the results show that GPT-4.1 is only significantly better for localization recall while o4-mini is beter than o3-mini for overall localization but not for detection.

\begin{table}[!tbh]
\centering
\resizebox{\textwidth}{!}{%
\begin{tabular}{@{}lccc@{}}
\toprule
\textbf{Model Comparison} & \textbf{Detection} & \textbf{Localization P} & \textbf{Localization R} \\
\midrule
\textsc{MetaLint} (SFT) vs \textsc{MetaLint} (SFT+RS-DPO) & 7192.0 (1.92e-12) & 0.0 (2.49e-20) & 0.0 (2.92e-18) \\
\textsc{MetaLint} w CoT (RS-SFT) vs \textsc{MetaLint} w CoT (RS-SFT+RS-DPO) & 839.5 (2.18e-03) & 740.0 (3.95e-03) & 523.0 (2.34e-05) \\
\textsc{MetaLint} (SFT) vs \textsc{MetaLint} w CoT (RS-SFT) & 528.0 (6.91e-14) & 11.0 (1.38e-15) & 113.0 (7.95e-12) \\
\textsc{MetaLint} (SFT+RS-DPO) vs \textsc{MetaLint} w CoT (RS-SFT+RS-DPO) & 8140.0 (5.55e-01) & 2568.5 (8.42e-01) & 2544.0 (4.44e-01) \\
\bottomrule
\end{tabular}
}
\caption{Wilcoxon signed-rank test results comparing MetaLint variants. Each cell reports test statistic (p-value). All the \textsc{MetaLint} models are trained Qwen3-4B variants. We use the Bonferroni corrected significance threshold $\alpha=0.0125$.}
\label{tab:metalint_stat_tests}
\end{table}

\begin{table}[!tbh]
\centering
\resizebox{\textwidth}{!}{%
\begin{tabular}{@{}lccc@{}}
\toprule
\textbf{Model Comparison} & \textbf{Detection} & \textbf{Localization P} & \textbf{Localization R} \\
\midrule
Qwen3-4B vs Qwen3-4B \textsc{MetaLint} (SFT) & 560.5 (8.64e-03) & 363.5 (1.82e-02) & 238.0 (4.85e-04) \\
Qwen3-4B vs Qwen3-4B \textsc{MetaLint} (SFT+RS-DPO) & 7260.0 (4.13e-09) & 411.0 (1.99e-15) & 979.5 (7.30e-09) \\
Qwen3-4B w CoT vs Qwen3-4B \textsc{MetaLint} w CoT (RS-SFT) & 1224.0 (7.22e-01) & 937.0 (6.12e-01) & 918.5 (5.39e-01) \\
Qwen3-4B w CoT vs Qwen3-4B \textsc{MetaLint} w CoT (RS-SFT+RS-DPO) & 1728.0 (1.83e-02) & 1011.0 (1.53e-03) & 966.0 (1.02e-03) \\
\bottomrule
\end{tabular}
}
\caption{Wilcoxon signed-rank test results comparing \textsc{MetaLint} models against their untrained counterparts, with Bonferroni-adjusted significance threshold $\alpha = 0.0125$. Each cell reports the test statistic (p-value).}
\label{tab:metalint_vs_untrained}
\end{table}

\begin{table}[!tbh]
\centering
\resizebox{\textwidth}{!}{%
\begin{tabular}{@{}lccc@{}}
\toprule
\textbf{Model Comparison} & \textbf{Detection} & \textbf{Localization P} & \textbf{Localization R} \\
\midrule
Qwen3-8B vs \textsc{MetaLint} (SFT+RS-DPO) & 8140.5 (7.21e-03) & 1309.5 (2.19e-08) & 2067.0 (2.12e-03) \\
Qwen3-8B w CoT vs \textsc{MetaLint} w CoT (RS-SFT+RS-DPO) & 2070.0 (9.17e-01) & 1974.5 (1.88e-01) & 2161.0 (6.58e-01) \\
Qwen3-14B vs \textsc{MetaLint} (SFT+RS-DPO) & 7304.0 (4.96e-01) & 2816.5 (3.15e-02) & 3159.0 (2.88e-02) \\
Qwen3-14B w CoT vs \textsc{MetaLint} w CoT (RS-SFT+RS-DPO) & 2392.0 (2.78e-01) & 2749.5 (1.26e-01) & 2319.0 (1.32e-03) \\
Qwen3-32B vs \textsc{MetaLint} (SFT+RS-DPO) & 7175.0 (4.48e-01) & 3262.0 (1.93e-02) & 2818.5 (5.56e-03) \\
Qwen3-32B w CoT vs \textsc{MetaLint} w CoT (RS-SFT+RS-DPO) & 2677.5 (9.95e-03) & 3479.0 (8.38e-02) & 3180.5 (5.64e-04) \\
R1-Distill-Qwen-7B vs \textsc{MetaLint} (SFT+RS-DPO) & 8244.0 (2.65e-08) & 555.0 (8.12e-15) & 1924.0 (1.91e-04) \\
R1-Distill-Qwen-7B vs \textsc{MetaLint} w CoT (RS-SFT+RS-DPO) & 2907.0 (7.07e-10) & 915.5 (1.04e-12) & 1569.5 (8.64e-06) \\
R1-Distill-Qwen-14B vs \textsc{MetaLint} (SFT+RS-DPO) & 9877.0 (3.99e-06) & 2582.5 (9.36e-06) & 3085.0 (2.00e-03) \\
R1-Distill-Qwen-14B vs \textsc{MetaLint} w CoT (RS-SFT+RS-DPO) & 2660.0 (9.11e-08) & 1703.0 (2.98e-06) & 1791.0 (3.97e-05) \\
R1-Distill-Qwen-32B vs \textsc{MetaLint} (SFT+RS-DPO) & 8677.5 (2.51e-01) & 5767.5 (1.93e-01) & 3705.0 (6.67e-06) \\
R1-Distill-Qwen-32B vs \textsc{MetaLint} w CoT (RS-SFT+RS-DPO) & 3125.0 (3.11e-02) & 3641.5 (6.35e-02) & 2175.0 (4.99e-06) \\
Qwen2.5-3B vs \textsc{MetaLint} (SFT+RS-DPO) & 8001.0 (1.41e-15) & 0.0 (1.24e-22) & 68.5 (1.26e-19) \\
Qwen2.5-3B vs \textsc{MetaLint} w CoT (RS-SFT+RS-DPO) & 949.0 (4.96e-23) & 0.0 (4.24e-22) & 0.0 (9.89e-20) \\
Qwen2.5-7B vs \textsc{MetaLint} (SFT+RS-DPO) & 7312.5 (3.37e-10) & 208.0 (1.44e-18) & 610.0 (1.99e-12) \\
Qwen2.5-7B vs \textsc{MetaLint} w CoT (RS-SFT+RS-DPO) & 1187.5 (1.14e-14) & 226.0 (2.60e-18) & 406.5 (5.69e-14) \\
Qwen2.5-14B vs \textsc{MetaLint} (SFT+RS-DPO) & 8677.5 (2.51e-01) & 3045.5 (5.70e-04) & 4006.0 (3.83e-01) \\
Qwen2.5-14B vs \textsc{MetaLint} w CoT (RS-SFT+RS-DPO) & 4123.0 (4.86e-01) & 3228.0 (1.51e-03) & 4383.0 (9.89e-01) \\
Qwen2.5-32B vs \textsc{MetaLint} (SFT+RS-DPO) & 8640.0 (1.76e-04) & 1492.5 (3.12e-08) & 2971.5 (4.55e-02) \\
Qwen2.5-32B vs \textsc{MetaLint} w CoT (RS-SFT+RS-DPO) & 1792.0 (8.16e-06) & 1166.0 (2.43e-08) & 1983.5 (4.71e-03) \\
Qwen2.5Coder-3B vs \textsc{MetaLint} (SFT+RS-DPO) & 11184.0 (8.64e-03) & 953.5 (3.73e-12) & 1716.0 (3.00e-07) \\
Qwen2.5Coder-3B vs \textsc{MetaLint} w CoT (RS-SFT+RS-DPO) & 3683.5 (6.45e-03) & 1126.0 (4.39e-11) & 1403.5 (1.32e-08) \\
Qwen2.5Coder-7B vs \textsc{MetaLint} (SFT+RS-DPO) & 8001.0 (1.41e-15) & 0.0 (7.69e-23) & 0.0 (2.02e-20) \\
Qwen2.5Coder-7B vs \textsc{MetaLint} w CoT (RS-SFT+RS-DPO) & 949.0 (4.96e-23) & 0.0 (2.61e-22) & 0.0 (6.55e-20) \\
Qwen2.5Coder-14B vs \textsc{MetaLint} (SFT+RS-DPO) & 9123.5 (1.04e-12) & 289.0 (8.19e-20) & 736.0 (7.32e-14) \\
Qwen2.5Coder-14B vs \textsc{MetaLint} w CoT (RS-SFT+RS-DPO) & 1112.0 (1.82e-19) & 159.5 (7.63e-20) & 408.5 (3.02e-15) \\
Qwen2.5Coder-32B vs \textsc{MetaLint} (SFT+RS-DPO) & 6833.5 (2.86e-01) & 4500.0 (9.59e-01) & 2651.5 (7.07e-05) \\
Qwen2.5Coder-32B vs \textsc{MetaLint} w CoT (RS-SFT+RS-DPO) & 1039.5 (1.16e-02) & 2655.0 (9.39e-01) & 1235.5 (1.44e-05) \\
o3-mini vs \textsc{MetaLint} (SFT+RS-DPO) & 7520.0 (4.83e-02) & 4986.0 (5.20e-01) & 4427.5 (2.87e-01) \\
o3-mini vs \textsc{MetaLint} w CoT (RS-SFT+RS-DPO) & 1944.0 (7.15e-04) & 3169.0 (5.16e-01) & 2683.0 (4.23e-01) \\
\bottomrule
\end{tabular}
}
\caption{Wilcoxon signed-rank test statistics and p-values comparing MetaLint variants against baseline models. All the \textsc{MetaLint} variants are Qwen3-4B variants and Qwen2.5 and Qwen2.5Coder variants are instruction tuned checkpoints. We use the Bonferroni corrected significance threshold $\alpha=0.0017$.}
\label{tab:metalint_vs_baselines}
\end{table}

\begin{table}[!tbh]
\centering
\resizebox{\textwidth}{!}{%
\begin{tabular}{@{}lccc@{}}
\toprule
\textbf{Model Comparison} & \textbf{Detection} & \textbf{Localization P} & \textbf{Localization R} \\
\midrule
Qwen3-4B vs Qwen3-4B w CoT & 1260.0 (3.99e-08) & 425.0 (1.91e-09) & 624.5 (2.25e-05) \\
Qwen3-8B vs Qwen3-8B w CoT & 1924.0 (4.27e-03) & 1132.5 (8.69e-06) & 1005.0 (1.15e-04) \\
Qwen3-14B vs Qwen3-14B w CoT & 1691.0 (2.01e-01) & 2127.0 (4.27e-04) & 2398.5 (1.26e-01) \\
Qwen3-32B vs Qwen3-32B w CoT & 1572.5 (2.75e-01) & 1767.0 (6.35e-06) & 2596.5 (7.31e-02) \\
\bottomrule
\end{tabular}
}
\caption{Wilcoxon signed-rank test results measuring the effect of Chain-of-Thought (CoT) prompting across Qwen3 model scales. Each cell reports test statistic (p-value). We use the Bonferroni corrected significance threshold $\alpha=0.0125$.}
\label{tab:cot_effect}
\end{table}

\begin{table}[!tbh]
\centering
\resizebox{\textwidth}{!}{%
\begin{tabular}{@{}lccc@{}}
\toprule
\textbf{Model Comparison} & \textbf{Detection} & \textbf{Localization P} & \textbf{Localization R} \\
\midrule
Qwen3-4B vs Qwen3-8B & 546.0 (2.35e-08) & 618.0 (1.37e-04) & 572.5 (1.72e-03) \\
Qwen3-8B vs Qwen3-14B & 1350.0 (2.11e-03) & 1503.5 (3.96e-05) & 1008.5 (2.69e-07) \\
Qwen3-14B vs Qwen3-32B & 875.0 (2.22e-02) & 3129.5 (7.94e-01) & 2061.0 (4.14e-01) \\
Qwen3-4B w CoT vs Qwen3-8B w CoT & 1468.5 (1.90e-02) & 1578.5 (1.07e-01) & 1081.5 (2.60e-03) \\
Qwen3-8B w CoT vs Qwen3-14B w CoT & 904.5 (1.40e-01) & 1248.0 (1.66e-03) & 1099.5 (1.82e-05) \\
Qwen3-14B w CoT vs Qwen3-32B w CoT & 850.0 (3.78e-02) & 1834.5 (4.87e-01) & 2396.0 (4.40e-01) \\
R1-Distill-Qwen-7B vs R1-Distill-Qwen-14B & 4278.0 (2.67e-01) & 1431.0 (4.66e-03) & 1962.5 (6.01e-01) \\
R1-Distill-Qwen-14B vs R1-Distill-Qwen-32B & 2432.0 (1.44e-12) & 1475.5 (8.07e-11) & 843.5 (6.38e-15) \\
Qwen2.5Coder-3B vs Qwen2.5Coder-7B & 1541.0 (4.56e-14) & 0.0 (3.46e-10) & 0.0 (7.07e-10) \\
Qwen2.5Coder-7B vs Qwen2.5Coder-14B & 8.0 (7.89e-04) & 0.0 (2.04e-03) & 0.0 (2.14e-03) \\
Qwen2.5Coder-14B vs Qwen2.5Coder-32B & 423.0 (2.83e-27) & 100.5 (2.00e-22) & 43.5 (9.29e-23) \\
Qwen2.5-3B vs Qwen2.5-7B & 18.0 (2.43e-08) & 0.0 (3.43e-05) & 0.0 (3.58e-05) \\
Qwen2.5-7B vs Qwen2.5-14B & 960.5 (1.70e-13) & 500.5 (1.65e-11) & 574.0 (2.62e-10) \\
Qwen2.5-14B vs Qwen2.5-32B & 1925.0 (1.87e-04) & 1858.5 (6.60e-03) & 2108.5 (1.43e-02) \\
\bottomrule
\end{tabular}
}
\caption{Wilcoxon signed-rank test results measuring the effect of increasing model scale across families and CoT settings. Each cell shows the test statistic (p-value). All Qwen2.5 and Qwen2.5Coder variants are instruction-tuned checkpoints. We use the Bonferroni corrected significance threshold $\alpha=0.0036$.}
\label{tab:model_scale_effect}
\end{table}

\begin{table}[!tbh]
\centering
\begin{tabular}{@{}lccc@{}}
\toprule
\textbf{Model Comparison} & \textbf{Detection} & \textbf{Localization P} & \textbf{Localization R} \\
\midrule
GPT-4o vs GPT-4.1 & 2550.0 (9.21e-01) & 3961.0 (5.83e-01) & 2207.5 (4.48e-06) \\
o3-mini vs o4-mini & 575.0 (6.68e-01) & 1171.0 (1.44e-03) & 1079.0 (1.47e-04) \\
\bottomrule
\end{tabular}
\caption{Wilcoxon signed-rank test results comparing GPT model variants. Each cell shows the test statistic (p-value). We use the Bonferroni corrected significance threshold $\alpha=0.025$.}
\label{tab:gpt_model_comparison}
\end{table}

\subsection{Failure Analysis of MetaLint CoT Model VS Non CoT Model}
\label{sec:appendix:results:cot_model_failure_analysis}
We observe that a significant portion of the lower detection recall of the CoT \textsc{MetaLint} Qwen3-4B model, relative to its non CoT counterpart, can be attributed to its higher tendency to predict \texttt{NO VIOLATIONS FOUND} in cases that do, in fact, contain violations. Specifically, the CoT model fails to flag violations in 89 additional instances compared to the non CoT model, amounting to nearly 17\% of the evaluation set (89 out of 536 examples).

\begin{figure}[!tbh]
    \centering
    \includegraphics[width=0.8\textwidth]{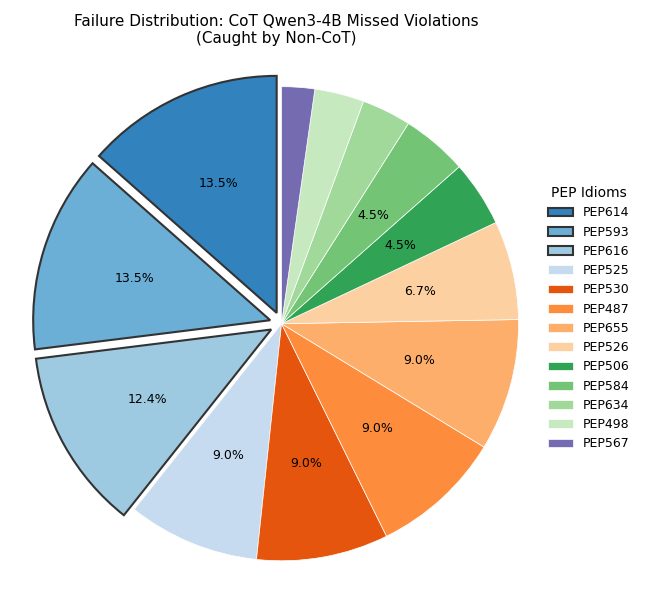}
    \caption{Distribution of comparative failures of the CoT \textsc{MetaLint} Qwen3-4B model relative to its non-CoT variant. While errors span a long tail across many PEPs, the majority are concentrated in three: PEP614, PEP593, and PEP616, which motivates our focused analysis on these cases.}
    \label{fig:cot_model_failure_pep_dist}
\end{figure}

The idiom wise distribution of these missed violations is shown in Figure~\ref{fig:cot_model_failure_pep_dist}. While the failure distribution follows a somewhat long tail pattern, the most significant drops occur for PEP 614, PEP 616, and PEP 593. Notably, if the CoT model matched the non CoT model’s performance on just these three PEPs, its detection recall would rise to 0.605, surpassing that of all open source baselines evaluated.

Upon inspecting CoT traces for these and other idioms (see examples in Table~\ref{tab:cot_model_failure_CoT_egs}), we identify several recurring failure modes:  
1) Ambiguity in interpreting the idiom specification. For example, in PEP 614, which targets decorators with complex expressions, the CoT model often labels expressions that humans consider complex as simple.  
2) Overthinking and repetitive reasoning traces, particularly for PEP 616.  
3) Skipping or entirely missing lines that contain violations, again observed in PEP 616.  
4) Underspecified idioms. For instance, in PEP 593, which recommends using the \texttt{Annotated} type from the \texttt{typing} module to attach metadata to type hints, the spec lacks clarity and concrete examples, making it hard to learn what constitutes a violation.

We also find similar issues in idioms like PEP 487, which discourages the use of metaclasses for simple customization tasks that could be handled via \texttt{\_\_init\_subclass\_\_} or \texttt{\_\_set\_name\_\_}. The CoT model often misclassifies such ``simple'' use cases as complex.

Overall, these patterns suggest that the CoT model applies the idiom specifications more conservatively, resulting in higher precision but at the cost of reduced recall.

\begin{table}[!tbh]
\centering
\resizebox{\textwidth}{!}{%
\begin{tabular}{@{}lll@{}}
\toprule
\textbf{PEP} & \textbf{Chain of Thought} & \textbf{Reason for Failure} \\ \midrule
614 & \begin{tabular}[c]{@{}l@{}}... I need to check the code file for violations of the 614 idiom, \\ which is about using complex expressions in decorators. ...\\ ... The register variable is defined on line 7 as partial(register, \\ server=hippa). So, the decorator is a partial function. ...\\ ... the idiom allows for complex expressions in the decorator. \\ For example, if the decorator was something like \\ @lambda func: timed(...)(func), that would be allowed. ...\\ ... the register is a partial function, which is a simple expression. ...\\ ... the decorator is just a variable that's a partial function. ...\end{tabular} & \begin{tabular}[c]{@{}l@{}}Ambiguity about what constitutes \\ a complex expression. The idiom \\ targets decorators with complex \\ expressions that make code \\ more verbose\end{tabular} \\ \midrule
525 & \begin{tabular}[c]{@{}l@{}}... The user ... wants me to check for ... idiom, 525, which \\ is about using asynchronous generators with "async for" \\ syntax ... the violation would be if the code uses \\ synchronous generators (like regular for loops) instead of \\ asynchronous generators (using async for) ... Wait, the \\ example given in the idiom is a class that uses \_\_aiter\_\_ \\ and \_\_anext\_\_ to create an asynchronous iterator. The fix \\ is to use a function with async def and yield. So, I need to \\ check if there are any instances where the code is using \\ synchronous generators (like for loops) instead of \\ asynchronous ones. ... Looking through the code, I \\ don't see any for loops that are using 'async for'. ...\end{tabular} & \begin{tabular}[c]{@{}l@{}}CoT accidentally looks for the \\ incorrect construct at a crucial \\ point (async for instead of for)\end{tabular} \\ \midrule
487 & \begin{tabular}[c]{@{}l@{}}... using \_\_init\_subclass\_\_ and \_\_set\_name\_\_ for class \\ customization without metaclasses ... using metaclasses \\ for simple customization tasks makes code unnecessarily \\ complex ... there's a class called FirestoreClientMeta, \\ which is a metaclass ... used for registration of transports \\ ... the problem is whether this is a simple customization ... \\ not just for a single customization but for managing \\ multiple transports ... the idiom is about using metaclasses \\ for simple customization ... but in this case ... handling a \\ more complex task ... maybe this is not a violation ...\end{tabular} & \begin{tabular}[c]{@{}l@{}}Ambiguity around what constitutes \\ a simple class. The idiom targets \\ instances where  metaclasses \\ are used for simple customization \\ tasks that could have been handled \\ with "\_\_init\_subclass\_\_" or \\ "\_\_set\_name\_\_")\end{tabular} \\ \bottomrule
\end{tabular}
}
\caption{Example chains of thought for various PEPs where the CoT model incorrectly flags \texttt{NO VIOLATIONS FOUND} instead of the non CoT model.}
\label{tab:cot_model_failure_CoT_egs}
\end{table}

\begin{table}[!tbh]
\centering
\resizebox{\textwidth}{!}{%
\begin{tabular}{@{}lrrrrrrr@{}}
\toprule
\multirow{2}{*}{\textbf{Model}} & \multirow{2}{*}{\textbf{Transfer}} & \multicolumn{3}{r}{\textbf{Detection}} & \multicolumn{3}{r}{\textbf{Localization}} \\ \cmidrule(l){3-8} 
 &  & \textbf{$P_{Det}$} & \textbf{$R_{Det}$} & \textbf{$F_{Det}$} & \textbf{$P_{Det}$} & \textbf{$R_{Det}$} & \textbf{$F_{Det}$} \\ \midrule
Llama3.2-3B-Instruct & \multirow{3}{*}{PMD $\rightarrow$ PMD} & 0.0457 & 0.0079 & 0.0134 & 0.0015 & 0.0022 & 0.0017 \\
Llama3.2-3B-Instruct + SFT & & 0.2251 & 0.4421 & 0.2983 & 0.2822 & 0.2778 & 0.28 \\
Llama3.2-3B-Instruct + SFT + RS-DPO &  & 0.4395 & 0.8908 & 0.5886 & 0.593 & 0.5969 & 0.5949 \\ \midrule
Llama3.1-8B-Instruct & \multirow{3}{*}{PMD $\rightarrow$ PMD} & 0.3656 & 0.4015 & 0.3827 & 0.1253 & 0.131 & 0.1281 \\
Llama3.1-8B-Instruct + SFT & & 0.2264 & 0.4508 & 0.3014 & 0.3201 & 0.3152 & 0.3177 \\
Llama3.1-8B-Instruct + SFT + RS-DPO &  & 0.4427 & 0.9191 & 0.5976 & 0.6506 & 0.6709 & 0.6606 \\ \midrule
Llama3.2-3B-Instruct & \multirow{3}{*}{PMD $\rightarrow$ JEP} & 0.3855 & 0.0096 & 0.0187 & 0.0005 & 0.0004 & 0.0005 \\
Llama3.2-3B-Instruct + SFT & & 0.2286 & 0.4072 & 0.2928 & 0.1626 & 0.1336 & 0.1467 \\
Llama3.2-3B-Instruct + SFT + RS-DPO & & 0.4903 & 0.8338 & 0.6175 & 0.4216 & 0.3333 & 0.3721 \\ \midrule
Llama3.1-8B-Instruct & \multirow{3}{*}{PMD $\rightarrow$ JEP} & 0 & 0 & 0 & 0 & 0 & 0 \\
Llama3.1-8B-Instruct + SFT & & 0.2166 & 0.3724 & 0.2739 & 0.1455 & 0.1142 & 0.128 \\
Llama3.1-8B-Instruct + SFT + RS-DPO & & 0.4964 & 0.8047 & 0.614 & 0.4615 & 0.3395 & 0.3912 \\ \midrule
Llama3.2-3B-Instruct & \multirow{3}{*}{JEP $\rightarrow$ JEP} & 0.3855 & 0.0096 & 0.0187 & 0.0005 & 0.0004 & 0.0005 \\
Llama3.2-3B-Instruct + SFT &  & 0.9567 & 0.8411 & 0.8952 & 0.7837 & 0.754 & 0.7686 \\
Llama3.2-3B-Instruct + SFT + RS-DPO &  & 0.9406 & 0.86 & 0.8985 & 0.7859 & 0.7651 & 0.7753 \\ \midrule
Llama3.1-8B-Instruct & \multirow{3}{*}{JEP $\rightarrow$ JEP} & 0 & 0 & 0 & 0 & 0 & 0 \\
Llama3.1-8B-Instruct + SFT &  & 0.9658 & 0.8466 & 0.9023 & 0.809 & 0.7844 & 0.7965 \\
Llama3.1-8B-Instruct + SFT + RS-DPO &  & 0.9308 & 0.8686 & 0.8986 & 0.8131 & 0.7756 & 0.7939 \\ \midrule
Llama3.2-3B-Instruct & \multirow{3}{*}{JEP $\rightarrow$ PMD} & 0.0457 & 0.0079 & 0.0134 & 0.0015 & 0.0022 & 0.0017 \\
Llama3.2-3B-Instruct + SFT & & 0.3722 & 0.2708 & 0.3152 & 0.0574 & 0.0869 & 0.0692 \\
Llama3.2-3B-Instruct + SFT + RS-DPO &  & 0.4322 & 0.4054 & 0.4183 & 0.0878 & 0.1222 & 0.1022 \\ \midrule
Llama3.1-8B-Instruct & \multirow{3}{*}{JEP $\rightarrow$ PMD} & 0.3656 & 0.4015 & 0.3827 & 0.1253 & 0.131 & 0.1281 \\
Llama3.1-8B-Instruct + SFT &  & 0.3514 & 0.2229 & 0.2728 & 0.0383 & 0.0753 & 0.0508 \\
Llama3.1-8B-Instruct + SFT + RS-DPO &  & 0.436 & 0.4898 & 0.4613 & 0.0831 & 0.1351 & 0.1029 \\ \bottomrule
\end{tabular}
}
\caption{\textbf{Cross-Idiom Generalization on JEP \& PMD Best Practice Idioms:} Effect of different \textsc{MetaLint} training setups (SFT and RS-DPO) on Llama3.2-3B-Instruct (Table~\ref{tab:gen_synth_jep_pmd_experiments}). The transfer column indicates training and test data on the left and right side of the arrow. Best score across the compared training setups per model are bolded.}
\label{tab:gen_synth_jep_pmd_experiments}
\end{table}

\begin{table}[!tbh]
\centering
\renewcommand{\arraystretch}{1.2}
\resizebox{\textwidth}{!}{%
\begin{tabular}{@{}l p{0.65\linewidth} r@{}}
\toprule
\textbf{Error Type} & \textbf{Definition} & \textbf{Ocurrences (\%)} \\ \midrule
Adjacent &
Predicts a line immediately adjacent to the true violating line, but not the correct one. &
12\% \\
Before/After &
Flags a line that follows the best practice rather than the line that violates it. &
4\% \\
Definition vs.\ Initialization &
Flags an initialization instead of the definition that actually constitutes the violation. &
4\% \\
Plausible-Looking &
Flags a superficially plausible but ultimately incorrect line due to surface-level similarity or insufficient semantic nuance. &
46\% \\
Random &
Selects an apparently arbitrary line, suggesting a collapse in localization despite correct detection. &
28\% \\
Repetition &
Repeats the same incorrect line multiple times, indicating a decoding or sampling artifact. &
6\% \\ \bottomrule
\end{tabular}
}
\caption{\textbf{Egregious Localization Error Types:} Breakdown of recurring error categories observed in a manual analysis of 50 egregious localization failures made by the Qwen3-4B non-CoT model trained with SFT+RS-DPO. An egregious localization error is defined as predicting a non-empty set of line numbers that is completely disjoint from the non-empty ground-truth set, indicating correct detection but failed localization.}
\label{tab:egregious_localization_errors}
\end{table}

\section{Code Repository}
\label{sec:appendix:code_repository}
The supplementary code repository contains all scripts and data necessary to reproduce the experiments in this paper. The repository is organized as follows. Training and evaluation data for Ruff Python idioms is provided under \texttt{data/ruff\_meta\_linting/}, with train/test splits that will be publicly released on HuggingFace upon acceptance; the manually curated PEP hard best-practice benchmark, including code files, ground-truth line annotations, few-shot examples, and idiom specifications, is under \texttt{data/pep\_benchmark/} and \texttt{data/pep\_idiom\_specs/}. Model training uses the alignment-handbook framework; per-model SFT and DPO configuration files are in \texttt{alignment-handbook/recipes/}. Inference scripts for both the Ruff transfer evaluation and the PEP benchmark are in \texttt{src/model/}, and the corresponding detection and localization metric scripts are in \texttt{src/metrics/}. DPO preference pair construction from multi-sample outputs is handled by \texttt{src/dpo/convert\_dpo\_samples\_to\_pairs.py}. A step-by-step walkthrough of the full training and evaluation workflow — from SFT through DPO to PEP benchmark evaluation — is provided in the \texttt{README.md}.

\end{document}